\documentclass{article}

\usepackage{arxiv}

\usepackage[utf8]{inputenc} 
\usepackage[T1]{fontenc}    
\usepackage{hyperref}       
\usepackage{url}            
\usepackage{booktabs}       
\usepackage{amsfonts}       
\usepackage{nicefrac}       
\usepackage{microtype}      
\usepackage{amsthm}
\usepackage{mathrsfs}
\usepackage{bm}
\usepackage{mathtools}
\usepackage{graphics}
\usepackage{dutchcal}
\usepackage{enumitem}
\usepackage[dvipsnames]{xcolor}
\usepackage{pdflscape}

\def\y{\mathbf{y}}
\def\yo{\y_0}
\def\yy{\bm{\mathit{y}}}
\def\x{\mathbf{x}}
\def\xo{\x_0}
\def\xs{\x^\ast}

\def\v{\mathbf{v}}

\def\h{\mathbf{h}}

\def\u{\mathbf{u}}
\def\us{\u^\ast}
\def\uo{\u_0}

\def\N{\mathbb{N}}
\def\Nor{\mathcal{N}}
\def\O{\mathcal{O}}
\def\Op{\mathcal{O}_p}
\def\op{\mathcal{o}_p}
\def\z{\mathbf{z}}
\def\B{\mathcal{B}}
\def\g{\mathbf{g}}
\def\gc{\mathbcal{g}}
\def\h{\mathbf{h}}
 
\def\E{\mathbb{E}}
\def\a{\mathbf{a}}
\def\b{\mathbf{b}}
\def\J{\mathbf{J}}
\def\K{\mathbf{K}}
\def\Ks{\mathcal{K}}

\def\M{\mathbf{M}}
\def\w{\mathbf{w}}
\def\wo{\w_0}
\def\Q{\mathbf{Q}}
\def\U{\mathcal{U}}

\def\G{\mathcal{G}}

\def\uQ{\underline{Q}}
\def\oQ{\overline{Q}}
\def\f{\mathbf{f}}
\def\bI{\mathbf{I}}
\def\X{\mathbf{X}}

\def\P{\mathbf{P}}
\def\H{\mathbf{H}}

\def\D{\mathbf{D}}
\def\C{\mathcal{C}}
\def\bC{\mathbf{C}}

\def\bA{\mathbf{A}}

\def\bS{\bm{\Sigma}}

\def\bt{\bm{\theta}}

\def\hbt{\hat{\bt}}

\def\bto{\bt_0}

\def\bT{\bm{\Theta}}

\def\bO{\bm{\Omega}}

\def\bp{\bm{\psi}}
\def\bP{\bm{\Psi}}
\def\bvP{\bm{\varphi}}
\def\bpi{\bm{\pi}}
\def\hbpi{\hat{\bpi}}

\def\bpio{\bpi_0}
\def\bPi{\bm{\Pi}}

\def\bx{\bm{\xi}}

\def\bX{\bm{\Xi}}

\def\be{\bm{\epsilon}}
\def\bb{\bm{\beta}}
\def\hbb{\hat{\bb}}
\def\bbo{\bb_0}
\def\bD{\bm{\Delta}}
\def\Xs{\bT^\circ}
\def\Ys{\bPi^\circ}

\def\R{{\rm I\!R}}
\def\wcp{\overset{p}{\rightarrow}}
\def\0{\mathbf{0}}
\def\eqd{\overset{d}{=}}
\def\iid{identically and independently distributed}
\def\1{\mathbf{1}}
\def\d{\mathop{}\!\mathrm{d}}

\DeclareMathOperator{\diag}{diag}
\DeclareMathOperator*{\argzero}{argzero}
\DeclareMathOperator*{\argmin}{argmin}
\DeclareMathOperator*{\erf}{erf}
\DeclareMathOperator{\id}{id}
\DeclareMathOperator{\rank}{rank}
\DeclareMathOperator{\trace}{trace}
\DeclareMathOperator{\sign}{sign}
\DeclareMathOperator{\vect}{vec}

\newtheorem{remark}{Remark}
\newcounter{dummy}
\newtheorem{theorem}[dummy]{Theorem}
\newtheorem{lemma}[dummy]{Lemma}
\newtheorem{corollary}[dummy]{Corollary}
\newtheorem{proposition}[dummy]{Proposition}
\newtheorem{example}[dummy]{Example}
\newtheorem{definition}[dummy]{Definition}
\newtheorem{assumption}[dummy]{Assumption}

\def\boxit#1{\vbox{\hrule\hbox{\vrule\kern3pt
          \vbox{\kern3pt#1\kern3pt}\kern3pt\vrule}\hrule}}

\definecolor{pinegreen}{rgb}{0.0, 0.47, 0.44}

\title{A simple recipe for making accurate parametric inference in finite sample}

\author{%
  St\'ephane~Guerrier \\
  Department of Statistics\\
  Pennsylvania State University\\
  University Park, PA 16802, USA \\
  \texttt{szg279@psu.edu} \\
  \And%
  Mucyo~Karemera \\
  Department of Statistics\\
  Pennsylvania State University\\
  University Park, PA 16802, USA \\
  \texttt{mxk1257@psu.edu}
  \And%
  Samuel~Orso\\
  Geneva School of Economics and Management \\
  University of Geneva \\
  Geneva, Switzerland \\
  \texttt{Samuel.Orso@unige.ch}
  \And%
  Maria-Pia~Victoria-Feser \\
  Geneva School of Economics and Management \\
  University of Geneva \\
  Geneva, Switzerland \\
  \texttt{Maria-Pia.VictoriaFeser@unige.ch}
}

\begin{document}
\maketitle



\section{Introduction}
The algorithmic principle of the bootstrap method is quite simple: reiterate the mechanism
that produces an estimator on pseudo-samples. But when it comes to estimators that are
numerically complicated to obtain, 
the bootstrap is less attractive to use due to the numerical burden. 
If one estimator is hard to find, reiterating compounds this issue.
Paraphrasing Emile
in the French comedy \textit{La Cit\'e de la Peur}: we can implement the bootstrap when
the estimator is simple to obtain or we can compute a numerically complex point estimator,
but it is too computationally cumbersome to do both.

Although this limitation is purely practical and tends to be reduced by the ever increasing
computational power at our disposal, everyone would agree that it is nonetheless attractive
to have a method that frees the user from the computational burden, or at least provides 
an answer within a reasonable time.
In this chapter, we explore a special case of the efficient method of moments (\cite{gallant1996moments}) that encompasses both  
the computation of numerically complex estimators and of a ``bootstrap distribution'' at 
a reduced cost. The idea deviates from the algorithmic principle of the bootstrap:
the proposed method no longer attempts at reproducing the sample mechanism that lead to an estimator, 
but instead, tries to find every estimators that may have produced the observed sample,
or more often, some statistics on the sample.

The idea is not new though, several methods follow this pattern.
The indirect inference method (\cite{gourieroux1993indirect,smith1993estimating})
similarly attempts at finding the point estimate that lead to statistics obtained
from simulated samples as close as possible to the same statistics on the observed sample. 
Mostly used in econometric and financial contexts, indirect inference has been successfully applied
to the estimation of stable distribution (\cite{garcia2011estimation}), 
stochastic volatility models (\cite{monfardini1998estimating,lombardi2009indirect}),
financial contingent claims (\cite{phillips2009simulation}),
dynamic panel models (\cite{gourieroux2010indirect}),
dynamic stochastic equilibrium models (\cite{dridi2007indirect}),
continuous time models (\cite{gallant2010simulated}),
diffusion processes (\cite{broze1998quasi});
but it has also been used in queueing theory (\cite{heggland2004estimating}),
robust estimation of generalized linear latent variable models (\cite{moustaki2006bounded}),
robust income distribution (\cite{guerrier2018simulation}),
high dimensional generalized linear model and penalized regression (\cite{guerrier2018properties}).
Often presented as the Bayesian counterpart of the indirect inference,
the approximate Bayesian computation (\cite{tavare1997inferring,pritchard1999population})
aims at finding the values that match the 
statistics computed on simulated samples and the statistics on the observed sample, with a certain degree approximation.
The method has however grown in a different context of applications.
For example, it has been successfully employed in 
population genetics (\cite{beaumont2002approximate}), in ecology (\cite{beaumont2010approximate}),
in evolutionary biology (\cite{cornuet2008inferring,wilkinson2010dating}).
Less popular, R.A. Fisher's fiducial inference 
(see for instance~\cite{fisher1922mathematical,fisher1930inverse,fisher1933concepts,fisher1935fiducial,fisher1956statistical})
and related methods such as the generalized fiducial inference (\cite{hannig2009generalized,hannig2013generalized,hannig2016generalized}),
D.A.S. Fraser's structural inference (\cite{fraser1968structure}, see also~\cite{dawid1973marginalization}),
Dempster-Shafer theory (\cite{shafer1976mathematical,dempster2008dempster}) and inferential models
(\cite{martin2013inferential,martin2015conditional,martin2015plausibility})
follow a similar pattern, the main idea being to find all possible values
that permit to generate simulated sample as close as possible to the observed sample,
but without specifying any prior distribution.

Regardless of the difference in philosophy of the aforementioned methods, they
have in common that they are usually very demanding in computational resources when implemented
for non-trivial applications.
This is a major difference with the approach we endorse in this chapter.
By letting the statistics be the solution of an estimating function of the same
dimension as the quantity of interest, we demonstrate that it is possible
to bypass the computation of the same statistics on simulated sample
by directly estimating the quantity of interest within the estimating function,
resulting thereby in a potential significant gain in computational time.
In Section~\ref{ch2:sec:equiv}, we demonstrate in finite sample that under some weak conditions
the estimators resulting from our approach is equivalent to the estimators
one would have obtained using certain forms of indirect inference, approximate
Bayesian computation or fiducial inference approaches, whereas it is
different than parametric bootstrap estimators, except in the case of a location parameter.
This section innovates on two aspects. First, it implicates that our approach
can be employed in practice to solve problems that relate to indirect inference,
approximated Bayesian compuation and fiducial inference in a computationally efficient manner.
Second, it proves or disproves formally the link between the aforementioned methods, 
and this in the most general situation as the results remain true for any sample size.

Contructing tests or confidence regions that controls over the error rates in the long-run 
is probably one of the most important problem in statistics ever since at least 
Neyman-Pearson famous article \cite{neyman1933ix}.
Yet, the theoretical justification for most methods in statistics is asymptotic.
The bootstrap for example, despite its simplicity and its widespread usage is 
an asymptotic method (\cite{hall1992boot}); for the other methods, 
see for example~\cite{frazier2018asymptotic} for 
approximate Bayesian computation,~\cite{gourieroux1993indirect} for indirect inference
and~\cite{hannig2016generalized} for generalized fiducial inference.
There are in general no claim about the exactness of the inferential procedures
in finite sample (see~\cite{martin2015plausibility} for one of the exceptions).
In Section~\ref{ch2:sec:fs}, we study theoretically the frequentist error rates of
confidence regions constructed on the distribution issued from our proposed approach.
In particular, we demonstrate under some strong, but frequently encountered, conditions that
the confidence regions have exact coverage probabilities in finite sample.
Asymptotic justification is nonetheless provided in Section~\ref{ch2:sec:ap}.
In addition, we bear the comparison with the asymptotic properties of indirect inference method
to conclude that, surprisingly, both approaches reach the same conclusion
but under distinct conditions. Some leads are evoked,
but we lack to elucidate the fundamental reason behind such discrepancy.

Although the proposed method is first and foremost computational, surprisingly in some situations 
explicit closed-form solutions may be found. We gather a non-exhaustive number of such examples, some important,
in Section~\ref{ch2:sec:st}. The numerical study in Section~\ref{ch2:sec:sim} ends this chapter.
We study via Monte Carlo simulations the coverage probabilities obtained from our approach
and compare with others on a variety of problems. We conclude that in most situations, exact coverage probability
computed within a reasonable computational time can be claimed with our method.

\section{Setup}
Let $\N$ ($\N^+$) be the sets of all positive integers including (excluding) 0.
For any positive integer $n$, let $\N_n$ be the set whose elements are the integers 
$0,1,2,\dots,n$; similarly $\N^+_n=\{1,2,\dots,n\}$. 

We consider a sequence of random variables $\{\x_i:i\in\N^+_n\}$, possibly multivariate, 
to follow an assumely known distribution $F_{\bt}$,
indexed by a vector of parameters $\bt\in\bT\subset\R^p$.
We suppose that it is easy to generate artificial samples $\xs$ from $F_{\bt}$.
Specifically, we generate the random variable $\x$ with a known algorithm
that associates $\bt$ and a random variable $\u$. We denote the generating
mechanism as follows:
\begin{equation*}
	\x = \g(\bt,\u).
\end{equation*}
The random variable $\u$ follows a known model $F_{\u}$ that does not depend on $\bt$.
Using this notation, the observed sample is $\xo = \g(\bto,\uo)$ and the 
artificial sample is $\xs = \g(\bt,\us)$, where $\uo$ and $\us$ are realizations of $\u$.
\begin{example}[Normal]\label{ch2:ex:gaus}
	Suppose $\x\sim\Nor(\theta,1)$, then four examples of possible generating mechanism are:
	\begin{enumerate}
		\item $\g(\bt,\u) = \bt + \u$ where $\u\sim\Nor(0,1)$,
		\item $\g(\bt,\u) = \bt + \sqrt{2}\erf^{-1}(2\u-1)$ where $\u\sim\U(0,1)$ 
			and $\erf(z)=\frac{2}{\sqrt{\pi}}\int_0^ze^{-t^2}\d t$ is the error function,
		\item $\g(\bt,\u) = \bt + \sqrt{-2\ln(\u_{1})}\cos(2\pi \u_{2})$ where 
			$\u=(\u_1,\u_2)^T$, $\u_{1}\sim\U(0,1)$ and $\u_{2}\sim\U(0,1)$,
		\item $\g(\bt,\u) = \bt + \u_{2}\sqrt{\frac{-2\ln(\u_{3})}{\u_{3}}}$ where $\u=(\u_{1},\u_{2},\u_{3})$, 
			$\u_{3}=\u_{1}+\u_{2}$, $\u_{1}\sim\U(0,1)$, $\u_{2}\sim\U(0,1)$.
	\end{enumerate}
	A possible counter-example is the following: $\g(\bt,\u) = \u - \bt$ where $\u\sim\Nor(2\bt,1)$.
	Clearly $\x=\g(\bt,\u)$, but this $\g$ is not adequate because the distribution of $\u$
	depends on $\bt$. 
\end{example}
We now define the estimators we wish to study.
\begin{definition}[SwiZs]\label{ch2:def:swizs2}
	We consider the following sequence of estimators:
	\begin{equation*}
		\hbpi_n\in\bPi_n = \argzero_{\bpi\in\bPi}\frac{1}{n}\sum_{i=1}^n\bp\left(\g\left(\bto,\u_{0i}\right),\bpi\right)
				 = \argzero_{\bpi\in\bPi}\bP_n\left(\bto,\uo,\bpi\right),
	\end{equation*}
	\begin{equation*}
		\hbt_n^{(s)}\in\bT^{(s)}_n = \argzero_{\bt\in\bT}\frac{1}{n}\sum_{i=1}^n\bp\left(\g\left(\bt,\us_{si}\right),\hbpi_n\right)
				    	   = \argzero_{\bt\in\bT}\bP_n\left(\bt,\us_s,\hbpi_n\right),
	\end{equation*}
	where $\bp$ is an estimating function and $s\in\N^+_S$.
	The estimators $\hbpi_n$ are referred as the \textit{auxiliary estimators}.
	Any sequence of estimators $\{\hbt_n^{(s)}:s\in\N^+_S\}$ is called \textit{Switched Z-estimators}, 
	or in short, \textit{SwiZs}.
	The collection of the solutions is $\bT_n=\cup_{s\in\N^+_S}\bT^{(s)}_n$.
\end{definition}
\begin{remark}
	The SwiZs in the Definition~\ref{ch2:def:swizs2} may arguably be viewed as a special
	case of the Efficient Method of Moment (EMM) estimator proposed by~\cite{gallant1996moments}.
	Indeed, to have an EMM estimator the only modification to the Definition~\ref{ch2:def:swizs2}
	is 
	\begin{equation*}
		\hbt_{\text{EMM},n}^{(s)}\in\bT^{(s)}_{\text{EMM},n}
		= \argzero_{\bt\in\bT}\frac{1}{H}\sum_{h=1}^H
		\bP_n\left(\bt,\us_{sh},\hbpi_n\right),
	\end{equation*}
	where $H\in\N^+$. Ergo, the SwiZs and EMM coincide whenever $H=1$.
	Note that in general the EMM is defined with $H$ large and $S=1$.
\end{remark}

\section{Equivalent methods}\label{ch2:sec:equiv}
As already remarked, the SwiZs does not appear to be a new estimator.
The SwiZs in fact offers a new point of view to different existing methods
as it federates several techniques under the same hat.
In this Section, we show the equivalence or disequivalence of the SwiZs to other existing
methods, for any sample size $n$, to conclude that the distribution obtained by the 
SwiZs is (approximatively) a Bayesian posterior, and thereby that it is valid for the purpose
of inference.

The EMM and the indirect inference estimator of~\cite{smith1993estimating,gourieroux1993indirect}
are known to have the same asymptotic distribution when $\dim(\bpi)=\dim(\bt)$ 
(see Proposition 4.1 in~\cite{gourieroux1996simulation}).
In the next result, we demonstrate that the SwiZs and a certain form of indirect inference
estimator are equivalent for any $n$.
\begin{definition}[indirect inference estimators]\label{ch2:def:iie}
	Let $\hbpi_n$ and $\{\u_j:j\in\N\}$ be defined as in the Definition~\ref{ch2:def:swizs2}.
	We consider the following sequence of estimators, for $s\in\N^+_S$:
	\begin{equation*}
		\hbpi^{(s)}_{\text{II},n}(\bt)\in\bPi^{(s)}_{\text{II},n} 
		= \argzero_{\bpi\in\bPi}\bP_n\left(\bt,\us_s,\bpi\right),\quad\bt\in\bT,
	\end{equation*}
	\begin{equation*}
		\hbt^{(s)}_{\text{II},n}\in\bT^{(s)}_{\text{II},n}
		= \argzero_{\bt\in\bT} d\left(\hbpi_n,\hbpi_{\text{II},n}^{(s)}(\bt)\right),
		\quad\hbpi_n\in\bPi_n,\quad\hbpi_{\text{II},n}^{(s)}\in\bPi_n^{(s)},
	\end{equation*}
	where $d$ is a metric. 
	We call $\{\hbt_{\text{II},n}^{(s)}:s\in\N^+_S\}$ the \textit{indirect inference estimators}.
	The collections of solutions are denoted $\bPi_{\text{II},n}=\cup_{s\in\N^+_S}\bPi_{\text{II},n}^{(s)}$ and 
	$\bT_{\text{II},n}=\cup_{s\in\N^+_S}\bT_{\text{II},n}^{(s)}$.
\end{definition}
\begin{remark}
	In Definition~\ref{ch2:def:iie}, we are implicitly assuming that $\bT$
	contains at least one of, possibly many zeros, of the distance between the
	auxiliary estimators on the sample and the pseudo-sample. Therefore, the 
	theory is the same for any measure of distance that we denote generically by $d$.
\end{remark}
\begin{remark}
	The indirect inference estimators in Definition~\ref{ch2:def:iie} is a special
	case of the more general form
	\begin{equation*}
		\hbt^{(s)}_{\text{II},B,m}\in\bT^{(s)}_{\text{II},B,m}
		= \argzero_{\bt\in\bT} d\left(\hbpi_n,\frac{1}{B}\sum_{b=1}^B\hbpi_{\text{II},b,m}^{(s)}(\bt)\right),
	\end{equation*}
	$B\in\N^+$, $m\geq n$. In Definition~\ref{ch2:def:iie} we fixed $B=1$ and $m=n$.
	\cite{gourieroux1993indirect} considered two cases: first, $B$ large, $m=n$ and $S=1$,
	second, $B=1$, $m$ large and $S=1$. For both cases, the $\ell_2$-norm was used as 
	the measure of distance (see the preceding remark).
\end{remark}
\begin{assumption}[uniqueness]\label{ch2:ass:uniq}
	For all $(\bt,s)\in\bT\times\N_S$, $\argzero_{\bpi\in\bPi}\bP_n(\bt,\u_s,\bpi)$ has a unique solution
\end{assumption}
\begin{theorem}[Equivalence SwiZs/indirect inference]\label{ch2:thm:equiv}
	If Assumption~\ref{ch2:ass:uniq} is satisfied, then the following holds for any $s\in\N^+_S$:
	\begin{equation*}
		\bT_n^{(s)}=\bT_{\text{II},n}^{(s)}.
	\end{equation*}
\end{theorem}
Theorem~\ref{ch2:thm:equiv} is striking because it concludes that a certain form of EMM, the SwiZs, and 
indirect inference estimators (as in Definition~\ref{ch2:def:iie}) are actually the very same estimators,
not only asymptotically, but for any sample size, and under a very mild condition.
Indeed, Assumption~\ref{ch2:ass:uniq} requires the roots of the estimating function to 
be well separated so there exists a unique solution. This requirement is unrestrictive
and it is typically satisfied.
One may even wonder what would be the purpose
of an estimating function for which Assumption~\ref{ch2:ass:uniq} would not hold.
In this spirit, Assumption~\ref{ch2:ass:uniq} may be qualified as the ``minimum 
criterion'' for choosing an estimating function.

Even if the optimizer is perfect, Theorem~\ref{ch2:thm:equiv} does not
imply that the exact same values are found using the SwiZs or the indirect inference estimators,
but that they belong to the same set of solutions, and thereby that they share the same
statistical properties.
Hence, Theorem~\ref{ch2:thm:equiv} offers us two different ways of computing the 
same estimators. Simple calculations however show that the SwiZs is computationally more
attractive. Indeed, if we let $k$ denotes the cost evaluation of $\bP_n$, $l$ the numbers of evaluations
of $\bP_n$ for obtaining an auxiliary estimator or the final estimator, then the SwiZs has
a total cost of roughly $\O(2kl)$ whereas it is $\O(kl+kl^2)$ for the indirect inference estimator,
so a reduction in order of $\O(kl^2)$. This computational efficiency of the SwiZs accounts
for the fact that it is not necessary to compute $\hbpi_{\text{II},n}$, and thus avoids
the numerical problem of the indirect inference estimator of having an optimization nested within an optimization. 
This discrepancy is also, quite surprisingly, reflected in the theory we develop in Section~\ref{ch2:sec:fs}
for the finite sample properties and in Section~\ref{ch2:sec:ap} for the asymptotic properties.

At first glance, the SwiZs may appear similar to the parametric bootstrap
(see the Definiton~\ref{ch2:def:pb} below).
If we strengthen our assumptions and think of the auxiliary estimator as an unbiased
estimator of $\bt$, it is natural to think of the SwiZs and the parametric bootstrap as being equivalent.
In any cases, both methods use the exact same ingredients, so we may wonder whether 
actually they are the same. The next result demonstrates that in fact, they will 
be seldom equivalent.
\begin{definition}[parametric bootstrap]\label{ch2:def:pb}
	Let $\hbpi_n$ and $\{\u_j:j\in\N\}$ be defined as in Definition~\ref{ch2:def:swizs2}.
	We consider the following sequence of estimators:
	\begin{equation*}
		\hbt_{\text{Boot},n}^{(s)}\in\bT^{(s)}_{\text{Boot},n}=\argzero_{\bt\in\bT}
		\bP_n\left(\hbpi_n,\us_s,\bt\right), \quad s\in\N^+_S.
	\end{equation*}
	The collection of the solutions is $\bT_{\text{Boot},n}=\cup_{s\in\N^+_S}\bT^{(s)}_{\text{Boot},n}$.
\end{definition}
\begin{remark}\label{ch2:rem:pb}
	For the solutions $\bT^{(s)}_{\text{Boot},n}$ in Definition~\ref{ch2:def:pb}
	to be nonempty, the parametric bootstrap requires that $\bPi_n\subset\bT$.
	The SwiZs has not such requirement.
\end{remark}
\begin{assumption}\label{ch2:ass:zero}
	The zeros of the estimating functions are symmetric on $(\bt,\bpi)$, that is
	\begin{equation*}
		\bP_n(\bt,\u_s,\bpi)=\bP_n(\bpi,\u_s,\bt)=\0.
	\end{equation*}
\end{assumption}
\begin{theorem}[equivalence SwiZs/parametric bootstrap]\label{ch2:thm:pb}
	If and only if Assumption~\ref{ch2:ass:zero} is satisfied, then
	it holds that
	\begin{equation*}
		\bT^{(s)}_n = \bT^{(s)}_{\text{Boot},n}.
	\end{equation*}
\end{theorem}
Assumption~\ref{ch2:ass:zero} is very restrictive, so Theorem~\ref{ch2:thm:pb} suggests
that in general the SwiZs and the parametric bootstrap are not equivalent. This may appear
as a surprise as only the argument $\bt$ and $\bpi$ are interchanged in the estimating function.
Then, if they are different, the question of which one should be preferred naturally arises.
We do not attempt at answering this question, but we rather prefer to stimulate debates by giving motivations for using the SwiZs.
Popularized by~\cite{efron1979bootstrap}, the bootstrap has been a long-standing technique for (frequentist) statistician, 
it is relatively straightforward to implement and has a well-established theory (see for instance~\cite{hall1992boot}).
On the other hand, although the idea of the SwiZs has been arguably around for decades (see the comparison
with the fiducial inference at the end of this section),
we lack evidence of its widespread usage, at least not under the form presented here.
When facing situations where $\hbpi_n$ is an unbiased estimator of $\bto$, compared to the parametric
bootstrap, the SwiZs is more demanding for the implementation and is generally less numerically efficient
(see Section~\ref{ch2:sec:sim}) suggesting that solving $\bP_n(\bt,\bpi)$ in $\bt$ 
is computationally more involved than in $\bpi$. However, in all the other situations where for example
$\hbpi_n$ may be an (asymptotically) biased estimator of $\bto$, a sample statistic or a consistent
estimator of a different model, the parametric bootstrap cannot be invoked directly, at least not
with the same form as in Definition~\ref{ch2:thm:pb}. Indeed, the parametric bootstrap
requires $\hbpi_n$ to be a consistent estimator of $\bto$. Therefore, when considering complex model
for which a consistent estimator is not readily available at a reasonable cost, the SwiZs may be
computationally more attractive. The rest of this section aims at demonstrating that
the distribution of the SwiZs is valid for the purpose of inference, whereas the following section
theorizes the inferential properties of the SwiZs in finite sample for which
Sections~\ref{ch2:sec:st} and~\ref{ch2:sec:sim} gather evidences. But before, having
emphasized their differences, we would like to share a rather common problem on which the 
parametric bootstrap and the SwiZs are equivalent.

The condition under which the SwiZs and the parametric bootstrap are equivalent (Assumption~\ref{ch2:ass:zero}) is very strong
and generally not met. There is one situation however where this condition holds,
if the inferential problem is on the parameter of a location family as formalized in the 
next Proposition~\ref{ch2:thm:loc}.
\begin{proposition}[equivalence SwiZs/parametric bootstrap in location family problems]\label{ch2:thm:loc}
	Suppose that $x$ is a univariate random variable \iid\; according to a
	location family, that is $x\eqd\theta+y$, where $\theta\in\R$ is the location
	parameter. If the auxiliary parameter is estimated by the sample average and $x$ 
	is symmetric around 0, that is $x\eqd-x$, then
	\begin{equation*}
		\bT^{(s)}_n = \bT^{(s)}_{\text{Boot},n}.
	\end{equation*}
\end{proposition}
The conditions which satisfies Proposition~\ref{ch2:thm:loc} are restrictive. Indeed,
they are satisfied for location families for which the centered random
variable is symmetric. Proposition~\ref{ch2:thm:loc} holds for example with a Gaussian,
a Student, a Cauchy and a Laplace random variables (variance and degrees of freedom known), but not,
for example, for a generalized extreme value, a skewed Laplace and a skewed $t$ random variables (even with non-location parameters being fixed).
The proof uses an average as the auxiliary estimator, but it should be easily extended to other
estimator of location such as the trimmed mean.
Proposition~\ref{ch2:thm:loc} is illustrated with a Cauchy random variable in Example~\ref{ch2:ex:cauchy} of
Section~\ref{ch2:sec:st}.

Although the parametric bootstrap and the SwiZs will lead rarely to the same estimators,
in spite of the similitude of their forms, the next result demonstrates that the distribution
of the SwiZs corresponds in fact to (some sort of) a Bayesian posterior.
Likewise the indirect inference, the approximate Bayesian computation (ABC) techniques
were proposed to respond to complex problems. The two techniques are often presented
to be respectively the frequentist and the Bayesian approaches to a same problem and have even
been mixed sometimes (see~\cite{drovandi2015bayesian}). We now show under what
conditions the SwiZs and the ABC are equivalent, but before, we need to give more precision
on what type of ABC\@. Often dated back to~\cite{diggle1984monte}, the ABC has evolved and covers now 
a broad-spectrum of techniques such as rejection sampling (see e.g.~\cite{tavare1997inferring,pritchard1999population}),
the Markov chain  Monte Carlo (see e.g.~\cite{marjoram2003markov,bortot2007inference}),
the sequential Monte Carlo sampling (see e.g.~\cite{sisson2007sequential,beaumont2009adaptive,toni2009approximate}) among others 
(see~\cite{marin2012approximate} for a review). The equivalence between the SwiZs and the
ABC is demonstrated with a rejection sampling presented in the next definition.
However, the note of~\cite{sisson2010note} suggests that this result may be extended
to Markov chain Monte Carlo and sequential Monte Carlo sampling algorithms.
We leave such rigorous demonstration for further research.
\begin{definition}[Approximate Bayesian Computation (ABC) estimators]\label{ch2:def:abc}
	Let $\hbpi_n$ and $\{\u_j:j\in\N\}$ be defined as in Definition~\ref{ch2:def:swizs2}.
	Let $\hbpi^{(s)}_{\text{II},n}(\bt)$ be defined as in Definition~\ref{ch2:def:iie}.
	We consider the following algorithm. For a given $\varepsilon\geq0$, for a given
	infinite sequence $\{\u_s:s\in\N^+_S\}$, for a given infinite sequence of empty sets
	$\{\bT^{(s)}_{\text{ABC},n}(\varepsilon):s\in\N^+_S\}$,
	for a given prior distribution $\mathcal{P}$ of $\bt$, 
	repeat (indefinitely) the following steps:
	\begin{enumerate}
		\item Generate $\bt^\star\sim\mathcal{P}$.
		\item Compute $\hbpi_{\text{II},n}^{(s)}\left(\bt^\star\right)$.
		\item If the following criterion is satisfied
			\begin{equation*}
				d\left(\hbpi_n,\hbpi^{(s)}_{\text{II},n}(\bt^\star)\right)\leq\varepsilon,
			\end{equation*}
			add $\bt^\star$ to the set $\bT^{(s)}_{\text{ABC},n}$, i.e. 
			$\bT^{(s)}_{\text{ABC},n}(\varepsilon)=\bT^{(s)}_{\text{ABC},n}(\varepsilon)\cup\{\bt^\star\}$.
	\end{enumerate}
	For a given $s\in\N^+_S$, we denote by $\hbt^{(s)}_{\text{ABC},n}(\varepsilon)$ an element of $\bT^{(s)}_{\text{ABC},n}(\varepsilon)$.
	The collection of the solutions is denoted 
	$\bT_{\text{ABC},n}(\varepsilon)=\cup_{s\in\N^+}\bT^{(s)}_{\text{ABC},n}(\varepsilon)$.
\end{definition}
\begin{remark}
	The ABC algorithm presented in Definition~\ref{ch2:def:abc} is a specific version of the
	simple accept/reject algorithm proposed by~\cite{tavare1997inferring,pritchard1999population},
	where the auxiliary estimators are the solution of an estimating function and the dimensions
	of $\bpi$ and $\bt$ are the same.
\end{remark}
\begin{definition}[posterior distribution]\label{ch2:def:post}
	The distribution of the infinite sequence $\{\hbt^{(s)}_{\text{ABC},n}(\varepsilon):s\in\N^+_S\}$
	issued from Definition~\ref{ch2:def:abc}
	is referred to as the $(\varepsilon,\hbpi_n)$-approximate posterior distribution.
	If $\varepsilon=0$, we have the $\hbpi_n$-approximate posterior distribution.
	If $\hbpi_n$ is a sufficient statistic, we have the $\varepsilon$-approximate posterior distribution.
	If both $\varepsilon=0$ and $\hbpi_n$ is sufficient, then we simply refer
	to the posterior distribution.
\end{definition}
\begin{remark}
	In Definition~\ref{ch2:def:post}, we mention two sources of approximation
	to the posterior distribution, $\varepsilon$ and $\hbpi_n$. 
	There is actually a third source of approximation stemming from the number of simulations $S$,
	if indeed $S<\infty$. Since it is common to every methods presented, it is left implicit.
\end{remark}
\begin{assumption}[existence of a prior]\label{ch2:ass:prior}
	For every $s\in\N^+_S$ and for all $n$, 
	there exists a prior distribution $\mathcal{P}$ such that
	\begin{equation*}
		\lim_{\varepsilon\downarrow 0}\Pr\left(d\left(\hbpi_n,\hbpi^{(s)}_{\text{II},n}(\bt^\star)\right)
		\leq\varepsilon\right)=1,\quad\bt^\star\sim\mathcal{P}.
	\end{equation*}
\end{assumption}
\begin{theorem}[Equivalence SwiZs/ABC]\label{ch2:thm:equiv2}
	If Assumptions~\ref{ch2:ass:uniq} and~\ref{ch2:ass:prior} are satisfied,
	then the following holds:
	\begin{equation*}
		\bT_n^{(s)} = \lim_{\varepsilon\downarrow 0}\bT_{\text{ABC},n}^{(s)}(\varepsilon).
	\end{equation*}
\end{theorem}

From Theorem~\ref{ch2:thm:equiv2} and Definition~\ref{ch2:def:post}, we have clearly
established that the distribution obtained by the SwiZs
is a $\hbpi_n$-approximate posterior distribution. 
Yet, the conclusion reached by Theorem~\ref{ch2:thm:equiv2} is surprising at two different levels:
first, Theorem~\ref{ch2:thm:equiv2} implies the possibility of obtaining an $\hbpi_n$-approximate
posterior distribution without specifying explicitly a prior distribution by using the SwiZs, second, 
whereas, for each $s\in\N^+_S$, it would in general require a very large number of sampled $\bt^\star$ for 
the ABC to approach an $\hbpi_n$-approximate posterior distribution ($\varepsilon=0)$, 
it is obtainable by the SwiZs at a much reduced cost. 
Indeed, for a given $s\in\N^+_S$, it demands in general a considerable number of attempts to sample
a $\bt^\star$ that satisfies the matching criterion with an error of $\varepsilon\approx0$,
whereas it is replaced by one optimization for the SwiZs, so it may be more computationally efficient
to use the SwiZs. Note also that in the situation where one has a prior knowledge on $\bt$,
the SwiZs may be modified, for example, by including an importance sampling weight,
in the same fashion that the ABC would be modified when the prior distribution is improper (see e.g.~\cite{del2006sequential}).
However, for some problems, the optimizations to obtain the SwiZs distribution 
may be numerically cumbersomes and the ABC may prove itself a facilitating alternative
(for example~\cite{fearnhead2012constructing} argued in this direction for some of their examples
when comparing the indirect inference and the ABC).

Switching between the SwiZS and the ABC algorithms for estimating a posterior poses the fundamental and
practical question of which prior distribution to use.
Assumption~\ref{ch2:ass:prior} stating that a prior distribution exists is very reasonable and 
widely accepted (although a frequentist fundamentalist may argue differently), but the result of Theorem~\ref{ch2:thm:equiv2}
brings at least three questions: which prior distribution satisfies both 
the SwiZs and the ABC at the same time, whether the prior distribution under 
which Theorem~\ref{ch2:thm:equiv2} holds is unique and 
whether there is an ``optimal'' prior in the numerical sense (that would produce $\bt^\star$ 
satisfying ``rapidly'' the matching criteria as defined at the point 3 of Definition~\ref{ch2:def:abc}).
We do not answer these questions because, firstly, the numerical problems we face 
in Section~\ref{ch2:sec:sim} are achievable quite efficiently by the SwiZs, secondly,
they would deserve much more attention than what 
we are able to conduct in the present. Thus, we content ourselves by mentioning only briefly studies
made on this direction. In order to approach this topic, we first need to present
an ultimate technique.

The possibility of obtaining an (approximate) Bayesian posterior without specifying 
explicitly a prior distribution on the parameters of interest
inescapably links the SwiZs to R.A. Fisher's controversial fiducial inference
(see for instance~\cite{fisher1922mathematical,fisher1930inverse,fisher1933concepts,fisher1935fiducial,fisher1956statistical}). 
Here we keep the SwiZs neutral and do not aim at reanimating any debate.
It is delicate to give an unequivocal definition of the fiducial inference as
it has changed on many occasion over time (see~\cite{zabell1992ra} for a comprehensive historical review)
and we rather give the presentation with the generalized fiducial inference
proposed by~\cite{hannig2009generalized} (see also~\cite{hannig2013generalized,hannig2016generalized})
which includes R.A. Fisher's fiducial inference. 
Other efforts to generalize R.A. Fisher's fiducial inference include
Fraser's structural inference (~\cite{fraser1968structure}, see also~\cite{dawid1973marginalization}),
the Dempster-Shafer theory (~\cite{shafer1976mathematical,dempster2008dempster}, see also~\cite{zhang2011dempster})
refined later with the concept of inferential models (\cite{martin2013inferential,martin2015conditional}).
As argued by~\cite{hannig2009generalized}, Fraser's structural inference
may be viewed as a special case of the generalized fiducial inference
where the generating function $\g$ has
a specific structure. The concept of inferential models is similar to the 
generalized fiducial inference in appearance but they differ in their respective theory.
The departure point of the inferential models is to conduct inference with the
conditional distribution of the pivotal quantity $\u$  
given $\xo$ after the sample has been observed. It is argued that keeping $\u\sim F_{\u}$ after
the sample has been observed makes the whole procedure subjective (\cite{martin2015conditional}),
but the idea is essentially a gain in efficiency of the estimators.
Also this idea is sound (see Lemma~\ref{ch2:thm:uniq} in the next section),
we do not see how it can be applied for the practical examples we use in Section~\ref{ch2:sec:sim},
and more fundamentally, we do not understand how such conditional distribution may be
built without some form of prior (and arguably subjective) knowledge on $\uo$. We
therefore leave such consideration for further research and limit the 
equivalence to the generalized fiducial inference given in the next definition.

\begin{definition}[Generalized fiducial inference]\label{ch2:def:gfd}
	The generalized fiducial distribution is given by 
	\begin{equation*}
		\hbt^{(s)}_{\text{GFD},n}\in\bT^{(s)}_{\text{GFD},n}
		= \argzero_{\bt\in\bT}d\left(\x,\g\left(\bt,\us_s\right)\right).
	\end{equation*}
\end{definition}
\begin{remark}
	The generalized fiducial distribution in Definition~\ref{ch2:def:gfd}
	is slightly more specific than usually defined in the literature.
	In Definition 1 in~\cite{hannig2016generalized}, it is given by
	\begin{equation*}
		\lim_{\varepsilon\downarrow0}\left[\argmin_{\bt\in\bT}\left\lVert\x-\g\left(\bt,\us_s\right)\right\rVert
			\Big\vert\min_{\bt}\left\lVert\x-\g\left(\bt,\us_s\right)\right\rVert\leq\varepsilon \right],
	\end{equation*}
	for any norm. Here, in addition, we assume that $\bT$ contains at least
	one of, possibly many, zeros.
\end{remark}
If we let the sample size equals the dimension of the parameter of interest, $n=p$,  then it is obvious
from their definitions that the generalized fiducial distribution and the indirect inference estimators are equivalent.
We formalize this finding for the sake of the presentation.
\begin{assumption}\label{ch2:ass:gfd}
	The followings hold:
	\begin{enumerate}[label=\roman*.]
		\item $\hbpi_n = \x$;
		\item $\hbpi_{\text{II},n}(\bt) = \g(\bt,\u)$.
	\end{enumerate}
\end{assumption}
\begin{proposition}\label{ch2:thm:equiv3}
	If Assumption~\ref{ch2:ass:gfd} is satisfied, then the following holds:
	\begin{equation*}
		\bT^{(s)}_{\text{II},n} = \bT^{(s)}_{\text{GFD},n}.
	\end{equation*}
\end{proposition}
Also the link between the indirect inference and the generalized fiducial inference seems
self-evident, it was, at the best of our knowledge, never mentioned in the literature.
It may be explained by the two different goals that each of these methods target,
that may respectively be loosely summarized as finding a point-estimate of a complex problem 
and making Bayesian inference without using a prior distribution. Having established this equivalence,
the connection with the SwiZs is direct from Theorem~\ref{ch2:thm:equiv} and
formalize in the next proposition.
\begin{proposition}\label{ch2:thm:equiv4}
	If Assumptions~\ref{ch2:ass:uniq} and~\ref{ch2:ass:gfd} are satisfied, then the following holds:
	\begin{equation*}
		\bT^{(s)}_n = \bT^{(s)}_{\text{GFD},n}.
	\end{equation*}
\end{proposition}
In the light of Proposition~\ref{ch2:thm:equiv4}, the SwiZs may appear equivalent
to the generalized fiducial inference under a very restrictive condition.
Indeed, the only possibility for Assumption~\ref{ch2:ass:gfd} to hold is 
that the sample size must equal the dimension of the problem.
But we would be willing to concede that this apparent rigidity is thiner as one may propose to use 
sufficient statistics with minimal reduction on the sample, thereby leaving $n$ greater than $p$,
and Proposition~\ref{ch2:thm:equiv3} would still hold. Such situation 
however is confined to problems dealing with exponential families 
as demonstrated by the Pitman-Koopman-Darmois theorem, so in general, when $n$ is greater
than $p$ and the problem at hand is outside of the exponential family, the SwiZs and
the generalized fiducial inference are not equivalent.

Although the link between the generalized fiducial inference and the indirect inference
has remained silent, the connection with the former to the ABC has been much more emphased.
Indeed, the algorithms 
proposed to solve the generalized fiducial inference problems are mostly borrowed from
the ABC literature (see~\cite{hannig2014computational}). Therefore, the discussion we 
conducted above on the numerical
aspects of the SwiZs and the ABC still holds here, the SwiZs may be an efficient
alternative to solve the generalized fiducial inference problem.

The generalized fiducial inference is also linked by~\cite{hannig2016generalized} to
what may be called ``non-informative'' prior approaches (see~\cite{kass1994formal} 
for a broad discussion of this concept).
More specifically, it appears that some distribution resulting  
from the generalized fiducial inference corresponds to the posterior distribution
obtained by~\cite{fraser2010default} based on a data-dependent prior proportional to the likelihood function
in the absence of information. This result enlarges the previous vision brought
by~\cite{lindley1958fiducial} that concluded that R.A. Fisher's fiducial inference
is ``Bayes inconsistent'' (in the sense that the Bayes' theorem cannot be invoked) apart
from problems on the Gaussian and the gamma distributions. \cite{lindley1958fiducial}'s
results relied on a narrower definition of fiducial inference than brought by the generalized fiducial
inference, so whether the generalized fiducial inference has become Bayes consistent
for broader problems nor~\cite{fraser2010default} approach with an uninformative
prior is Bayes inconsistent remains an open question. But most importantly, the strong
link between the generalized fiducial inference and this non-informative prior approach
reveals the common goal towards which of these approaches tends, which might be stated
as tackling the individual subjectivism in the Bayesian inference that has been one of the major
subject of criticism ever since at least~\cite{fisher1922mathematical}.

Last but not least, we complete the loop by the following Corollary which is a consequence
of Theorems~\ref{ch2:thm:equiv},~\ref{ch2:thm:pb} and~\ref{ch2:thm:equiv2}, and Propositions~\ref{ch2:thm:equiv3}
and~\ref{ch2:thm:equiv4}.
\begin{corollary}
	We have the followings:
	\begin{enumerate}[label=\roman*.]
		\item If Assumptions~\ref{ch2:ass:uniq} and~\ref{ch2:ass:prior} are satisfied, then
			$\bT_{\text{II},n}^{(s)} = \lim_{\varepsilon\downarrow 0}\bT_{\text{ABC},n}^{(s)}(\varepsilon)$;
		\item If Assumptions~\ref{ch2:ass:uniq}, \ref{ch2:ass:prior} and \ref{ch2:ass:zero} are satisfied, then
			$\bT_{\text{Boot},n}^{(s)} = \lim_{\varepsilon\downarrow 0}\bT_{\text{ABC},n}^{(s)}(\varepsilon)$;
		\item If Assumptions~\ref{ch2:ass:uniq} and~\ref{ch2:ass:zero} are satisfied, then
			$\bT_{\text{II},n}^{(s)} = \lim_{\varepsilon\downarrow 0}\bT_{\text{Boot},n}^{(s)}(\varepsilon)$;
		\item If Assumptions~\ref{ch2:ass:uniq}, \ref{ch2:ass:zero} and~\ref{ch2:ass:gfd} are satisfied, then
			$\bT_{\text{Boot},n}^{(s)} = \lim_{\varepsilon\downarrow 0}\bT_{\text{GFD},n}^{(s)}(\varepsilon)$;
		\item If Assumptions~\ref{ch2:ass:uniq}, \ref{ch2:ass:prior} and~\ref{ch2:ass:gfd} are satisfied, then
			$\bT_{\text{ABC},n}^{(s)} = \lim_{\varepsilon\downarrow 0}\bT_{\text{GFD},n}^{(s)}(\varepsilon)$.
	\end{enumerate}
\end{corollary}

\section{Exact frequentist inference in finite sample}\label{ch2:sec:fs}
Having demonstrated that the distribution of the SwiZs sequence, for a single experiment, is approximatively
a Bayesian posterior, we now turn our interest to the long-run statistical properties of the SwiZs.
Our point of view here is frequentist, that is 
we suppose that we have an indefinite number of independent trials with fixed sample size $n$
and fixed $\bto\in\bT$. For each experiment we compute an exact $\alpha$-credible set, as given
in the Definition~\ref{ch2:def:cred} below,
using the SwiZs independently: the knowledge acquired on an experiment is not used as a prior
to compute the SwiZs on another experiment. The goal of this Section is to demonstrate
under what conditions the SwiZs leads to exact frequentist inference when the sample size is fixed.

\begin{definition}[sets of quantiles]\label{ch2:def:quant}
	Let $F_{\hbt_n\vert\hbpi_n}$ be a $\hbpi_n$-approximate posterior cumulative distribution function.
	We define the following sets of quantiles:
	\begin{enumerate}
		\item Let $\uQ_\alpha=\left\{\hbt_n\in\bT_n,\alpha\in(0,1):F_{\hbt_n\vert\hbpi_n}(\hbt_n)\leq\alpha\right\}$ be the 
			set of all $\hbt_n$ for which $F_{\hbt_n\vert\hbpi_n}$ is below the threshold $\alpha$.
		\item Let $\oQ_\alpha=\left\{\hbt_n\in\bT_n,\alpha\in(0,1):F_{\hbt_n\vert\hbpi_n}(\hbt_n)\geq 1-\alpha\right\}$ be the
			set of all $\hbt_n$ for which $F_{\hbt_n\vert\hbpi_n}$ is above the threshold $1-\alpha$.
	\end{enumerate}
\end{definition}
\begin{definition}[credible set]\label{ch2:def:cred}
	Let $F_{\hbt_n\vert\hbpi_n}$ be a $\hbpi_n$-approximate posterior cumulative distribution function.
	A set $C_{\hbpi_n}$ is said to be an $\alpha$-credible set if
	\begin{equation}
		\Pr\left(\hbt_n\in C_{\hbpi_n}\vert\hbpi_n \right)\geq 1-\alpha, \quad\alpha\in(0,1),
		\label{def:cred:set:eq1}
	\end{equation}
	where 
	\begin{equation*}
		C_{\hbpi_n} = \bT_n\setminus\left\{\uQ_{\alpha_1}\cup\oQ_{\alpha_2}\right\},
		\quad \alpha_1+\alpha_2=\alpha.
	\end{equation*}
	If we replace ``$\geq$'' by the equal sign in~\eqref{def:cred:set:eq1}, we say that the coverage probability of $C_{\hbpi_n}$ is exact.
\end{definition}
Definition~\ref{ch2:def:cred} is standard in the Bayesian literature (see e.g.~\cite{robert2007bayesian}).
Note that an $\alpha$-credbile set can have an exact coverage only if the random variable
is absolutely continuous. Such credible set is referred to as an ``exact $\alpha$-credible set''.

The next result gives a mean to verify the exactness of frequentist coverage of an exact
$\alpha$-credible set.
\begin{proposition}[Exact frequentist coverage]\label{ch2:thm:exa}
	If a $\hbpi_n$-approximate posterior distribution evaluated at $\bto\in\bT_n$ 
	is a realization from a standard uniform variate \iid,
	$F_{\hbt_n\vert\hbpi_n}(\bto)=u$, $u\sim\U(0,1)$, 
	then every exact $\alpha$-credible
	set built from the quantiles of $F_{\hbt_n\vert\hbpi_n}$ 
	leads to exact frequentist coverage probability in the sense that
	$\Pr\left(C_{\hbpi_n}\ni\bto\right)=1-\alpha$ (unconditionally).
\end{proposition}
Proposition~\ref{ch2:thm:exa} states that if the cumulative distribution function (cdf),
obtained from the SwiZs, variates (across independent trials!) uniformly around $\bto$ (fixed!), 
so does any quantities computed from the percentiles of this cdf, leading to exact coverage
in the long-run. The proof relies on Borel's strong law of large number. 
Although this result may be qualified of unorthodox by mixing both Bayesian posterior and 
frequentist properties, it arises very naturally.
Replacing $\hbpi_n$-approximate posterior distribution by any conditional distribution on $\hbpi_n$
in Proposition~\ref{ch2:thm:exa} leads to the same result.
This proposition is similar in form to the concept of confidence distribution formulated 
by~\cite{schweder2002confidence} and later
refined by~\cite{singh2005combining,xie2011confidence,xie2013confidence}. 
The confidence distribution is however a concept entirely frequentist and could
not be directly exploited here.
The general theoretical studies on the finite sample frequentist properties are quite rare in the literature,
we should eventually mention the study of~\cite{martin2015plausibility},
although the theory developped is around inferential models and different than our, the author
uses the same criterion of uniformly distributed quantity to demonstrate the frequentist properties.

\begin{remark}\label{ch2:rem:exa}
	In Proposition~\ref{ch2:thm:exa}, we use a standard uniform variable as a mean to 
	verify the frequentist properties. With the current statement of the proposition, 
	other distributions with 
	support in $[0,1]$ may be candidates to verify the exactness of the frequentist coverage.
	However, if we restrain the frequentist exactness to be $\Pr(C_{\hbpi_n}\ni\bto)=1-\alpha$,
	$\Pr(\oQ_{\alpha_2}\ni\bto)=\alpha_2$ and $\Pr(\uQ_{\alpha_1}\ni\bto)=\alpha_1$, for 
	$\alpha = \alpha_1+\alpha_2$, then the uniform distribution would be the only candidate.
\end{remark}
In the light of Proposition~\ref{ch2:thm:exa}, we now give the conditions under which the distribution
of the sequence $\{\hbt_n^{(s)}:s\in\N^+\}$, $F_{\hbt_n\vert\hbpi_n}$, leads to exact frequentist
coverage probabilities. We begin with a lemma which is essential in the construction
of our argument.
\begin{lemma}\label{ch2:thm:uniq}
	If the mapping $\bpi\mapsto\bP_n$ has unique zero in $\bPi$
	and the mapping $\bt\mapsto\bP_n$ has unique zero in $\bT$,
	then the following holds
	\begin{equation*}
		\bto = \hbt_n = \argzero_{\bt\in\bT}\bP_n\left(\bt,\uo,\hbpi_n\right).
	\end{equation*}
\end{lemma}
The idea behind Lemma~\ref{ch2:thm:uniq} is that if one knew the
true pivotal quantity $\uo$ that generated the data,
then one could directly recover the true quantity of interest $\bto$
from the sample. Of course, both $\uo$ and $\bto$ are unknown (otherwise statisticians
would be an extinct species!), but
here we are exploiting the idea that, for a sufficiently large number
of simulations $S$, at some point we will generate $\u_s$
``close enough'' to $\uo$. This idea is reflected in the following assumption. 
\begin{assumption}\label{ch2:ass:piv}
	Let $\bT_n\subseteq\bT$ be the set of the solutions of the SwiZs in the Definition~\ref{ch2:def:swizs2}.
	We have the following:
	\begin{equation*}
		\bto\in\bT_n.
	\end{equation*}
\end{assumption}
%

The following functions are essential for convenient data reduction.
\begin{assumption}[data reduction]\label{ch2:ass:dr}
	We have:
	\begin{enumerate}[label=\roman*.]
		\item There exists a Borel measurable surjection such that $\b(\u)$ has the same dimension as $\x$. 
		\item There exists a Borel measurable surjection such that $\h\circ\b(\u)$ has the same dimension as $\bt$. 
	\end{enumerate}
\end{assumption}
\begin{remark}\label{ch2:rem:func}
	The function $\b$ allows to work with a random variable of the 
	same dimension as the observed variable. Indeed we have
	\begin{equation*}
		\x\eqd\g(\bt,\u)\eqd\gc\circ(\id_{\bT}\times\b)(\bt,\u)\eqd\gc(\bt,\v),
	\end{equation*}
	where $\v = \b(\u)$ has the same dimension as $\x$ and $\id_{\bT}$ is 
	the identity function on the set $\bT$. On the other hand,
	the function $\h$ allows us to deal with random variables of the same 
	dimension as $\bt$, and thus $\bpi$.	
\end{remark}
\begin{remark}\label{ch2:rem:func2}
	In Assumption~\ref{ch2:ass:dr}, by saying the functions $\h$ and $\b$ are Borel
	measurable, we want to emphasis thereby that after applying these functions we 
	still work with random variables, which is essential here.
\end{remark}
To fix ideas, we consider the following example:
\begin{example}[Explicit form for $\h$ and $\b$]\label{ch2:ex:hb}
	As in Example~\ref{ch2:ex:gaus}, suppose that $\x=x_1,\cdots,x_n$ is \iid\; according to $\Nor(\theta,\sigma^2)$,
	where $\sigma^2$ is known, and consider the generating function $\g\in\G$
	\begin{equation*}
		\g(\theta,\u,\sigma^2) = \theta + \sigma\sqrt{-2\ln(u_1)}\cos(2\pi u_2),
	\end{equation*}
	where $u_{1i},u_{2i}$, $i=1,\cdots,n$, are \iid\;according to $\U(0,1)$. Letting $\v \equiv \b(\u) = \sqrt{-2\ln(u_1)}\cos(2\pi u_2)$,
	we clearly have that $\v\sim\Nor(0,\bI_n)$ is a random variable of the same dimension as $\x$.
	Now, if we consider $\h$ as the function that averages its argument, we have 
	$w \equiv \h\circ\b(\u) = \nicefrac{1}{n}\sum_{i=1}^nv_i$, so by properties of Gaussian random variable
	we have that $w$ has a Gaussian distribution with mean 0 and variance $\nicefrac{1}{n}$. Since
	$w$ is a scalar, it has the same dimensions as $\theta$.
\end{example}
Example~\ref{ch2:ex:hb} shows explicit forms for functions in Assumption~\ref{ch2:ass:dr}. 
It is however not requested to have an explicit form as we will see.
Indeed, under Assumption~\ref{ch2:ass:dr}, we can construct the following
estimating function:
\begin{equation*}
	\bP_n\left(\bt,\us,\bpi\right) = \bvP_p\left(\bt,\w,\bpi\right),
\end{equation*}
where $\w = \h\circ\b(\us)$ is a $p$-dimensional random variable.
The index $p$ in the estimating function $\bvP_p$ aims at emphasing that 
$\w$ has the same dimensions as $\bt$ and $\bpi$, which is essential in our argument.
Since the sample size $n$ and dimension $p$ are fixed here, it is disturbing.
For some fixed $\bt_1\in\bT$ and $\bpi_1\in\bPi$, it clearly holds that:
\begin{align*}
	\hbpi_n &= \argzero_{\bpi\in\bPi}\bP_n\left(\bt_1,\us,\bpi\right) 
		= \argzero_{\bpi\in\bPi}\bvP_p\left(\bt_1,\w,\bpi\right), \\
	\hbt_n 	&= \argzero_{\bt\in\bT}\bP_n\left(\bt,\us,\bpi_1\right)
		= \argzero_{\bt\in\bT}\bvP_p\left(\bt,\w,\bpi_1\right).
\end{align*}
\begin{assumption}[characterization of $\bvP_p$]\label{ch2:ass:bvp}
	Let $\bT_n\subseteq\bT$ and $W_n$ be open subsets of $\R^p$.
	Let $\hbpi_n$ be the unique solution of $\bP_n(\bto,\uo,\bpi)$.
	Let $\bvP_{\hbpi_n}(\bt,\w)\equiv\bvP_p(\bt,\w,\hbpi_n)$ be the map where $\hbpi_n$ is fixed.
	We have the followings:
	\begin{enumerate}[label=\roman*.]
		\item $\bvP_{\hbpi_n}\in\C^{1}\left(\bT_n\times W_n,\R^p\right)$ is once continuously
			differentiable on $\left(\bT_n\times W_n\right)\setminus K_n$, where 
			$K_n\subset\bT_n\times W_n$ is at most countable,
		\item $\det\left(D_{\bt}\bvP_{\hbpi_n}(\bt,\w)\right)\neq0$, $\det\left(D_{\w}\bvP_{\hbpi_n}(\bt,\w)\right)\neq0$
			for every $(\bt,\w)\in\left(\bT_n\times W_n\right)\setminus K_n$,
		\item $\lim_{\lVert(\bt,\w)\rVert\to\infty}\left\lVert\bvP_{\hbpi_n}(\bt,\w)\right\rVert=\infty$.
	\end{enumerate}
\end{assumption}
\begin{assumption}[characterization of $\bvP_p$ II]\label{ch2:ass:bvp2}
	Let $\bT_n\subseteq\bT$, $W_n$ and $\bPi_n\subseteq\bPi$ be open subsets of $\R^p$.
	Let $\bvP_{\bt_1}(\w,\bpi)\equiv\bvP_p(\bt_1,\w,\bpi)$ be the map where $\bt_1\in\bT$ is fixed.
	Let $\bvP_{\w_1}(\bt,\bpi)\equiv\bvP_p(\bt,\w_1,\bpi)$ be the map where $\w_1\in W_n$ is fixed.
	We have the followings:
	\begin{enumerate}[label=\roman*.]
		\item $\bvP_{\bt_1}\in\C^{1}\left(W_n\times\bPi_n,\R^p\right)$ is once continuously
			differentiable on $\left(W_n\times\bPi_n\right)\setminus K_{1n}$, where 
			$K_{1n}\subset W_n\times\bPi_n$ is at most countable,
		\item $\bvP_{\w_1}\in\C^{1}\left(\bT_n\times\bPi_n,\R^p\right)$ is once continuously
			differentiable on $\left(\bT_n\times\bPi_n\right)\setminus K_{2n}$, where 
			$K_{2n}\subset \bT_n\times\bPi_n$ is at most countable,
		\item $\det\left(D_{\w}\bvP_{\bt_1}(\w,\bpi)\right)\neq0$, $\det\left(D_{\bpi}\bvP_{\bt_1}(\w,\bpi)\right)\neq0$
			for every $(\w,\bpi)\in\left(W_n\times\bPi_n\right)\setminus K_{1n}$,
		\item $\det\left(D_{\bt}\bvP_{\w_1}(\bt,\bpi)\right)\neq0$, $\det\left(D_{\bpi}\bvP_{\w_1}(\bt,\bpi)\right)\neq0$
			for every $(\bt,\bpi)\in\left(\bT_n\times\bPi_n\right)\setminus K_{2n}$,
		\item $\lim_{\lVert(\w,\bpi)\rVert\to\infty}\left\lVert\bvP_{\bt_1}(\w,\bpi)\right\rVert=\infty$,
		\item $\lim_{\lVert(\bt,\bpi)\rVert\to\infty}\left\lVert\bvP_{\w_1}(\bt,\bpi)\right\rVert=\infty$.
	\end{enumerate}
\end{assumption}
\begin{theorem}\label{ch2:thm:freq}
	If Assumptions~\ref{ch2:ass:dr} and~\ref{ch2:ass:piv} and one of Assumptions~\ref{ch2:ass:bvp} 
	or~\ref{ch2:ass:bvp2} are satisfied, then the followings hold:
	\begin{enumerate}
		\item There is a $\C^1$-diffeomorphism map $\a:W_n\to\bT_n$ such
			that the distribution function of $\hbt_n$ given $\hbpi_n$ is  
		\begin{equation*}
			\int_{\bT_n} f_{\hbt_n\vert\hbpi_n}\left(\hbt_n\vert\hbpi_n\right)\d\bt = 
			\int_{W_n} f\left(\a(\w)\vert\hbpi_n\right)
			\left\lvert J(\w\vert\hbpi_n)\right\rvert\d\w,
		\end{equation*}
		where
		\begin{equation*}
			J(\w\vert\hbpi_n) = \frac{\det\left(D_{\bt}\bvP_{\hbpi_n}(\a(\w),\w)\right)}
			{\det\left(D_{\w}\bvP_{\hbpi_n}(\a(\w),\w)\right)}.
		\end{equation*}
		\item For all $\alpha\in(0,1)$,	every exact $\alpha$-credible set built 
			from the percentiles of the distribution function have exact 
			frequentist coverage probabilities.
	\end{enumerate}
\end{theorem}
Theorem~\ref{ch2:thm:freq} is very powerful as it concludes that 
the SwiZs (Assumptions~\ref{ch2:ass:dr},~\ref{ch2:ass:piv} and~\ref{ch2:ass:bvp}) and the indirect inference
estimators (Assumption~\ref{ch2:ass:dr},~\ref{ch2:ass:piv} and~\ref{ch2:ass:bvp2}) 
have exact frequentist coverage probabilities in finite sample.
Our argument is based on the possibility of changing variables from $\hbt_n$
to $\w$, but also from $\w$ to $\hbt_n$ (hence the diffeomorphism). 
This argument may appear tautological, but this is actually
because we are able to make this change-of-variable in both directions that the conlcusion
of Theorem~\ref{ch2:thm:freq} is possible (see the parametric bootstrap
in Examples~\ref{ch2:ex:unif} and~\ref{ch2:ex:gamma} for counter-examples).
The result is very general because we do not suppose that we know
explicitly the estimators $\hbt_n$ and $\hbpi_n$, neither the random variable $\w$.
Because of their unknown form, we employ a global implicit function theorem for our proof
which permits to characterize the derivative of these estimators through their estimating function.
One of the conclusion of the global implicit function theorem is the existence
of a unique and global invertible function $\a$. It seems not possible to reach the conclusion
of Theorem~\ref{ch2:thm:freq} with a local implicit function theorem (usually encountered in textbooks),
but it may be of interest for further research as some conditions may accordingly be relaxed.

Although powerful, Theorem~\ref{ch2:thm:freq}'s conditions are restrictive or difficult
to inspect, but not hard to believe as we now explain. First, the existence of the random variable $\w$ depends
on the possibility to have data reduction as expressed in Assumption~\ref{ch2:ass:dr}.
We do not need to know explicitly $\w$ and $\w$ does not need to be unique,
so essentially Assumption~\ref{ch2:ass:dr} holds for every problem 
for a which a maximum likelihood estimator exists (see e.g.~\cite{hogg2005introduction}, Theorem 2 in Chapter 7);
see also~\cite{fraser2010second,martin2015conditional} for the construction of $\w$ by conditioning.
Yet, it remains unclear if this condition holds in the situations when the likelihood function
does not exist. The indirect inference and ABC literatures are overflowing with examples where
the likelihood is not tractable, but one should keep in mind that such situation does not exclude the existence
of a maximum likelihood, it is simply impractical to obtain one.
Second, Assumption~\ref{ch2:ass:piv} states that the true value $\bto$ belongs to the 
set of solutions. This condition can typically only be verified in simulations when controlling
all the parameters of the experiment, although
it is not critical to believe such condition holds when making a very large number of simulations $S$.
We interpret the inclusion of the set of solutions to $\bT$ as follows: once $\bto\in\bT$ is fixed, it is
not necessary to explore the whole set $\bT$ (that would require $S$ to be extremly large), but an area
sufficiently large of $\bT$ such that it includes $\bto$.
Third, Assumptions~\ref{ch2:ass:bvp} and~\ref{ch2:ass:bvp2} are more technical and concerns
the finite sample behavior of the estimating functions of, respectively, the SwiZs and the indirect inference estimators.
Although we cannot conclude that Assumption~\ref{ch2:ass:bvp} is weaker than
Assumption~\ref{ch2:ass:bvp2}, it seems easier to deal with the former.

Assumption~\ref{ch2:ass:bvp} (\textit{i})
requires the estimating function to be once continuously differentiable in $\bt$ and $\w$ almost everywhere.
The estimators $\hbt_n$ and $\hbpi_n$ are not known in an explicit form, but they can be characterized
by their derivatives using an implicit function theorem argument.
Since $\bt$ and $\w$ appears in the generating function $\gc$,
this assumption may typically be verified with the example at hand using a chain rule argument: 
the estimating function must be once continuously differentiable in the observations represented by $\gc$,
and $\gc$ must be once continuously differentiable in both its arguments. Discrete random variables
are automatically ruled out by this last requirement, but this should not appear as a surprise as
exactness of the coverage cannot be claimed in general for discrete distribution (see e.g.~\cite{cai2005one}).
The smoothness requirement on the estimating function excludes for example estimators based
on order statistics. In general, relying on non-smooth estimating function leads to less efficient
estimators and less stable numerical solutions, but they may be an easier estimating function to choose in situations
where it is not clear which one to select.
Although, non-smooth estimating functions and discrete random variables
are dismissed, the condition may nearly be satisfied when considering a $n$ large enough.
Assumption~\ref{ch2:ass:bvp2} (\textit{i}, \textit{ii}) requires in addition the estimating
equation to be once continuously differentiable in $\bpi$.

Assumption~\ref{ch2:ass:bvp} (\textit{ii}), as well as Assumption~\ref{ch2:ass:bvp2} (\textit{iii}, \textit{iv}),
essentially necessitate the estimating function to be ``not too flat'' globally.
It is one of the weakest condition to have invertibility of the Jacobian matrices.
Usually only one of the Jacobian has such requirement for an implicit function theorem,
but since we are targeting a $\C^1$-diffeomorphism, we strenghten the assumption on both Jacobians.
Once verified the first derivative of the estimating function as explained in the preceding paragraph,
the non-nullity of determinant may be appreciated, it typically depends on the model and the choosen estimating function.
An example for which this condition is not globally satisfied is when considering robust estimators as
the estimating function is constant on an uncountable set once exceeding some threeshold. This consideration gives
raise to the question on whether this condition may be relaxed to hold only locally,
condition which would be satisfied by the robust estimators, but
Example~\ref{ch2:ex:lomax} with the robust Lomax distribution in the Section~\ref{ch2:sec:sim}
seems to indicate the opposite direction.

Assumption~\ref{ch2:ass:bvp} (\textit{iii}), as well as Assumption~\ref{ch2:ass:bvp2} (\textit{v}, \textit{vi}),
is a necessary and sufficient condition to invoke Palais' global inversion theorem (\cite{palais1959natural})
which is a key component of the global implicit function theorem of~\cite{cristea2017global} we use. It can
be verified in two steps by, first, letting $\gc$ diverges in the estimating function, and then
letting $\bt$ and $\w$ diverges in $\gc$. Once again, robust estimators do not fulfill this requirement
as their estimating functions do not diverge with $\gc$ but rather stay constant.

Theorem~\ref{ch2:thm:freq} is derived under sufficient conditions. In its actual form, 
although very general, it excludes some specific estimating functions and non-absolutly continuous
random variable. It is of both practical and theoretical interest to develop results
for a wider-range of situations. Such considerations are left for further research.

We finish this section by considering a special, though maybe common, case
where the auxiliary estimator is known in an explicit form. 
Suppose $\hbpi_n = \h(\xo)$ where $\h$ is a known (surjective) function of the observations
(see Assumption~\ref{ch2:ass:dr}).
We can define a (new) indirect inference estimator as follows:
\begin{equation}\label{ch2:eq:ii2}
	\hbt_{\text{II},n}^{(s)}\in\bT^{(s)}_{\text{II},n} = \argzero_{\bt\in\bT} d\left[\h(\xo),\g(\bt,\w_s)\right].
\end{equation}
\begin{remark}
	The estimator defined in Equation~\ref{ch2:eq:ii2} is a special case of the indirect inference
	estimators as expressed in Definition~\ref{ch2:def:iie},
	and thus of the SwiZs by Theorem~\ref{ch2:thm:equiv}, where
	the auxiliary estimators $\hbpi_n$ and $\hbpi_{\text{II},n}$
	are known in an explicit form.
\end{remark}
\begin{assumption}[characterization of $\gc$]\label{ch2:ass:g}
	Let $\bT_n\subseteq\bT$, $W_n$ be subsets of $\R^p$ and $K_n\subset\bT_n\times W_n$ be at most countable.
	The followings hold:
	\begin{enumerate}[label=\roman*.]
		\item $\gc\in\C^{1}\left(\bT_n\times W_n,\R^p\right)$ is once continuously
			differentiable on $(\bT_n\times W_n)\setminus K_n$,
		\item $\det(D_{\bt}\gc(\bt,\w))\neq0$ and $\det(D_{\w}\gc(\bt,\w))\neq0$ for every
	$(\bt,\w)\in(\bT_n\times W_n)\setminus K_n$,
		\item $\lim_{\lVert(\bt,\w)\rVert\to\infty}\lVert\gc(\bt,\w)\rVert=\infty$.
	\end{enumerate}
\end{assumption}
\begin{proposition}\label{ch2:thm:freq3}
	If Assumptions~\ref{ch2:ass:dr},~\ref{ch2:ass:piv} and~\ref{ch2:ass:g} are satisfied,
	then the conclusions (1) and (2) of Theorem~\ref{ch2:thm:freq} hold.
	In particular, the distribution function is:
	\begin{equation*}
		\int_{\bT_n} f_{\hbt_n\vert\hbpi_n}\left(\bt\vert\h(\xo)\right)\d\bt = 
		\int_{W_n} f\left(\a(\w)\vert\h(\xo)\right)\left\lvert J(\w\vert\h(\xo))\right\rvert\d\w,
	\end{equation*}
	where
	\begin{equation*}
		J(\w\vert\h(\xo)) = \frac{\det\left(D_{\bt}\gc(\a(\w),\w)\right)}{\det\left(D_{\w}\gc(\a(\w),\w)\right)}.
	\end{equation*}
\end{proposition}
The message of Proposition~\ref{ch2:thm:freq3} is fascinating:
once the auxiliary estimator is known in an explicit form, the conditions
to reach the conclusion of Theorem~\ref{ch2:thm:freq} simplify accounting
for the fact that the implicit function theorem is no longer necessary. 
The discussion we have after Theorem~\ref{ch2:thm:freq} still holds,
but the verification process of the conditions is reduced to inspecting
the generating function.

%
%
\section{Asymptotic properties}\label{ch2:sec:ap}

When $n\to\infty$, different assumptions than in Section~\ref{ch2:sec:fs} may be considered
to derive the distribution of the SwiZs.
By Theorem~\ref{ch2:thm:equiv}, the SwiZs in Definition~\ref{ch2:def:swizs2}
and the indirect inference estimators in Definition~\ref{ch2:def:iie} are equivalent for any $n$.
Yet, due to their different forms, the conditions to derive their asymptotic properties differ, at least
in appearance. We treat both the asymptotic properties of the SwiZs and the 
indirect inference estimators in an unified fashioned and highlight their differences.
We do not attempt at giving the weakest conditions possible as our goal is primarly
to demonstrate in what theoretical aspect the SwiZs is different from the indirect inference
estimators. The asymptotic properties of the indirect inference estimators were already derived by several authors in
the literature, and we refer to~\cite{gourieroux1996simulation}, Chapter 4, for the comparison.

The following conditions are sufficient to prove the consistency of any estimator 
$\hbt_n^{(s)}$ in Defintions~\ref{ch2:def:swizs2} and~\ref{ch2:def:iie}.
When it is clear from the context, we simply drop the suffix and denote $\hbt_n$ for any of these estimators.
\begin{assumption}\label{ch2:ass:aux}
	The followings hold:
	\begin{enumerate}[label=\roman*.]
		\item The sets $\bT, \bPi$ are compact,
		\item For every $\bpi_1,\bpi_2\in\bPi$, $\bt\in\bT$ and $\u\sim F_{\u}$, there exists
			a random value $A_n = \Op(1)$ such that, for a sufficiently large $n$,
			\begin{equation*}
				\left\lVert\bP_n(\bt,\u,\bpi_1)-\bP_n(\bt,\u,\bpi_2)\right\rVert
				\leq A_n\left\rVert\bpi_1-\bpi_2\right\rVert,
			\end{equation*}
		\item For every $\bt\in\bT$, $\bpi\in\bPi$, the estimating function $\bP_n\left(\bt,\u,\bpi\right)$ 
			converges pointwise to $\bP(\bt,\bpi)$.
		\item For every $\bt\in\bT$, $\bpi_1,\bpi_2\in\bPi$, we have
			\begin{equation*}
				\bP\left(\bt,\bpi_1\right)=\bP\left(\bt,\bpi_2\right),
			\end{equation*}
			if and only if $\bpi_1=\bpi_2$.
	\end{enumerate}
\end{assumption}
\begin{assumption}[SwiZs]\label{ch2:ass:sw}
	The followings hold:
	\begin{enumerate}[label=\roman*.] 
		\item For every $\bt_1,\bt_2\in\bT$, $\bpi\in\bPi$ and $\u\sim F_{\u}$, there exists
			a random value $B_n = \Op(1)$ such that, for sufficiently large $n$,
			\begin{equation*}
				\left\lVert\bP_n(\bt_1,\u,\bpi)-\bP_n(\bt_2,\u,\bpi)\right\rVert\leq B_n\left\rVert\bt_1-\bt_2\right\rVert,
			\end{equation*}
		\item For every $\bt_1,\bt_2\in\bT$, $\bpi\in\bPi$, we have
			\begin{equation*}
				\bP\left(\bt_1,\bpi\right)=\bP\left(\bt_2,\bpi\right),
			\end{equation*}
			if and only if $\bt_1=\bt_2$.
	\end{enumerate}
\end{assumption}
\begin{assumption}[IIE]\label{ch2:ass:iie}
	The followings hold:
	\begin{enumerate}[label=\roman*.]
		\item For every $\bt_1,\bt_2\in\bT$, there exists a random 
			value $C_n=\Op(1)$ such that, for sufficiently large $n$,
			\begin{equation*}
				\left\lVert\hbpi_{\text{II},n}(\bt_1)-\hbpi_{\text{II},n}(\bt_2)
				\right\rVert\leq C_n\left\lVert\bt_1-\bt_2\right\rVert;
			\end{equation*}
		\item Let $\bpi(\bt)$ denotes the mapping towards which $\hbpi_{\text{II},n}(\bt)$ converges
			pointwise for every $\bt\in\bT$. For every $\bt_1,\bt_2\in\bT$, we have
			\begin{equation*}
				\bpi(\bt_1) = \bpi(\bt_2),
			\end{equation*}
			if and only if $\bt_1=\bt_2$.
	\end{enumerate}
\end{assumption}
\begin{theorem}[consistency]\label{ch2:thm:con}
	Let $\{\hbpi_n\}$ be a sequence of estimators of $\{\bP_n(\bpi)\}$.
	For any fix $\bt\in\bT$, let $\{\hbpi_{\text{II},n}(\bt)\}$ be the sequence
	of estimators of $\{\bP_n(\bt,\bpi)\}$.
	Let $\{\hbt_n\}$ be a sequence of estimators of $\{\bP_n(\bt)\}$.
	We have the following:
	\begin{enumerate}
		\item If Assumption~\ref{ch2:ass:aux} holds,
		then any sequence $\{\hbpi_n\}$ converges in probability to $\bpio$
		and any sequence $\{\hbpi_{\text{II},n}(\bt)\}$ converges in probability
		to $\bpi(\bt)$;
		\item Moreover, if one of Assumptions~\ref{ch2:ass:sw} or~\ref{ch2:ass:iie}
		holds, then any sequence $\{\hbt_n\}$ converges in probability to $\bto$.
	\end{enumerate}
\end{theorem}
Theorem~\ref{ch2:thm:con} demonstrates the consistency of $\hbt_n$ under two sets of conditions.
Assumptions~\ref{ch2:ass:aux} and~\ref{ch2:ass:iie}, or the conditions that are implied by these Assumptions,
are regular in the literature of the indirect inference estimators (see~\cite{gourieroux1996simulation}, Chapter 4).
More specifically, the mapping $\bt\mapsto\bpi$, usually referred to as the ``binding'' function (see e.g.~\cite{gourieroux1993indirect})
or the ``bridge relationship'' (see~\cite{jiang2004indirect}), is central in the argument and is required
to have a one-to-one relationship (Assumption~\ref{ch2:ass:iie} (\textit{ii})). 
Surprisingly, in Theorem~\ref{ch2:thm:con}, such requirement may be substitued by
the bijectivity of the deterministic estimating function with respect to $\bt$ (Assumption~\ref{ch2:ass:sw} (\textit{ii})).
Whereas the bijectivity of $\bpi(\bt)$ can typically only be assumed (if $\bt\mapsto\bpi$ was known explicitly,
then one would not need to use the indirect inference estimator unless of course one would be willing to lose
statistical efficiency and numerical stability for no gain), there is more hope
for Assumption~\ref{ch2:ass:sw} (\textit{ii}) to be verifiable. Since both Assumptions~\ref{ch2:ass:sw}
and~\ref{ch2:ass:iie} leads to the same conclusion, one would expect some strong connections between
them. Since $\bpi(\bt)$ may be interpreted as the implicit solution of $\bP(\bt,\bpi(\bt))=\0$, 
it seems possible to link both Assumptions with the help of an implicit function theorem, but it typically requires
further conditions on the derivatives of $\bP$ that are not necessary for obtaining the consistency results,
and we thus leave such considerations for further research.

Proving the consistency of an estimator relies on two major conditions: 
the uniform convergence of the stochastic objective function and the bijectivity 
of the deterministic objective function (Assumption~\ref{ch2:ass:aux} (\textit{iv}), 
Assumption~\ref{ch2:ass:sw} (\textit{ii}), Assumption~\ref{ch2:ass:iie} (\textit{ii})).
This second condition is referred to as the identifiability condition. It can sometimes be verified,
or sometimes it is only assumed to hold, but it is typically appreciated in accordance with the chosen probabilistic model.
Discrepancy among approaches mainly occurs on the demonstration of the uniform convergence.
Here we rely on a stochastic version of the classical Arzel\`a-Ascoli theorem,
see~\cite{van1998asymptotic} for alternative approaches based on the theory of empirical processes. 
To satisfy this theorem, we require the parameter sets to be compact (Assumption~\ref{ch2:ass:aux} (\textit{i})),
the stochastic objective function to converges pointwise (Assumption~\ref{ch2:ass:aux} (\textit{iii}))
and the stochastic objective function to be Lipschitz (Assumption~\ref{ch2:ass:aux} (\textit{ii}), 
Assumption~\ref{ch2:ass:sw} (\textit{i}), Assumption~\ref{ch2:ass:iie} (\textit{i})). 
Note that the last requirement is in fact for the objective function to be stochastically equicontinuous,
requirement verified by the Lipschitz condition,
see also~\cite{potscher1994generic} for a broad discussion on this condition and alternatives. 
Some authors proposed to relax the compactness condition, see for example~\cite{huber1967behavior},
but this is generally not a sensitive issue in practice.
The pointwise convergence of the stochastic objective function may be appreciated up to further details
depending on the context. For \iid\;observations, typically the weak law of large numbers may be employed,
thus requiring the stochastic objective function to have the same finite expected value across the observations. 
Other law of large numbers results may be used for serially dependent processes (see the Chapter 7 of~\cite{hamilton1994time})
and for non-identically distributed processes (see~\cite{andrews1988laws}), each results having its own conditions to satisfy.

We now turn our interest to the asymptotic distribution of an estimator $\hbt_n$.
Likewise the consistency result, the following sufficient conditions, are separated
to outline the difference between the SwiZs and the indirect inference estimators.
\begin{assumption}\label{ch2:ass:norm}
	The followings hold:
	\begin{enumerate}[label=\roman*.]
		\item Let $\Xs,\Ys$, the interior sets of $\bT,\bPi$, be open and convex subsets of $\R^p$,
		\item $\bto\in\Xs$ and $\bpio\in\Ys$,
		\item $\bP_n\in\C^{1}\left(\Xs\times\Ys,\R^p\times\R^p\right)$ when $n$ is sufficiently large,
		\item For every $\bt\in\Xs,\bpi\in\Ys$, $D_{\bpi}\bP_n(\bt,\u,\bpi),D_{\bt}\bP_n(\bt,\u,\bpi)$ 
			converge pointwise to $D_{\bpi}\bP(\bt,\bpi)\equiv\K(\bt,\bpi),D_{\bt}\bP(\bt,\bpi)\equiv\J(\bt,\bpi)$,
		\item $\K\equiv\K(\bto,\bpio),\J\equiv\J(\bto,\bpio)$ are nonsingular,
		\item $n^{1/2}\bP_n(\bto,\u,\bpio)\rightsquigarrow\Nor\left(\0,\Q\right)$, $\lVert\Q\rVert_{\infty}<\infty$.
	\end{enumerate}
\end{assumption}
\begin{assumption}[SwiZs II]\label{ch2:ass:sw2}
	For every $\bpi_1,\bpi_2\in\Ys$, $\bt\in\Xs$ and $\u\sim F_{\u}$, there exists
	a random value $E_n = \Op(1)$ such that, for sufficiently large $n$,
	\begin{equation*}
		\left\lVert D_{\bt}\bP_n(\bt,\u,\bpi_1)- D_{\bt}\bP_n(\bt,\u,\bpi_2)\right\rVert
		\leq E_n\left\rVert\bpi_1-\bpi_2\right\rVert.
	\end{equation*}
\end{assumption}
\begin{assumption}[IIE II]\label{ch2:ass:iie2}
	The followings hold:
	\begin{enumerate}[label=\roman*.]
		\item $\hbpi_{\text{II},n}\in\C^{1}(\Xs,\R^p)$ for sufficiently large $n$;
		\item For every $\bt\in\Xs$, $D_{\bt}\hbpi_{\text{II},n}(\bt)$ converges
			pointwise to $D_{\bt}\bpi(\bt)$.
	\end{enumerate}
\end{assumption}
\begin{theorem}[asymptotic normality]\label{ch2:thm:asynorm}
	If the conditions of Theorem~\ref{ch2:thm:con} are satisfied,
	we have the following additional results:
	\begin{enumerate}
		\item If Assumption~\ref{ch2:ass:norm} holds, then
			\begin{equation*}
				n^{1/2}\left(\hbpi_n-\bpio\right)\rightsquigarrow\Nor\left(\0,\K^{-1}\Q\K^{-T}\right),
			\end{equation*}
			and
			\begin{equation*}
				n^{1/2}\left(\hbpi_{\text{II},n}(\bt)-\bpi(\bt)\right)\rightsquigarrow\Nor\left(\0,\K^{-1}\Q\K^{-T}\right);
			\end{equation*}
		\item Moreover, if Assumption~\ref{ch2:ass:sw2} or~\ref{ch2:ass:iie2} holds, then
			\begin{equation*}
				n^{1/2}\left(\hbt_n-\bto\right)\rightsquigarrow\Nor\left(\0,2\J^{-1}\Q\J^{-T}\right).
			\end{equation*}
	\end{enumerate}
\end{theorem}
Theorem~\ref{ch2:thm:asynorm} gives the asymptotic distribution
of both the auxiliary estimator and the estimator of interest. The conditions
to derive the asymptotic distribution of the auxiliary estimator as expressed
in Assumption~\ref{ch2:ass:norm} is regular for most estimators in the statistical
literature. The proof of the first statement relies on the possibility 
to apply a delta method,
which requires the estimating function to be once continuously differentiable
(Assumption~\ref{ch2:ass:norm} (\textit{i}), (\textit{ii}) and (\textit{iii})).
The case where this condition is not met is typically when $\bto$ is a boundary
point of $\bT$. Not devoid of interest, this case is atypical and deserve to be treated on its own,
this situation is therefore excluded by Assumption~\ref{ch2:ass:norm} (\textit{ii}).
In contrast, relaxing the smoothness requirement on the estimating function has received
a much larger attention in the literature (see~\cite{huber1967behavior,newey1994large,van1998asymptotic} among others).
Here we content ourselves with the stronger smooth condition on the estimating function 
(Assumption~\ref{ch2:ass:norm} (\textit{iii})), maybe because it is largely admitted, but
also maybe because the smoothness of the estimating function is already required
when $n$ is finite by Theorem~\ref{ch2:thm:freq} to demonstrate the exact coverage probabilities,
a situation that encourages us to consider smooth estimating function in the practical examples.
The conditions for the Jacobian matrices to exist (Assumption~\ref{ch2:ass:norm} (\textit{iv}))
and to be invertible (Assumption~\ref{ch2:ass:norm} (\textit{v})) are regular ones.
The last condition is that a central limit theorem is applicable on the estimation equation
(Assumption~\ref{ch2:ass:norm} (\textit{vi})). This statement is very general and its validity
depends upon the context. For \iid~observations, one typically needs to verify Lindeberg's conditions 
(\cite{lindeberg1922neue}), which essentially requires that the two first moments exist and are finite.
The requirements are similar if the observations are non-identically observed (see e.g.~\cite{billingsley2012probability}).
The conditions are also similar for stationary processes (see e.g.~\cite{wu2011asymptotic}, for a review).
Note eventually that, also as minor as it might be, the delta method (which is essentially a mean value theorem)
largely in use in the statistical literature has recently been shown to be wrongly used by many for 
vector-valued function (\cite{feng2013mean}), this flaw has been taken into account in the present. 

The proof of the second statement of Theorem~\ref{ch2:thm:asynorm} on the asymptotic distribution
of the estimator of interest is more specific to the indirect inference literature. Compared to the
proof of the first statement, it requires in addition that, for $n$ large enough,
the binding function to be asymptotically differentiable
with respect to $\bt$ for the indirect inference estimator (Assumption~\ref{ch2:ass:iie2}) or the derivative of the 
estimating function with respect to $\bt$ to be stochastically Lipschitz for the SwiZs (Assumption~\ref{ch2:ass:sw2}).
For the same arguments we presented after the consistency Theorem~\ref{ch2:thm:con}, 
it may be more practical to verify Assumption~\ref{ch2:ass:sw2} as 
the verification of Assumption~\ref{ch2:ass:iie2} is impossible, at least directly,
as the binding function is unknown. This is actually not entirely true as
one may express the derivative of the binding function by invoking an implicit function theorem,
the condition then may be verified on the resulting explicit derivative.
The proof we use under Assumption~\ref{ch2:ass:iie2} uses this mechanism,
the derivative of the binding function is thus given by
\begin{equation*}
	D_{\bt}\bpi(\bt) = -\K^{-1}\J,
\end{equation*}
for every $\bt$ in a neighborhood of $\bto$ (see the proof in Appendix for more details).
It is only by using this implicit function theorem argument that the exact same explicit distribution
for both the SwiZs and the indirect inference estimators may be obtained.
The same idea may be used then to find the derivative of $\hbpi_{\text{II},n}(\bt)$ and 
verify Assumption~\ref{ch2:ass:iie2}. Note eventually
that~\cite{gourieroux1996simulation} have an extra condition not required here (but that would as well be required)
because they include a stochastic covariate with their indirect inference estimator.

Having demonstrated the asymptotic properties of one of the SwiZs estimators, $\hbt_n^{(s)}$,
$s\in\N^+_S$, we finish this section by giving the property of the average of the SwiZs sequence. 
The mean is an interesting estimator on its own and it is often considered
as a point estimate in a Bayesian context.
\begin{proposition}\label{ch2:thm:asynorm2}
	Let $\bar{\bt}_n$ be the average of $\{\hbt_n^{(s)}:s\in\N^+_S\}$.
	If the conditions of Theorem~\ref{ch2:thm:asynorm} are satisfied, then it holds
	that 
	\begin{equation*}
		n^{1/2}\left(\bar{\bt}_n-\bto\right)\rightsquigarrow\Nor\left(\0,\gamma\J^{-1}\Q\J^{-T} \right),
	\end{equation*}
	where the factor $\gamma=1+1/S$.	
\end{proposition}
The discussion of the proof and the condition to obtain Theorem~\ref{ch2:thm:asynorm}
are also valid for Proposition~\ref{ch2:thm:asynorm2}. The only point
that deserves further explanations is on the factor $\gamma$. This factor accounts
for the numerical approximation of the $\hbpi_n$-approximate posterior when $S$ is finite.
It is not surprising though for someone familiar with the indirect inference literature.
What may appear unclear is how this factor pass from 2 for one the SwiZs estimate in Theorem~\ref{ch2:thm:asynorm}
to $\gamma<2$ for the mean in Proposition~\ref{ch2:thm:asynorm}.
If the $\{\hbt_n^{(s)}:s\in\N^+_S\}$ are independent, then it is well-known
from the properties of the convolution of independent Gaussian random variables that $\gamma$ should equal 2.
In fact, the pivotal quantities $\{\u_s:s\in\N^+_S\}$ are indeed independent, but each of the $\{\hbt_n^{(s)}:s\in\N^+_S\}$
shares a ``common factor'', namely $\hbpi_n$, and thus this common variability may be reduced by increasing $S$.
Note eventually that the average estimator in Proposition~\ref{ch2:thm:asynorm2} has
the same asymptotic distribution as the two indirect inference estimators considered by~\cite{gourieroux1993indirect}
(given that the dimension of $\bt$ and $\bpi$ matches and that our implicit function theorem argument is used).

\section{Examples}\label{ch2:sec:st}
In this section, we illustrate the finite sample results of the Section~\ref{ch2:sec:fs}
with some examples for which explicit solutions exist. Indeed, for all the examples,
we are able to demonstrate analytically that the SwiZs' $\hbpi_n$-approximate posterior distribution
follows a uniform distribution when evaluated at the true value $\bto$, and thus 
concluding by Proposition~\ref{ch2:thm:exa} that any confidence regions built
from the percentiles of this posterior have exact coverage probabilities in the long-run.
In addition, and maybe more surprisingly, for most examples we are able to derive
the explicit posterior distribution that the SwiZs targets. This message is formidable,
one may not even need computations to characterize the distribution of $\hbt_n$
given $\hbpi_n$, but as one may foresee, these favorable situations are limited in numbers.
Lastly, we illustrate Proposition~\ref{ch2:thm:loc} on the equivalence between
the SwiZs and the parametric bootstrap with a Cauchy random variable in Example~\ref{ch2:ex:cauchy}
to conclude that they are indeed the same.
Since the SwiZs and the parametric bootstrap are seldom equivalent (see the discussion after Theorem~\ref{ch2:thm:pb}), 
we also demonstrate the nonequivalence of the two methods 
in the case of uniform random variable with unknown upper bound (Example~\ref{ch2:ex:unif})
and a gamma random variable with unknown rate (Example~\ref{ch2:ex:gamma}).
The considerations of this section are not only theoretical but also practical as 
we treat the linear regression (Example~\ref{ch2:ex:reg}) and the 
geometric Brownian motion when observed irregularly (Example~\ref{ch2:ex:gbm}),
two models widely use.

\begin{example}[Cauchy with unknown location]\label{ch2:ex:cauchy}
	Let $x_i\sim\text{Cauchy}(\theta,\sigma)$, $\sigma>0$ known, $i=1,\ldots,n$,
	be \iid. Consider the generating function $g(\theta,u)=\theta+u$
	where $u\sim\text{Cauchy}(0,\sigma)$ and the average as the (explicit) auxiliary estimator,
	$\hat{\pi}_n = \bar{x}$. We have $$\hat{\pi}_{\text{II},n}(\theta)=\frac{1}{n}\sum_{i=1}^n
	g(\theta,u_i)=\theta + w,$$ where $w = \frac{1}{n}\sum_{i=1}^n u_i$.
	By the properties of the Cauchy distribution, we have that $w\sim\text{Cauchy}(0,\sigma)$,
	that is the average of independent Cauchy variables has the same distribution of one of its components.
	Let $\hat{\theta}_n$ be the solution of $d(\hat{\pi}_n,\hat{\theta}_n + w)=0$, hence
	we have the explicit solution $\hat{\theta}_n = \hat{\pi}_n - w$.
	Note that by symmetry of $w$ around 0 we have $w\eqd-w$, so $\hat{\theta}_n=\hat{\pi}_n+w$.
	We therefore have that 
	\begin{align*}
		\Pr\left(\hat{\theta}_n\leq\theta_0\vert\hat{\pi}_n\right)
		&= \Pr\left(\hat{\pi}_n+w\leq\theta_0\vert\hat{\pi}_n\right) \\
		&= \Pr\left(\theta_0 - w_0 + w\leq\theta_0\vert\theta_0,w_0 \right) \\
		&= \Pr\left(w\leq w_0\right)\sim\mathcal{U}(0,1),
	\end{align*}
	and by Proposition~\ref{ch2:thm:exa} the coverage obtained on the percentiles
	of the distribution of $\hat{\theta}_n\vert\hat{\pi}_n$ are exact in the long-run (frequentist).
	
	The distribution of $\hat{\theta}_n\vert\hat{\pi}_n$ can be known in an explicit form.
	From the solution of $\hat{\theta}_n$, we let $w=a(\theta)=\hat{\pi}_n+\theta$.
	Following Proposition~\ref{ch2:thm:freq3}, we have
	\begin{equation*}
		f_{\hat{\theta}_n}\left(\theta\vert\hat{\pi}_n\right) 
		= f_{w}\left(a(\theta)\vert\hat{\pi}_n\right)
		\left\lvert\frac{\frac{\partial}{\partial\theta}g(\theta,w)}
		{\frac{\partial}{\partial w}g(\theta,w)}\right\rvert.
	\end{equation*}
	Since $g(\theta,w)=\theta+w$, the scaling factor is 1 and 
	$\hat{\theta}_n\vert\hat{\pi}_n\sim\text{Cauchy}(\hat{\pi}_n,\sigma)$.
	
	Eventually, we illustrate Theorem~\ref{ch2:thm:pb}, more specifically Proposition~\ref{ch2:thm:loc}, 
	by showing that the parametric bootstrap is equivalent.
	The bootstrap estimators is $\hat{\theta}_{\text{Boot},n}=\frac{1}{n}\sum_{i=1}^n
	g(\hat{\pi}_n,u_i) = \hat{\pi}_n + w$. It follows immediately that
	$\hat{\theta}_n=\hat{\theta}_{\text{Boot},n}$ and both estimators are equivalently
	distributed.
\end{example}
\begin{example}[uniform with unknown upper bound]\label{ch2:ex:unif}
	Let $x_i\sim\mathcal{U}(0,\theta)$, $i=1,\ldots,n$, be \iid.  
	Consider the generating function $g(\theta,u)=u\theta$ where 
	$u\sim\mathcal{U}(0,1)$
	and the (explicit) auxiliary estimator $\max_i{x_i}$.
	Clearly, $\max_i{x_i}=\theta\max_i{u_i}$. 
	Denote $w=\max_i{u_i}$ so the auxiliary
	estimator on the sample is $\hat{\pi}_n = w_0\theta_0$.
	Now define the estimator $\hat{\theta}_n$ to be the solution such
	that $d(\hat{\pi}_n,\hat{\theta} w)=0$.
	An explicit solution exists and is given by $\hat{\theta}_n=\frac{\theta_0 w_0}{w}$.
	We therefore have that 
	\begin{equation*}
		\Pr\left(\hat{\theta}_n\leq\theta_0\vert\hat{\pi}_n\right)
		= \Pr\left(\frac{\theta_0 w_0}{w}\leq\theta_0\vert \theta_0, w_0\right)
		= \Pr\left(w^{-1}\leq w_0^{-1}\right)\sim\mathcal{U}(0,1),
	\end{equation*}
	and by Proposition~\ref{ch2:thm:exa} the coverage obtained on the percentiles
	of the distribution of $\hat{\theta}_n$ are exact in the frequentist sense.

	We can even go further by expliciting the distribution of $\hat{\theta}_n$ given
	$\hat{\pi}_n$.
	Let define the mapping $a(\theta)=\frac{\theta_0 w_0}{\theta}$.
	By the change-of-variable formula we obtain:
	\begin{equation*}
		f_{\hat{\theta}_n}(\theta\vert\hat{\pi}_n) = 
		f_{w}(a(\theta)\vert\hat{\pi}_n) 
		\left\lvert\frac{\partial}{\partial\theta}a(\theta)\right\rvert.
	\end{equation*}
	The maximum of $n$ standard uniform random variables has the density $f_{w}(w) = n w^{n-1}$.
	The derivative is given by $\partial a(\theta)/\partial\theta = -\theta_0 w_0/\theta^2$.
	Note that by Proposition~\ref{ch2:thm:freq3} we equivalently have 
	\begin{equation*}
		\frac{\frac{\partial }{\partial\theta}g(\theta,w)}{\frac{\partial }{\partial w}g(\theta,w)}\Big\vert_{w=a(\theta)}
		= \frac{w}{\theta} \Big\vert_{w=\theta_0 w_0/\theta}
		= \frac{\theta_0 w_0}{\theta^2}.
	\end{equation*}
	Hence, we eventually obtain:
	\begin{equation*}
		f_{\hat{\theta}_n}(\theta\vert\hat{\pi}_n) = \frac{n{\hat{\pi}_n}^{n}}{\theta^{n+1}}, \quad \hat{\pi}_n = \theta_0 w_0.
	\end{equation*}
	Note that $\hat{\pi}_n$ is a sufficient statistic.
	Therefore we have obtained that the posterior distribution of $\hat{\theta}_n$ given
	$\hat{\pi}_n$ is a Pareto distribution parametrized by $\hat{\pi}_n$, the minimum value of the support,
	and the sample size $n$, as the shape parameter.

	In view of the preceding display, it is not difficult to develop a similar result
	for the parametric bootstrap (see the Definition~\ref{ch2:def:pb}). The bootstrap
	estimator solution is simple, it is given by $\hat{\theta}_{\text{Boot},n} = \max_i u_i\hat{\pi}_n = \theta_0w_0w$.
	We thus obtain 
	\begin{equation*}
		\Pr\left(\hat{\theta}_{\text{Boot},n}\leq\theta_0\vert\hat{\pi}_n\right)
		= \Pr\left(\theta_0w_0w\leq\theta_0\vert\theta_0,w_0\right)
		= \Pr\left(w\leq w_0^{-1}\right),
	\end{equation*}
	so it cannot be concluded that $F_{\hat{\theta}_{\text{Boot},n}\vert\hat{\pi}_n}(\theta_0)$ follows
	a uniform distribution and we cannot invoke Proposition~\ref{ch2:thm:exa}. Note that
	however we cannot exclude that the parametric bootstrap leads to exact coverage probability 
	in virtue of Proposition~\ref{ch2:thm:exa} (see Remark~\ref{ch2:rem:exa}).
	The parametric bootstrap is well-known to be inadequate in such problem.
	This fact may be made more explicit as we give now the distribution of
	the parametric bootstrap estimators.
	Let define the mapping $w=b(\tilde{\theta})=\frac{\tilde{\theta}}{\theta_0w_0}$.
	Note that $b(\theta_0)=1/w_0\neq w_0$.
	We obtain by the change-of-variable formula
	\begin{equation*}
		f_{\hat{\theta}_{\text{Boot},n}}\left(\tilde{\theta}\vert\hat{\pi}_n\right)
		= f_{w}\left(b(\tilde{\theta})\vert\hat{\pi}_n\right)
		\left\lvert\frac{\partial}{\partial\tilde{\theta}}b(\tilde{\theta})\right\rvert
		= \frac{n\tilde{\theta}^{n-1}}{\hat{\pi}_n^n}.
	\end{equation*}
	This distribution is known to be the power-function distribution, a special case
	of the Pearson Type I distribution (see~\cite{johnson1994continuous}). More interestingly,
	we have the following relationship between the parametric bootstrap and the SwiZs estimates:
	\begin{equation*}
		\hat{\theta}_{\text{Boot},n} \eqd \frac{1}{\hat{\theta}_n}.
	\end{equation*}
	Ultimately, note that the support of the distribution of $\hat{\theta}_{\text{Boot},n}$
	is $(0,\hat{\pi}_n)$ whereas it is $(\hat{\pi}_n,+\infty)$ for the SwiZs, so 
	both distributions never cross! Since $\hat{\pi}_n$ is systematically bias downward 
	the true value $\theta_0$, the coverage of the parametric bootstrap is always null.
	We illustrate this fact in the next figure.
\end{example}
\begin{example}[exponential with unknown rate parameter]\label{ch2:ex:expo}
	Let $x_i\sim\mathcal{E}(\theta)$, $i=1,\ldots,n$, be \iid.
	Consider the generating function $g(\theta,u)=\frac{u}{\theta}$,
	where $u\sim\Gamma(1,1)$, and the inverse of the average as 
	auxiliary estimator, denoted $\bar{x}^{-1}$. Clearly we have
	$\bar{x}^{-1} = \theta/w$, where $w = \sum_{i=1}^n u_i /n$,
	so $\hat{\pi}_n = \theta_0/w_0$.	
	The solution of $d(\hat{\pi}_n,\theta/w)=0$ in $\theta$
	is denoted $\hat{\theta}_n$, it is given by $\hat{\theta}_n = \theta_0 w/w_0=w\hat{\pi}_n$.
	We therefore have
	\begin{equation*}
		\Pr\left(\hat{\theta}_n\leq\theta_0\vert\hat{\pi}_n\right) = \Pr\left(w\leq w_0\right)\sim\mathcal{U}(0,1).
	\end{equation*}
	It results from Proposition~\ref{ch2:thm:exa} that any intervals built from the percentiles
	of the distribution of $\hat{\theta}_n$ has exact frequentist coverage. The distsribution
	can be found in explicit form. We have by the additive property of the Gamma distribution
	that $w\sim\Gamma(n,1/n)$ (shape-rate parametrization). 
	It immediately results from the change-of-variable formula
	that 
	\begin{equation*}
		\hat{\theta}_n\vert\hat{\pi}_n\sim\Gamma\left(n,\sum_{i=1}^{n}x_i\right).
	\end{equation*}
	Note that $\hat{\pi}_n$ is a sufficient statistic so the obtained distribution
	is a posterior distribution.
\end{example}
This last example on an exponential variate can be (slightly) generalized to a gamma random variable as follows.
\begin{example}[gamma with unknown rate parameter]\label{ch2:ex:gamma}
	Consider the exact same setup as in Example~\ref{ch2:ex:expo}
	with the exception that $x_i\sim\Gamma(\alpha,\theta)$ and $u\sim\Gamma(\alpha,1)$, 
	where $\alpha>0$ is a known shape parameter. Following the same steps
	as in Example~\ref{ch2:ex:expo} we find the following posterior distribution:
	\begin{equation*}
		\hat{\theta}_n\vert\hat{\pi}_n\sim\Gamma\left(\alpha n,\sum_{i=1}^n x_i\right).
	\end{equation*}
	We also have that any intervals built from the percentiles of the posterior have exact
	frequentist coverage probabilities.

	In view of this display and Example~\ref{ch2:ex:expo}, we can derive
	the distribution of the parametric bootstrap. The estimator is obtained as follows:
	\begin{equation*}
		\hat{\theta}_{\text{Boot},n} = \frac{n}{\sum_{i=1}^n g(\hat{\pi}_n,u_i)} = \frac{\hat{\pi}_n}{w},
	\end{equation*}
	where $w\sim\Gamma\left(n\alpha,1/n\right)$. It follows by the inverse of gamma variate and 
	the change-of-variable formula that
	\begin{equation*}
		\hat{\theta}_{\text{Boot},n}\sim\Gamma^{-1}\left(n\alpha,\sum_{i=1}^nx_i\right),
	\end{equation*}
	so $\hat{\theta}_{\text{Boot},n}\eqd 1/\hat{\theta}_n$. Since $\hat{\pi}_n=\theta_0/w_0$,
	we can also conclude that the parametric bootstrap is not uniformly distributed:
	\begin{equation*}
		\Pr\left(\hat{\theta}_{\text{Boot},n}\leq\theta_0\vert\hat{\pi}_n\right)
		= \Pr\left(\frac{\theta_0}{w_0w}\leq\hat{\pi}_n\vert\theta_0,w_0\right)
		= \Pr\left(\frac{1}{w}\leq w_0\right).
	\end{equation*}
\end{example}
The posterior distribution we obtained for the SwiZs in the last example coincides
with the fiducial distribution~\cite[see Table 1][]{veronese2015fiducial},
\cite[see Example 21.2][]{kendall1961advanced}. This correspondance is not 
surprising in view of the discussion held after Proposition~\ref{ch2:thm:equiv4}.
Indeed the gamma distribution is a member of the exponential family and we use
a sufficient statistics as the auxiliary estimator, so the SwiZs and the
generalized fiducial distribution are equivalent.

We now turn our attention to more general examples where $\bt$ is not a scalar.
\begin{example}[normal with unknown mean and unknown variance]\label{ch2:ex:norm2}
	Let $x_i\sim\Nor(\mu,\sigma^2)$ be \iid\; and consider $g(\mu,\sigma^2,u) = \mu + \sigma u$
	where $u\sim\Nor(0,1)$. Take the following auxiliary estimator,
	$\hbpi_n = {(\bar{x},ks^2)}^T = \h(x)$, where $\bar{x}=\frac{1}{n}\sum_{i=1}^n x_i$,
	$s^2 = \sum_{i=1}^n{\left(x_i-\bar{x}\right)}^2$ and $k\in\R$ is any constant.
	Note for example that $k<0$, so the auxiliary estimator of the variance may be negative.
	Indeed the SwiZs accepts situation for which $\bPi\cap\bT=\emptyset$, it is clearly 
	not the case of the parametric bootstrap for example (see Remark~\ref{ch2:rem:pb}).
	We have that
	\begin{equation*}
		\w = \h(u) = \begin{pmatrix}
			\frac{1}{n}\sum_{i=1}^n u_i \\
			\sum_{i=1}^n {\left(u_i-\frac{1}{n}\sum_{j=1}^n u_j \right)}^2
		\end{pmatrix}.
	\end{equation*}
	An explicit solution exists for $d(\hbpi_n,g(\mu,\sigma^2,\w))=0$ in $(\mu,\sigma^2)$ and is given
	by 
	\begin{equation*}
		\hbt_n = \begin{pmatrix} \hat{\mu} \\ \hat{\sigma}^2 \end{pmatrix}
			= \begin{pmatrix} \bar{x}_0 - \hat{\sigma} w_1 \\ \frac{s^2_0}{w_2} \end{pmatrix}
				= \a(\w).
	\end{equation*}
	Note that $\bar{x}_0=\mu_0+\sigma_0 w_{0,1}$ and $s^2_0 = \sigma^2_0 w_{0,2}$.
	We obtain the following
	\begin{align*}
		\Pr\left(\hbt_n\leq\bto\right) &= \Pr\left( \begin{pmatrix}\mu_0 + \sigma_0 w_{0,1} - 
			\sigma_0 w_1 \sqrt{\frac{w_{0,2}}{w_2}} \\ \sigma^2_0\frac{w_{0,2}}{w_2} \end{pmatrix}
			\leq\begin{pmatrix}\mu_0 \\ \sigma^2_0 \end{pmatrix}\right) \\
				&= \Pr\left(\begin{pmatrix}\frac{w_1}{\sqrt{w_2}} \\ \frac{1}{w_2} \end{pmatrix}
					\leq\begin{pmatrix}\frac{w_{0,1}}{\sqrt{w_{0,2}}} \\ \frac{1}{w_{0,2}}\end{pmatrix}\right)
						\sim\mathcal{U}(0,1).
	\end{align*}
	Therefore, by Proposition~\ref{ch2:thm:exa}, any region built from the percentiles of the 
	posterior distribution of $\hbt_n$ has exact frequentist coverage. This posterior
	distribution has a closed form.

	Note that $w_1\sim\Nor(0,1/n)$. Once realized that $u_i-\frac{1}{n}\sum_{j=1}^n u_j\sim\Nor(0,(n-1)/n)$,
	it is not difficult to obtain that $w_2\sim\Gamma(n/2,n/2(n-1))$, a gamma random variable (shape-rate parametrization).
	It is straightforward to remark that
	\begin{equation*}
		\hat{\mu}\vert(\hat{\sigma}^2,\hbpi_n)\sim\Nor\left(\bar{x}_0, \frac{\hat{\sigma}^2}{n}\right),
		\quad \hat{\sigma}^2\sim\Gamma^{-1}\left(\frac{n}{2},\frac{s^2_0 n}{2(n-1)}\right),
	\end{equation*}
	where $\Gamma^{-1}$ represents the inverse gamma distribution. The joint distribution
	is known in the Bayesian literature as the normal-inverse-gamma distribution (see~\cite{koch2007introduction}).
	We thus have the following joint distribution
	\begin{equation*}
		\hbt_n\vert\hbpi_n\sim\Nor\text{-}\Gamma^{-1}\left(\bar{x}_0,n,\frac{n}{2},\frac{s^2_0 n}{2(n-1)} \right).
	\end{equation*}
	The distribution of $\hat{\mu}$ unconditionnaly on $\hat{\sigma}^2$ is a non-standardized $t$-distribution
	with $n$ degrees of freedom,
	\begin{equation*}
		\hat{\mu}\vert\hbpi_n\sim t\left(\bar{x}_0,\frac{s^2_0 n}{n-1},n \right).
	\end{equation*}
\end{example}
The results on the normal distribution (Example~\ref{ch2:ex:norm2})
can be generalized to the linear regression.
\begin{example}[linear regression]\label{ch2:ex:reg}
	Consider the linear regression model $\y=\X\bb+\be$
	where $\be\sim\Nor(\0,\sigma^2\bI_n)$ and $\dim(\bb)=p$.
	Suppose the matrix $\X^T\X$ is of full rank.
	A natural generating function is $\g(\bb,\sigma^2,\X) = \X\bb+\sigma\u$
	where $\u\sim\Nor(\0,\bI_n)$ (see Example~\ref{ch2:ex:gaus} for other suggestions). 
	Take the ordinary least squares as
	the auxiliary estimator so we have the following explicit form:
	\begin{equation*}
		\hbpi_n = \begin{pmatrix} \hbpi_1 \\ \hat{\pi}_2 \end{pmatrix}
			= \begin{pmatrix} {\left(\X^T\X\right)}^{-1}\X^T\yo \\ k\yo^T\P\yo \end{pmatrix},
	\end{equation*}
	where $\P = \bI_n - \H$ is the projection matrix, $\H=\X{\left(\X^T\X\right)}^{-1}\X^T$
	is the hat matrix, $\yo$ denotes the observed responses and $k\in\R$ is any constant. 
	Note that $\P$ and $\H$ are symmetric idempotent matrices and that $\P\X=\0$.
	An explicit solution exists for $\hbt_n={(\hbb^T\,\hat{\sigma}^2)}^T$. 
	To find it, we use the indirect inference estimator, which by Theorem~\ref{ch2:thm:equiv}
	is the equivalent to the SwiZs estimator. Using $\y\eqd\X\bb+\sigma\u$, we have
	\begin{equation*}
		\hbpi_{\text{II},n}(\bt) = \begin{pmatrix} \hbpi_1(\bt) \\ \hat{\pi}_2(\bt) \end{pmatrix}
		= \begin{pmatrix} {\left(\X^T\X\right)}^{-1}\X^T\left(\X\bb+\sigma\u\right) \\
			k\sigma^2\u^T\P\u \end{pmatrix}.
	\end{equation*}
	Since $\hat{\pi}_2(\bt)$ depends only on $\sigma^2$, solving $d(\hat{\pi}_2,\hat{\pi}_2(\bt))=0$ in $\sigma^2$
	leads to 
	\begin{equation*}
		\hat{\sigma}^2 = \frac{\yo^T\P\yo}{\u^T\P\u}.	
	\end{equation*}
	On the other hand, solving $d(\hbpi_1,\hbpi_1(\bt))=\0$ in $\bb$ leads to 
	\begin{equation*}
		\hbb = {\left(\X^T\X\right)}^{-1}\X^T\left(\yo+\hat{\sigma}\u\right).
	\end{equation*}
	Since $\yo=\X\bbo+\sigma_0\uo$, we obtain the following:
	\begin{align*}
		\Pr\left(\hbt_n\leq\bto\right) 
		&= \Pr\left(\hbb\leq\bbo, \hat{\sigma}^2\leq{\sigma}^2_0\right) \\
		&= \Pr\left({\left(\X^T\X\right)}^{-1}\X^T\left(\X\bbo+\sigma_0\uo+\hat{\sigma}\u\right)\leq\bbo,
		\frac{{\left(\X\bbo+\sigma_0\uo\right)}^T\P\left(\X\bbo+\sigma_0\uo\right)}{\u^T\P\u}\leq\sigma_0^2\right) \\
		&= \Pr\left({\left(\X^T\X\right)}^{-1}\X^T\left(\sigma_0\uo-\hat{\sigma}\u\right)\leq\0,
		\frac{\sigma_0^2\uo^T\P\uo}{\u^T\P\u}\leq\sigma_0^2\right) \\
		&= \Pr\left(\frac{\X^T\u}{\sqrt{\u^T\P\u}}\leq\frac{\X^T\uo}{\sqrt{\uo^T\P\uo}},
		\frac{1}{\u^T\P\u}\leq\frac{1}{\uo^T\P\uo}\right)\sim\mathcal{U}(0,1).
	\end{align*}
	Note that at the third equality we use the fact that $\u\eqd-\u$ since $\u$ is symmetric around $\0$.
	The last development, together with Proposition~\ref{ch2:thm:exa}, demonstrates that
	any region built on the percentiles of the distribution of $\hbt_n$ leads to exact frequentist coverage
	probabilities. The distribution of $\hbt_n$ can be obtained in an explicit form.

	Since $\P$ is symmetric and idempotent, it is well known that $\u^T\P\u\sim\chi^2_{n-p}$~\cite[see Theorem 5.1.1][]{mathai1992quadratic}.
	Hence we obtain that
	\begin{equation*}
		\hbb\vert(\hat{\sigma}^2,\hbpi_n)\sim\Nor\left(\hbpi_1,\hat{\sigma}^2{\left(\X^T\X\right)}^{-1}\right),\quad
		\hat{\sigma}^2\vert\hbpi_n\sim\Gamma^{-1}\left(\frac{n-p}{2},\frac{\yo^T\P\yo}{2}\right).
	\end{equation*}
	As shown in Example~\ref{ch2:ex:norm2}, it follows that the joint distribution of 
	$\hbt_n$ conditionally on $\hbpi_n$ is a normal-inverse-gamma distribution
	\begin{equation*}
		\hbt_n\vert\hbpi_n\sim\Nor\text{-}\Gamma^{-1}\left({\left(\X^T\X\right)}^{-1}\X^T\yo,
		{\left(\X^T\X\right)}^{-1},\frac{n-p}{2},\frac{\yo^T\P\yo}{2}\right),
	\end{equation*}
	and the distribution of $\hbb$, unconditionally on $\hat{\sigma}^2$, is a multivariate non-standardized $t$ distribution
	with $n-p$ degrees of freedom
	\begin{equation*}
		\hbb\vert\hbpi_n\sim t\left({\left(\X^T\X\right)}^{-1}\X^T\yo,\frac{\yo^T\P\yo}{n-p}{\left(\X^T\X\right)}^{-1},n-p\right).
	\end{equation*}
\end{example}
In this last example on the linear regression, we employed the OLS
as the auxiliary estimator, which is known to be an unbiased estimator.
In fact, it is not a necessity to have unbiased auxiliary estimator.
The next example illustrate this point.
\begin{example}[ridge regression]\label{ch2:ex:ridge}
	Consider the same setup as in Example~\ref{ch2:ex:reg},
	$\y=\X\bb+\be$, $\be\sim\Nor(\0,\sigma^2\bI_n)$ and 
	$\rank(\X^T\X)=p$. Take the ridge estimator as the auxiliary estimator, 
	so for the regression coefficients we have 
	\begin{equation*}
		\hbpi_1^R = {\left(\X^T\X + \lambda\bI_p\right)}^{-1}\X^T\yo,
	\end{equation*}
	for some constant $\lambda\in\R$. 
	Consider the squared residuals as an estimator of the variance, 
	so after few manipulations, we obtain
	\begin{equation*}
		\hat{\pi}_2^R = k\yo^T\P_{\lambda}\P_{\lambda}\yo,
	\end{equation*}
	where $\P_{\lambda} \equiv \bI_n - \H_{\lambda}$, $\H_{\lambda} \equiv \X{\left(\X^T\X+\lambda\bI_p\right)}^{-1}\X^T$,
	$k\in\R$ is any constant. Note that $\P_{\lambda}$ is symmetric but not idempotent.
	As in Example~\ref{ch2:ex:reg}, let's use the indirect inference estimator with $\y\eqd\X\bb+\sigma\u$.
	We obtain 
	\begin{equation*}
		\hbpi_{\text{II},n}^R(\bt) = \begin{pmatrix} \hbpi_1^R(\bt) \\ \hat{\pi}^R_2(\bt) \end{pmatrix} 
			= \begin{pmatrix} {\left(\X^T\X + \lambda\bI_p\right)}^{-1}\X^T\left(\X\bb+\sigma\u\right) \\
				k{\left(\X\bb+\sigma\u\right)}^T\P_{\lambda}\P_{\lambda}\left(\X\bb+\sigma\u\right)
			\end{pmatrix}.
	\end{equation*}
	Let $\tilde{\bb}$ denotes the solution of $d(\hbpi_1^R,\hbpi_1^R(\bt))=0$ in $\bb$.
	We have the explicit solution given by
	\begin{equation*}
		\tilde{\bb} = {\left(\X^T\X\right)}^{-1}\X^T\left(\yo-\tilde{\sigma}\u\right).
	\end{equation*}
	Using $\tilde{\bb}$ in $\hat{\pi}^R_2(\bt)$ leads to
	\begin{equation*}
		\hat{\pi}^R_2(\tilde{\bt}) = k{\left(\H\yo-\tilde{\sigma}\P\u\right)}^T\P_{\lambda}
		\P_{\lambda}\left(\H\yo-\tilde{\sigma}\P\u\right),
	\end{equation*}
	where $\H\equiv\X{\left(\X^T\X\right)}^{-1}\X$ and $\P\equiv\bI_n-\H$.
	We have the followings: $\H\H_{\lambda}=\H_{\lambda}$, $\P\P_{\lambda}=\P$ and $\P\H=\0$.
	Finding $\tilde{\sigma}^2$ such that $d(\hat{\pi}_2^R,\hat{\pi}_2^R(\tilde{\bt}))=0$ gives
	\begin{equation*}
		\tilde{\sigma}^2\u^T\P\u + \yo^T\H\P_{\lambda}\P_{\lambda}\H\yo - \yo^T\P_{\lambda}\P_{\lambda}\yo = 0,
	\end{equation*}
	which leads to the following solution:
	\begin{equation*}
		\tilde{\sigma}^2 = \frac{\yo^T\P\yo}{\u^T\P\u}.
	\end{equation*}
	Therefore, $\tilde{\sigma}^2$ is the same as $\hat{\sigma}^2$ we found
	in Example~\ref{ch2:ex:reg}, and we directly have that
	$\tilde{\bb}=\hbb$. As a consequence, the distribution of
	$\tilde{\bt}$ is exactly the same as $\hbt_n$ in Example~\ref{ch2:ex:reg}
	and the frequentist coverage probabilities are exact.
\end{example}
From Example~\ref{ch2:ex:norm2} on the normal distribution, 
the derivation to closely related distribution is straightforward,
as we see now with the log-normal distribution.
\begin{example}[log-normal with unknown mean and unknown variance]\label{ch2:ex:logn}
	Let $x_i\sim\log\text{-}\Nor(\mu,\sigma^2)$ be \iid\; and consider
	$g(\mu,\sigma^2,u)=e^{\mu}e^{\sigma u}$ where $u\sim\Nor(0,1)$.
	If we take the maximum likelihood estimator as the auxiliary estimator,
	we have
	\begin{equation*}
		\hbpi_n= \begin{pmatrix} \hat{\pi}_1 \\ \hat{\pi}_2 \end{pmatrix} =
		\begin{pmatrix} \frac{1}{n}\sum_{i=1}^n\ln(x_i) \\
			\sum_{i=1}^n{\left(\ln(x_i)-\frac{1}{n}\sum_{j=1}^n\ln(x_j)\right)}^2
		\end{pmatrix}
	\end{equation*}
	The solution is the following
	\begin{equation*}
		\hbt_n = \begin{pmatrix} \hat{\mu} \\ \hat{\sigma}^2 \end{pmatrix}
			= \begin{pmatrix} \hat{\pi}_1 - \hat{\sigma}w_1 \\
				\frac{\hat{\pi}_2}{w_2},
			\end{pmatrix}
	\end{equation*}
	where $w_1=\frac{1}{n}\sum_{i=1}^n u_i$ and $w_2 = \sum_{i=1}^n{(u_i - \frac{1}{n}\sum_{j=1}^n u_j)}^2$.
	It is the same solution as Example~\ref{ch2:ex:norm2}, hence the posterior distribution
	of $\hbt_n$ is normal-inverse-gamma and any $\alpha$-credible region built on this posterior
	have exact frequentist coverage.
\end{example}
Having illustrated the theory for random variable that are \iid, 
we now show a last example on time series data. Note that (variations of) this
example is numerically studied in~\cite{gourieroux1993indirect}.
\begin{example}[irregularly observed geometric Brownian motion 
	with unknown drift and unknown volatility]\label{ch2:ex:gbm}
	Consider the stochastic differential equation
	\begin{equation*}
		dy_t = \mu y_t dt + \sigma y_t dW_t,
	\end{equation*}
	where $\{W_t:t\geq0\}$ is a Wiener process and $\bt={(\mu\;\sigma^2)}^T$ are the drift and volatility
	parameters. An explicit solution to It\^o's integral exists and is given by
	\begin{equation*}
		y_t = y_0\exp\left[\left(\mu-\frac{1}{2}\sigma^2\right)t+\sigma W_t\right].
	\end{equation*}
	Suppose we observe the process at $n$ points in time: $t_1<t_2<\ldots<t_n$, $\forall i$ $t_i\in\R^+$.
	Define the difference in time by $\Delta_i = t_i - t_{i-1}$, so we have $n-1$ time differences.
	Note that all the time differences are positive, $\Delta_i>0$, and we allow the process to be irregularly observed,
	$\Delta_i\neq\Delta_j, i\neq j$. Instead of working directly with the process $\{y_{t_i}:i\geq1\}$, it is 
	more convenient to work with the following transformation of the process $\{x_{t_i}=\ln(y_{t_i}/y_{t_{i-1}}):i\geq2\}$.
	Indeed, we have
	\begin{equation*}
		x_{t_i} = \left(\mu-\frac{1}{2}\sigma^2\right)\Delta_i + \sigma\left(W_{t_i}-W_{t_{i-1}}\right).
	\end{equation*}
	By the properties of the Wiener process, we have $W_{t_i}-W_{t_{i-1}}\sim\Nor(0,\Delta_i)$
	and $W_{t_i}-W_{t_{i-1}}$ is independent from $W_{t_j}-W_{t_{j-1}}$ for $i\neq j$.
	Hence the vector $\x={(x_{t_2}\;\dots\;x_{t_n})}^T$ is independentely
	but non-identically distributed according to the joint normal distribution
	\begin{equation*}
		\x\sim\Nor\left( \left(\mu-\frac{1}{2}\sigma^2\right)\bD,\sigma^2\Sigma\right),
	\end{equation*}
	where $\bD={(\Delta_2\;\dots\;\Delta_n)}^T$ and $\Sigma=\diag(\bD)$.
	Note that $\bD=\Sigma\1_{n-1}$, where $\1_{n-1}$ is a vector of $n-1$ ones,
	and $\bD^T\1_{n-1}=\bD^{T/2}\bD^{1/2}$ since all the $\Delta$ are positives.
	
	We consider the following auxiliary estimators:
	\begin{equation*}
		\hbpi_n = \begin{pmatrix} \hat{\pi}_1 \\ \hat{\pi}_2 \end{pmatrix}
			= \begin{pmatrix} \x^T_0\1_{n-1} \\ \x^T_0\Sigma^{-1}\x_0 \end{pmatrix}.
	\end{equation*}
	Since $\x\eqd(\mu-\sigma^2/2)\bD+\sigma\Sigma^{1/2}\z$, where $\z\sim\Nor(\0,\bI_{n-1})$,
	we obtain the following indirect inference estimators (or equivalently SwiZs), 
	\begin{equation*}
		\hat{\pi}_1(\bt) = {\left[\left(\mu-\frac{1}{2}\sigma^2\right)\bD+\sigma\Sigma^{1/2}\z\right]}^T\1_{n-1}
				 =  \left(\mu-\frac{1}{2}\sigma^2\right)\bD^{T/2}\bD^{1/2} + \sigma\z^T\bD^{1/2},
	\end{equation*}
	and
	\begin{align*}
		\hat{\pi}_2(\bt) &= {\left[\left(\mu-\frac{1}{2}\sigma^2\right)\bD+\sigma\Sigma^{1/2}\z\right]}^T
		\Sigma^{-1}{\left[\left(\mu-\frac{1}{2}\sigma^2\right)\bD+\sigma\Sigma^{1/2}\z\right]} \\
		&= {\left(\mu-\frac{1}{2}\sigma^2\right)}^2\bD^{T/2}\bD^{1/2} + 
		2\sigma\left(\mu-\frac{1}{2}\sigma^2\right)\z^T\bD^{1/2} + 
		\sigma^2\z^T\z.
	\end{align*}
	Solving $d(\hat{\pi}_1,\hat{\pi}_1(\hbt))=0$ in $\hat{\mu}$ gives
	\begin{equation}\label{ch2:eq:muhat}
		\hat{\mu} = \frac{1}{2}\hat{\sigma}^2 - \hat{\sigma}\z^T\bD^{1/2}{\left(\bD^{T/2}\bD^{1/2}\right)}^{-1}
		+ \x_0^T\1_{n-1}{\left(\bD^{T/2}\bD^{1/2}\right)}^{-1}.
	\end{equation}
	Now solving $d(\hat{\pi}_2,\hat{\pi}_2(\hbt))=0$ in $\hat{\sigma}^2$ and substituing 
	$\hat{\mu}$ by the above expression in~\eqref{ch2:eq:muhat} leads to
	\begin{equation*}
		\hat{\sigma}^2 = \frac{\x_0^T\Q\x_0}{\z^T\P\z},
	\end{equation*}
	where $\P=\bI_{n-1}-\bD^{1/2}{\left(\bD^{T/2}\bD^{1/2}\right)}^{-1}\bD^{T/2}$ is symmetric and idempotent,
	and $\Q=\Sigma^{-1}-\1_{n-1}{\left(\bD^{T/2}\bD^{1/2}\right)}^{-1}\1_{n-1}^T$.
	By the properties of the rank of a matrix, we have $\rank(\P) = \trace(\P) = n-2$.
	Note that by independence $\z^T\Delta^{1/2}\eqd z(\Delta^{T/2}\Delta^{1/2})$, 
	where $z$ is a single standard normal random variable.
	Similarly to the example on the linear regression (Example~\ref{ch2:ex:reg}), we obtain
	the explicit distributions
	\begin{align*}
		&\hat{\mu}\vert\left(\hbpi_n,\hat{\sigma}^2\right)\sim
		\Nor\left(\frac{1}{2}\hat{\sigma}^2+\x_0^T\1_{n-1}{\left(\bD^{T/2}\bD^{1/2}\right)}^{-1},
		\hat{\sigma}^2{\left(\bD^{T/2}\bD^{1/2}\right)}^{-1}\right), \\ 
		&\hat{\sigma}^2\vert\hbpi_n\sim\Gamma^{-1}\left(\frac{n-2}{2},
		\frac{\x_0^T\Q\x_0}{2}\right).
	\end{align*}
	As with Example~\ref{ch2:ex:reg}, this findings suggest that $\hbt_n\vert\hbpi_n$ is jointly distributed
	according to a normal-inverse-gamma distribution. However, $\hat{\sigma}^2$ appears in the mean
	of $\hat{\mu}\vert(\hbpi_n,\hat{\sigma}^2)$ so such conclusion is not straightforward. 
	We leave the derivation of the joint distribution and the distribution of $\hat{\mu}$ unconditionnal
	on $\hat{\sigma}^2$ for further research.

	We now demonstrate that the $\hbpi_n$-approximate posterior distribution of $\hbt_n$ 
	leads to exact frequentist coverage probabilities.
	Once realized that $\Sigma^{-1}=\Sigma^{-1/2}\Sigma^{-1/2}$, $\Sigma^{1/2}\1_{n-1}=\bD^{1/2}$,
	and $\bD^T\Sigma^{-1}=\1_{n-1}$, it is not difficult to show that 
	$\bD^T\Q\bD=0$, $\bD^T\Q\Sigma^{1/2}=0$ and 
	$\Sigma^{1/2}\Q\Sigma^{1/2}=\P$.
	Since $\x_0 = (\mu_0-\sigma^2_0/2)\bD + \sigma_0\Sigma^{1/2}\z_0$, we obtain
	\begin{align*}
		&\hat{\sigma}^2 = \sigma^2_0\frac{\z_0^T\P\z_0}{\z^T\P\z} = \sigma^2_0\frac{w_0}{w}, \\
		&\hat{\mu} = \frac{\sigma^2_0}{2}\frac{w_0}{w} - \sigma_0\sqrt{\frac{w_0}{w}}z + \mu_0-\frac{1}{2}\sigma^2_0
		+ \sigma_0z_0.
	\end{align*}
	Therefore,
	\begin{align*}
		\Pr\left(\hat{\mu}\leq\mu_0,\;\hat{\sigma}^2\leq\sigma^2_0\right) &=
		\Pr\left(\frac{\sigma^2_0}{2}\frac{w_0}{w}-\sigma_0\sqrt{\frac{w_0}{w}}z+-\frac{1}{2}\sigma^2_0
		+ \sigma_0z_0\leq0,\;\frac{w_0}{w}\leq1\right) \\
		&= \Pr\left(\frac{k_0}{w}-\frac{z}{\sqrt{w}}\leq\frac{k_0}{w_0}-\frac{z_0}{\sqrt{w_0}},\;w^{-1}\leq w_0^{-1}\right)
		\sim\mathcal{U}(0,1),
	\end{align*}
	where $k_0 = \sigma_0\sqrt{w_0}/2$. Thus, any region on the joint distribution of 
	$\hbt_n$ leads to exact frequentist coverage by Proposition~\ref{ch2:thm:exa}.
\end{example}

\section{Simulation study}\label{ch2:sec:sim}
The main goal of this section is threefold. First, we illustrate the results of the 
Section~\ref{ch2:sec:fs} on the frequentist properties in finite sample of the SwiZs 
in the general case where no solutions are known in explicit forms, as opposed to the Section~\ref{ch2:sec:st},
and thus requiring numerical solutions. In order to achieve this point, we measure at different levels
the empirical coverage probabilities of the intervals built from the percentiles of the 
$\hbpi_n$-approximate posterior obtained by the SwiZs. Note that for $\dim(\bt)>1$,
we only considered marginal intervals to avoid a supplementary layer of numerical nuisance,
the coverage probabilities are not concerned by this choice, only the length of the intervals.
Second, we elaborate on the verification of the conditions of Theorem~\ref{ch2:thm:freq}
with the examples at hand. As already motivated, the emphasis is on the estimating function.
It seems easier to verify Assumption~\ref{ch2:ass:bvp} than Assumption~\ref{ch2:ass:bvp2},
since only one of them is necessary to satisfy Theorem~\ref{ch2:thm:freq}, we concentrate
our efforts on the former.
We also brighten the study up to situations where Assumption~\ref{ch2:ass:bvp} 
does not entirely hold or cannot be verified to measure its consequences empirically.
Third, we give the general idea on how to implement the SwiZs. Indeed, anyone familiar
with the numerical problem of solving a point estimator such as the maximum likelihood
estimator has a very good idea on how to obtain the auxiliary estimator $\hbpi_n$. 
Solving the estimating function for the parameters of interest is very similar,
it requires the exact same tools but has the inconvenient of needing further 
analytical derivations and implementations details. As already remarked, the parametric
bootstrap does not possess such inconvenient. The counterpart is that
the SwiZs may lead to exact coverage probabilities.
The motto ``no pain, no gain'' is particularly relevant here. 
For this purpose, the parametric bootstrap is proposed as the point of comparison
for all the examples of this section. We measure the computational time as experienced by
the user in order to appreciate the numerical burden. 
In case both the SwiZs and the parametric bootstrap
have very similar coverage probabilities, we also quantify the length of the intervals 
as a mean of comparison.

As a subsidiary goal of this section, we study the point estimates of the SwiZs.
Indeed, the indirect inference is also a method for reducing the small sample bias of an initial (auxiliary) estimator,
even in situations where it may be ``unnatural'' to call such method, as for example,
when a maximum likelihood estimator may be easily obtained (see~\cite{guerrier2018simulation}).
Since the SwiZs is a special case of indirect
inference, it would be interesting to gauge the ability of the SwiZs to correct the bias.
We explore the properties of the mean and the median of the SwiZs.
This choice is arbitrary but largely admitted.

There are common factors in the implementation of all the examples of this section so we start by mentioning them
by category. For the design, we use $M=10,000$ independent trials so we can appreciate 
the coverage probabilities up to the fourth digit. We evaluate numerically the 
$\hbpi_n$-approximate posterior distribution of the SwiZs 
and the parametric bootstrap distribution based on $S=10,000$ replicates.
We measure the coverage probabilities 
at $50\%,75\%,90\%,95\%$ and $99\%$ levels. Although sometimes we do not report 
all of them for more clarity of the presentation, they are
however shown in Appendix for more transparency. 

We select five different scenarii. First, we start with a toy example
by considering a standard Student's $t$-distribution with unknown degrees of freedom (Example~\ref{ch2:ex:stud}).
Although the Student distribution is ubiquitous in statistics since at least Gosset's Biometrika paper (\cite{student1908probable}), 
there are no simple tractable way to construct an interval of uncertainty around the degrees of freedom.
In addition, the degrees of freedom is a parameter that gauges the tail of the distribution
and is not particularly easy to handle. The existence of the moments of this distribution depends
upon the values that this parameter takes. We take a particular interest in small values of this parameter for which, 
for example the variance or the kurtosis are infinite.
\begin{example}[standard $t$-distribution with unknown degrees of freedom]\label{ch2:ex:stud}
	Let $x_i\sim t(\theta)$, $i=1,\cdots,n$, be \iid\;with density 
	\begin{equation}\label{ch2:eq:stud}
		f(x_i,\theta)=\frac{{\left(1+\frac{x_i^2}{\theta}\right)}^{-\frac{\theta+1}{2}}}
		{\sqrt{\theta}\B\left(\frac{1}{2},\frac{\theta}{2}\right)},
	\end{equation}
	where $\theta$ represents the degrees of freedom and $\B$ is the beta function.
	We consider the likelihood score function as the estimating function and 
	we take the MLE as the auxiliary estimator. In this situation, $\Theta$ and $\Pi$ are
	equivalent, and thus, there are no reasons to disqualify the parametric bootstrap. 
	Substituing $\theta$ by $\pi$ in the Equation~\ref{ch2:eq:stud},
	taking then the derivative with respect to $\pi$ of the log-density
	leads to the following
	\begin{equation*}
		\Phi_n(\theta,\u,\pi) = \psi\left(\frac{\pi+1}{2}\right)-\psi\left(\frac{\pi}{2}\right)
		-\frac{1}{n}\sum_{i=1}^n\ln\left(\frac{{g(\theta,\u_i)}^2+1}{\pi}\right)+
		\frac{1}{n}\sum_{i=1}^n\frac{{g(\theta,\u_i)}^2-1}{{g(\theta,\u_i)}^2+\pi},
	\end{equation*}
	where $\psi$ is the digamma function. We now verify 
	Assumption~\ref{ch2:ass:bvp} so Theorem~\ref{ch2:thm:freq} can be invoked.
	Suppose Assumption~\ref{ch2:ass:dr} holds so we can write the following scalar-valued function
	\begin{equation*}
		\varphi_{\hat{\pi}_n}(\theta,w)=\frac{1}{2}\psi\left(\frac{\hat{\pi}_n+1}{2}\right)-
		\frac{1}{2}\psi\left(\frac{\hat{\pi}_n}{2}\right)
		-\frac{1}{2}\ln\left(\frac{{g(\theta,w)}^2+1}{\hat{\pi}_n}\right)
		+\frac{1}{2}\frac{{g(\theta,w)}^2-1}{{g(\theta,w)}^2+\hat{\pi}_n},
	\end{equation*}
	where $\hat{\pi}_n$ is fixed.
	The first derivative with respect to $\theta$ is given by
	\begin{equation}\label{ch2:eq:stud:bvp}
		\frac{\partial}{\partial\theta}\varphi_{\hat{\pi}_n}(\theta,w) =
		g(\theta,w)\frac{\partial}{\partial\theta}g(\theta,w)\left[\frac{\hat{\pi}_n-1}
			{{\left({g(\theta,w)}^2+\hat{\pi}_n\right)}^2}-\frac{1}{{g(\theta,w)}^2+1}\right].
	\end{equation}
	Substituing $(\partial/\partial\theta)g$ by $(\partial/\partial w)g$ gives the
	first derivative with respect to $w$. The derivative exists everywhere so $K_n=\emptyset$.
	Therefore, if the generating function $g(\theta,w)$ is once continuously differentiable in both
	its arguments then Assumption~\ref{ch2:ass:bvp} (\textit{i}) is satisfied.

	The determinant here is $\lvert\frac{\partial}{\partial\theta}\varphi_{\hat{\pi}_n}(\theta,w)\rvert$.
	It will be zero on a countable set of points: if $g(\theta,w)=0$, if $(\partial/\partial\theta)g(\theta,w)=0$
	or if the rightest term of the Equation~\ref{ch2:eq:stud:bvp} is 0. 
	Substituing $(\partial/\partial\theta)g$ by $(\partial/\partial w)g$ gives the
	same analysis. Hence, the determinant of the derivatives of the estimating function
	is almost everywhere non-null and Assumption~\ref{ch2:ass:bvp} (\textit{ii}) is satisfied.
	
	Eventually, we clearly have that
	\begin{equation*}
		\lim_{\lvert g\rvert\to\infty}\left\lvert\varphi_{\hat{\pi}_n}(\theta,w)\right\rvert=+\infty.
	\end{equation*}
	As a consequence, given that $\lim_{\lVert(\theta,w)\rVert\to\infty}\lvert g(\theta,w)\rvert=\infty$,
	Assumption~\ref{ch2:ass:bvp} (\textit{iii}) is satisfied.

	In the light of these findings,
	the choice of generating function is crucial and there are many 
	candidates~\cite[see e.g.][]{devroye1986nonuniform}.
	The inverse cumulative distribution function is a natural choice, 
	but a numerically complicated one in this case. Indeed, 
	it can be obtained by
	\begin{equation*}
		g_1(\theta,u_1) = \sign\left(u_1-\frac{1}{2}\right){\left(\frac{\theta(1-z)}{z}\right)}^{1/2},
	\end{equation*}
	where $u_1\sim\U(0,1)$ and $z$ is equal to the incomplete beta function inverse parametrized
	by $\theta$ and depending on $u_1$. An alternative choice, numerically and analytically simpler,
	is to consider Bailey's polar algorithm~\cite{bailey1994polar}, which is given by
	\begin{equation*}
		g_2(\theta,\u_2) = u_{2,1}\sqrt{\frac{\theta}{u_{2,2}}\left(u_{2,2}^{-2/\theta}-1\right)},
	\end{equation*}
	where $u_{2,2}\eqd u_{2,1}^2+u_{2,3}^2$ if $u_{2,2}\leq1$ and $u_{2,1},u_{2,3}\sim\U(-1,1)$.
	Clearly $g_2(\theta,\u_2)$ is once continuously differentiable in each of its arguments and 
	the limit is $\lim_{(\theta,u_{2,1},u_{2,2})\to(\infty,1,1)}\lvert g_2(\theta,u_{2,1},u_{2,3})\rvert = \infty$.
	Hence, even if $w$ is unknown, these results strongly suggests that the conditions of Theorem~\ref{ch2:thm:freq} hold,
	and as a conclusion, any intervals built 
	on the percentiles of the distribution of $\hat{\theta}_n$
	given $\hat{\pi}_n$ have exact frequentist coverage.


	The coverage probabilities in the Table~\ref{ch2:tab:st1} below are computed 
	for three different values of $\theta_0=\{1.5,3.5,6\}$ and a sample size of
	$n=50$. When $\theta_0=1.5$, the variance of a Student's random variable
	is infinite and the skewness and kurtosis of the distribution are undefined.
	When $\theta_0=3.5$, the variance is finite and the kurtosis is infinite.
	When $\theta_0=6$, the first five moment exists.
	\begin{table}[h!]
		\centering
		\begin{tabular}{@{}ccccccccccccc@{}}
			\toprule
			&& \multicolumn{3}{c}{SwiZs} & \multicolumn{3}{c}{parametric bootstrap} 
			& \multicolumn{3}{c}{BCa bootstrap}\\
			$\theta_0$ & $\alpha$ & $\hat{c}$ & $\bar{I}$ & $\bar{s}$ & $\hat{c}$ & $\bar{I}$ & $\bar{s}$
			& $\hat{c}$ & $\bar{I}$ & $\bar{s}$  \\
			\midrule
			1.5 & 50\% &50.66\%&0.5129&0.1622&49.13\%&0.5794&0.0358&47.69\%&0.4906&0.0333\\
			    & 75\% &75.39\%&0.8839&&73.27\%&1.0504&&71.64\%&0.8607& \\
			    & 90\% &90.15\%&1.2861&&87.03\%&1.6734&&86.64\%&1.2815& \\
			    & 95\% &94.68\%&1.5540&&91.42\%&2.1935&&91.82\%&1.5800& \\
			    & 99\% &98.84\%&2.1052&&96.05\%&3.8820&&97.13\%&2.2714& \\[.4em]
			3.5 & 50\% &50.08\%&1.7594&0.2010&47.65\%&2.8832&0.0349&44.94\%&1.8716&0.0322 \\
			    & 75\% &74.62\%&3.2780&&70.36\%&6.6243&&68.80\%&3.7372& \\
			    & 90\% &90.39\%&5.2129&&84.50\%&20.665&&84.36\%&6.5202& \\
			    & 95\% &94.85\%&6.8416&&89.63\%&240.11&&90.62\%&9.6584& \\
			    & 99\% &98.73\%&10.788&&95.11\%&3104.1&&95.60\%&29.011& \\[.4em]
			6   & 50\% &48.61\%&4.2027&0.2093&46.54\%&11.463&0.0342&44.29\%&4.6886&0.0305 \\
			    & 75\% &74.39\%&8.3688&&68.34\%&245.75&&69.99\%&12.245& \\
			    & 90\% &89.56\%&16.087&&80.83\%&2586.4&&87.45\%&41.335& \\
			    & 95\% &94.61\%&26.250&&85.06\%&3376.8&&93.05\%&515.51& \\
			    & 99\% &98.90\%&361.28&&95.55\%&4827.0&&95.94\%&2261.8& \\[.4em]
			\bottomrule
		\end{tabular}
		\caption{$\hat{c}$: estimated coverage probabilities,
		$\bar{I}$: median interval length,
		$\bar{s}$: average time in seconds to compute the intervals for
		one trial.}
		\label{ch2:tab:st1}
	\end{table}

	The SwiZs is accurate at all the confidence levels with a maximum discrepancy of
	1.39\% in absolute value. This is very reasonable considering the numerical
	task we perform. In comparison, the parametric bootstrap has a minimum discrepancy
	of $0.87\%$ for an average of $4.44\%$. The SwiZs is also more efficient,
	it dominates the parametric bootstrap with a median interval length systematically
	smaller. The parametric bootstrap is however about six times faster than the SwiZs 
	to compute the intervals. The comparison is not totally fair in disfavor of the 
	SwiZs as we were able here to use directly the log-likelihood for the parametric bootstrap,
	which is numerically simpler to evaluate than the estimating functions.
	We also bear the comparison with the bias-corrected and accelerated (BCa) resampling bootstrap
	of \cite{efron1994introduction}.
	Performances of this bootstrap scheme are comparable to the parametric bootstrap.
	Finally, when considered in absolute value, 0.2 second do not seem to be a hard
	effort for obtaining interval which is nearly exact and shorter.
\end{example}

Second, we consider a more practical case with the two-parameters Lomax distribution (\cite{lomax1954business})
(Example~\ref{ch2:ex:lomax}), also known as the Pareto II distribution. 
This distribution has been used to characterise wealth and income
distributions as well as business and actuarial losses (see~\cite{kleiber2003statistical} and the references therein).
Because of this close relationship to the application, we also measure the coverage probabilities
of the Gini index, the value-at-risk and the expected shortfall, 
quantities that may be of interest for the practitioner. The maximum likelihood estimator has been
shown in~\cite{giles2013bias} to suffer from small sample bias when $n$ is relatively small and the parameters are close
to the boundary of the parameter space. We add their proposal for bias adjustment 
to the basket of comparative methods. To keep the comparison fair, we use a similar simulation scenario 
to the ones they proposed, which were also motivated by their closeness to situations encountered in practice. 
Situations where the Lomax distribution is employed has been shown
to suffer from influential outliers ever since at least~\cite{victoria1994robust}, we therefore
consider, in a second time, the weighted maximum likelihood (\cite{field1994robust}) as the auxiliary estimator
to gain robustness. Interestingly, the weighted maximum likelihood estimator is generally not a consistent estimator
(see~\cite{dupuis2002robust,moustaki2006bounded}) so the parametric bootstrap cannot be invoked
directly, whereas, on the countrary, the SwiZs may be employed without any particular care.

\begin{example}[two-parameters Lomax distribution]\label{ch2:ex:lomax}
	Let $x_i\sim\text{Lomax}(\bt)$, $i=1,\cdots,n$, $\bt=(b,q)$, be \iid\;
	with density
	\begin{equation}\label{ch2:eq:lomax}
		f(x_i,\bt) = \frac{q}{b}\left(1 + \frac{x_i}{b}\right)^{-q-1},\quad x_i>0,
	\end{equation}
	where $b,q>0$ are shape parameters.
	We consider the likelihood score function as the estimating function and 
	we take the MLE as the auxiliary estimator. The parameter sets $\bT$ and $\bPi$ are
	equivalent with this setup, and thus, the parametric bootstrap may be employed. 
	Substituing $\bt$ by $\bpi$ in the Equation~\ref{ch2:eq:lomax},
	taking then the derivative with respect to $\bpi$ of the log-density
	leads to the following
	\begin{equation*}
		\bP_n(\bt,\u,\bpi) = \begin{pmatrix} \frac{1}{\pi_2} - \sum_{i=1}^n\log\left(1+\frac{g(\bt,\u_i)}{\pi_1}\right) \\[1em]
			-\frac{1}{\pi_1} + \frac{(\pi_2+1)}{\pi_1}\sum_{i=1}^n\frac{g(\bt,\u_i)}{\pi_1+g(\bt,\u_i)} \end{pmatrix}.
	\end{equation*}
	We now verify Assumption~\ref{ch2:ass:bvp} so Theorem~\ref{ch2:thm:freq} can be invoked.
	Suppose Assumption~\ref{ch2:ass:dr} on the existence of a random variable with the same
	dimensions as $\bt$ holds, and let denote it by $\w={(w_1\; w_2)}^T$.
	Now assume that we can re-express the estimating function as follows 
	\begin{equation*}
		\bvP_{\hbpi_n}(\bt,\w) = 
		\begin{pmatrix} \frac{1}{\hat{\pi}_2} - \log\left(1+\frac{g(\bt,w_1)}{\hat{\pi}_1}\right) \\[1em]
			\frac{(\hat{\pi}_2+1)g(\bt,w_2)}{\hat{\pi}_1^2 + \hat{\pi}_1g(\bt,w_2)}-\frac{1}{\hat{\pi}_1}
		\end{pmatrix},
	\end{equation*}
	where $\hbpi_n$ is fixed.
	The Jacobian matrix with respect to $\bt$ is given by
	\begin{equation*}
		D_{\bt}\bvP_{\hbpi_n}(\bt,\w) = 
		\begin{pmatrix}
			\kappa_1(\bt)D_{\bt}g(\bt,w_1) \\[1em]
			\kappa_2(\bt)D_{\bt}g(\bt,w_2)
		\end{pmatrix},
	\end{equation*}
	where
	\begin{align*}
		\kappa_1(\bt) &= \frac{-1}{\hat{\pi}_1 + g(\bt,w_1)} \\
		\kappa_2(\bt) &= \frac{\hat{\pi}_1^2\left(\hat{\pi}_2+1\right)}{{\left(\hat{\pi}_1^2+\hat{\pi}_1g(\bt,w_2)\right)}^2}.
	\end{align*}
	Note that $\hbpi_n$ and $g(\bt,\w)$ are strictly positive, so $\kappa_1(\bt)<0$ and $\kappa_2(\bt)>0$.	
	Substituing $D_{\bt}g$ by $D_{\w}g$ leads to the Jacobian matrix with respect to $\w$, given by
	\begin{equation*}
		D_{\w}\bvP_{\hbpi_n}(\bt,\w) = 
		\begin{pmatrix}
			\kappa_1(\bt)\frac{\partial}{\partial w_1}g(\bt,w_1) & 0 \\[1em]
			0 & \kappa_2(\bt)\frac{\partial}{\partial w_2}g(\bt,w_2)
		\end{pmatrix}.
	\end{equation*}
	We see by inspection that the derivatives are defined everywhere and $\K_n=\{\emptyset\}$. 
	If $D_{\bt}g$ and $D_{\w}g$ exist and are continuous,
	then Assumption~\ref{ch2:ass:bvp} (\textit{i}) is satisfied.

	The determinants are given by
	\begin{align*}
		\det\left(D_{\bt}\bvP_{\hbpi_n}(\bt,\w)\right) &= \kappa(\bt,\w)
		\left[\frac{\partial}{\partial a}g(\bt,w_1)\frac{\partial}{\partial b}g(\bt,w_2)-
			\frac{\partial}{\partial a}g(\bt,w_2)\frac{\partial}{\partial b}g(\bt,w_1)\right] \\
		\det\left(D_{\bt}\bvP_{\hbpi_n}(\bt,\w)\right) &= \kappa(\bt,\w)
		\frac{\partial}{\partial w_1}g(\bt,w_1)\frac{\partial}{\partial w_2}g(\bt,w_2),
	\end{align*}
	where $\kappa(\bt,\w) = \kappa_1(\bt)\kappa_2(\bt)$ and $\kappa(\bt,\w)<0$.
	The only scenarii where these determinants are zero are whether all the partial derivatives
	are zero, or if $(\partial/\partial a)g(\bt,w_1) (\partial/\partial b)g(\bt,w_2)
	= (\partial/\partial a)g(\bt,w_2)\;(\partial/\partial b)g(\bt,w_1)$. 
	Since the Lomax random variables are absolutely continuous, it is impossible
	for the generating function to be flat on $\bt$ and on $\w$, except maybe in 
	extreme cases. Therefore, situations where the determinants are zero are countable,
	and Assumption~\ref{ch2:ass:bvp} (\textit{ii}) is satisfied.

	Suppose the generating function satisfies the following property:
	\begin{equation*}
		\lim_{\lVert(\bt,w_1)\rVert\to\infty}g(\bt,w_1)=\infty.
	\end{equation*}
	Since the limit of the natural logarithm tends to infinity when its argument diverges, 
	we clearly have that
	\begin{equation*}
		\lim_{\lVert(\bt,\w)\rVert\to\infty}\left\lVert\bvP_{\hbpi_n}(\bt,\w)\right\rVert=+\infty,
	\end{equation*}
	and as a consequence, Assumption~\ref{ch2:ass:bvp} (\textit{iii}) is satisfied.
	
	It remains to demonstrate that a generating function satisfies the above properties.
	A natural and computationally easy choice for the generating function is the
	inverse cdf, it is given by
	\begin{equation*}
		g(\bt,u) = b + bu^{-1/q},\quad u\sim\U(0,1).
	\end{equation*}
	Clearly the generating function is once continuously differentiable in each $(b,q,u)$.
	The only possibilities for the partial derivatives of $g$ to be zero are whether
	$q=\{+\infty\}$ or $u=\{0\}$. The generating function tends to infinity when
	$b$ diverges whereas it remains constant when $q$ or $u$ diverges.
	All these findings strongly suggest that Theorem~\ref{ch2:thm:freq} is applicable here,
	and as a conclusion that any intervals built on the percentiles of the SwiZs distribution
	lead to exact frequentist coverage probabilities.

	However, the situation is less optimistic with the weighted maximum likelihood. Indeed,
	the estimating function is typically modified as follows:
	\begin{equation*}
		\widetilde{\bP}_n\left(\bt,\u,\bpi\right) = \mathrm{w}(\bt,\u,\bpi,k)\bP_n\left(\bt,\u,\bpi\right),
	\end{equation*}
	where $\mathrm{w}(\bt,\u,\bpi,k)$ is some weight function typically taking values in $[0,1]$
	that depends upon a tuning constant $k$. Usual weight functions are Huber's type (\cite{huber1964robust}) and 
	Tukey's biweighted function (\cite{beaton1974fitting}); see~\cite{hampel2011robust} for a textbook
	on robust statistics. For an estimating function to be robust, the weight function either
	decreases to 0 or remains constant for large values of $x$. As a consquence, at least two out of the three hypothesis
	of Assumption~\ref{ch2:ass:bvp} do not hold. Indeed, the determinants will be zero
	on an uncountable set and $\lim_{\lVert(\bt,\w)\rVert\to\infty}\widetilde{\bP}_n<\infty$.
		
	For the simulations, we set $\bto={(2\;\;2.3)}^T$ and use $n=\{35,50,100,150,250,500\}$ 
	as sample sizes. As already mentioned, this setup is close to the ones proposed in~\cite{giles2013bias}, and
	we thus add their proposal for correcting the bias of the maximum likelihood estimator to the basket
	of the compared methods. The bias-adjustment estimator is given by
	\begin{equation*}
		\hbt_{\text{BA},n}^{(s)} = \hbpi_n - \mathbf{B}(\hbpi_n)\mathbf{A}(\hbpi_n)\vect\left(\mathbf{B}(\hbpi_n)\right),	
	\end{equation*}
	where
	\begin{equation*}
		\mathbf{A}(\bpi) = n\begin{pmatrix}
			\frac{2\pi_2}{\pi_1^3(\pi_2+2)(\pi_2+3)} & \frac{-1}{\pi_1^2(\pi_2+1)(\pi_2+2)}
			& \frac{\pi_2}{\pi_1^2{(\pi_2+2)}^2} & \frac{-1}{\pi_1{(\pi_2+1)}^2} \\
			\frac{-1}{\pi_1^2(\pi_2+1)(\pi_2+2)} & 0 & \frac{-1}{\pi_1{(\pi_2+1)}^2} & \frac{1}{\pi_2^3}
		\end{pmatrix},
	\end{equation*}
	and
	\begin{equation*}
		\mathbf{B}^{-1}(\bpi) = n\begin{pmatrix}
			\frac{\pi_2}{\pi_1^2(\pi_2+2)} & \frac{-1}{\pi_1(\pi_2+1)} \\
			\frac{-1}{\pi_1(\pi_2+2)} & \frac{1}{\pi_2^2}
		\end{pmatrix}.
	\end{equation*}
	\begin{figure}[htbp]
		\begin{center}
			\includegraphics[width=0.9\textwidth, keepaspectratio]{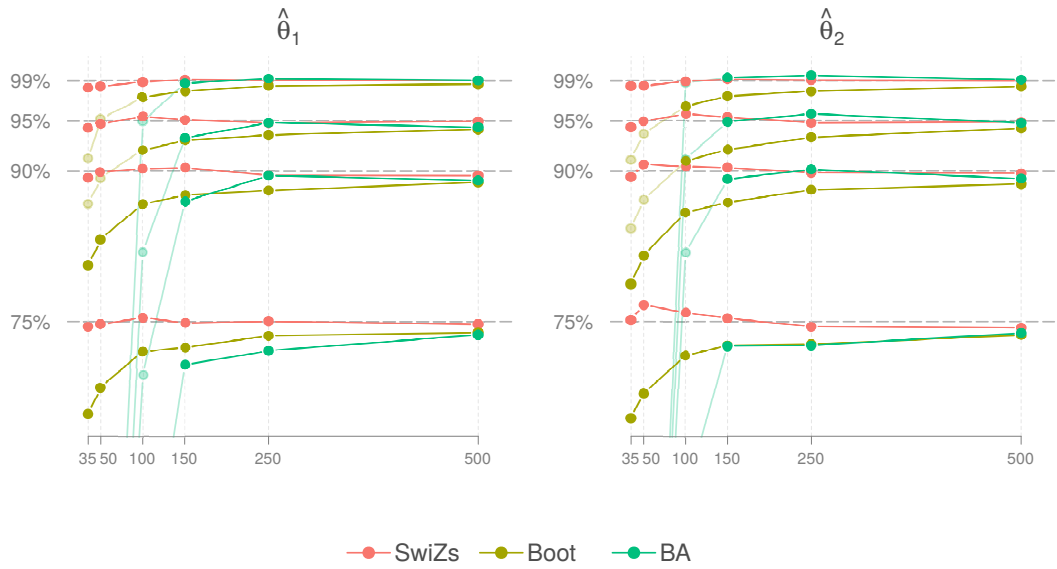}
		\end{center}
		\caption{Coverage probabilities of the SwiZs, the parametric bootstrap (Boot) and the 
			bias-adjustment (BA) proposal of~\cite{giles2013bias} for different sample sizes.
			\textit{On the left panel} is the coverage for the first estimator, and the second is on the \textit{right}.
			The gray horizontal dotted-lines indicate the perfect coverage probabilities. The closer
			to these lines is the better.}
		\label{ch2:fig:lomax1}
	\end{figure}
	\begin{figure}[htbp]
		\begin{center}
			\includegraphics[width=0.9\textwidth, keepaspectratio]{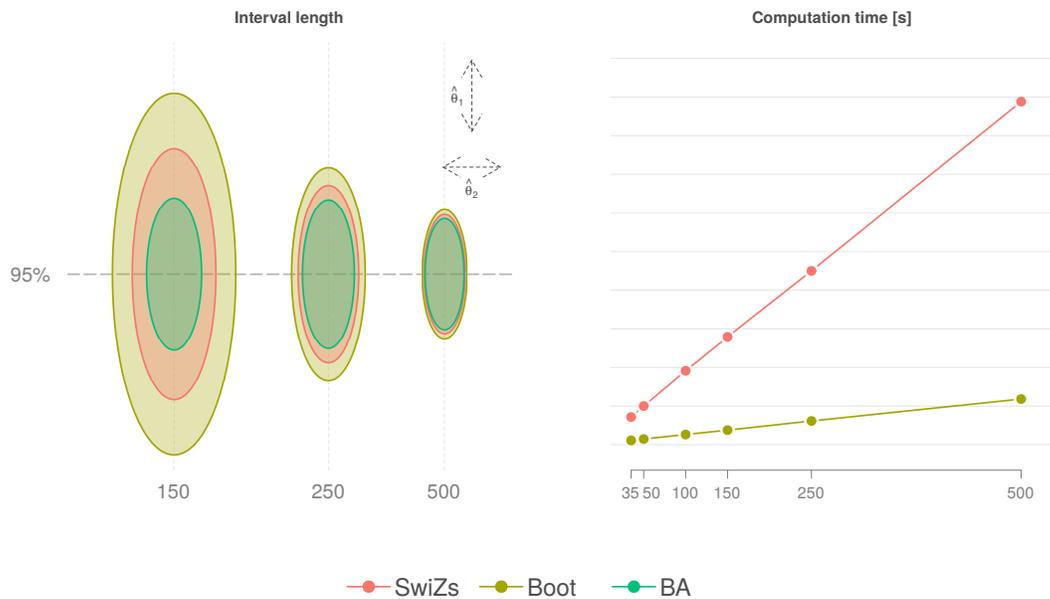}
		\end{center}
		\caption{\textit{On the left panel}: representation of the median interval lengths for a confidence level of 95\%
			for the SwiZs, the parametric bootstrap (Boot) and the bias-adjustment (BA) proposal of~\cite{giles2013bias}
			for three different sample sizes. The ellipses are just a representation and do not reflect the real shapes of
			the confidence regions. All the ellipses are on the same scale. The centre of the ellipses 
			is chosen for aesthetical reason and have no special meaning.
			The $y$-axis corresponds to the median interval length of the first parameter,
			the $x$-axis the one of the second parameter. The smaller the ellipse is, the better it is.
			\textit{On the right panel}: the average computational time in seconds of the 
			SwiZs and the Boot for the different sample sizes. Note that the computational
			time of the the BA (not on the figure) is quasi-identical to the Boot. The lower is the better.}
		\label{ch2:fig:lomax2}
	\end{figure}
	\begin{figure}[htbp]
		\begin{center}
			\includegraphics[width=0.9\textwidth, keepaspectratio]{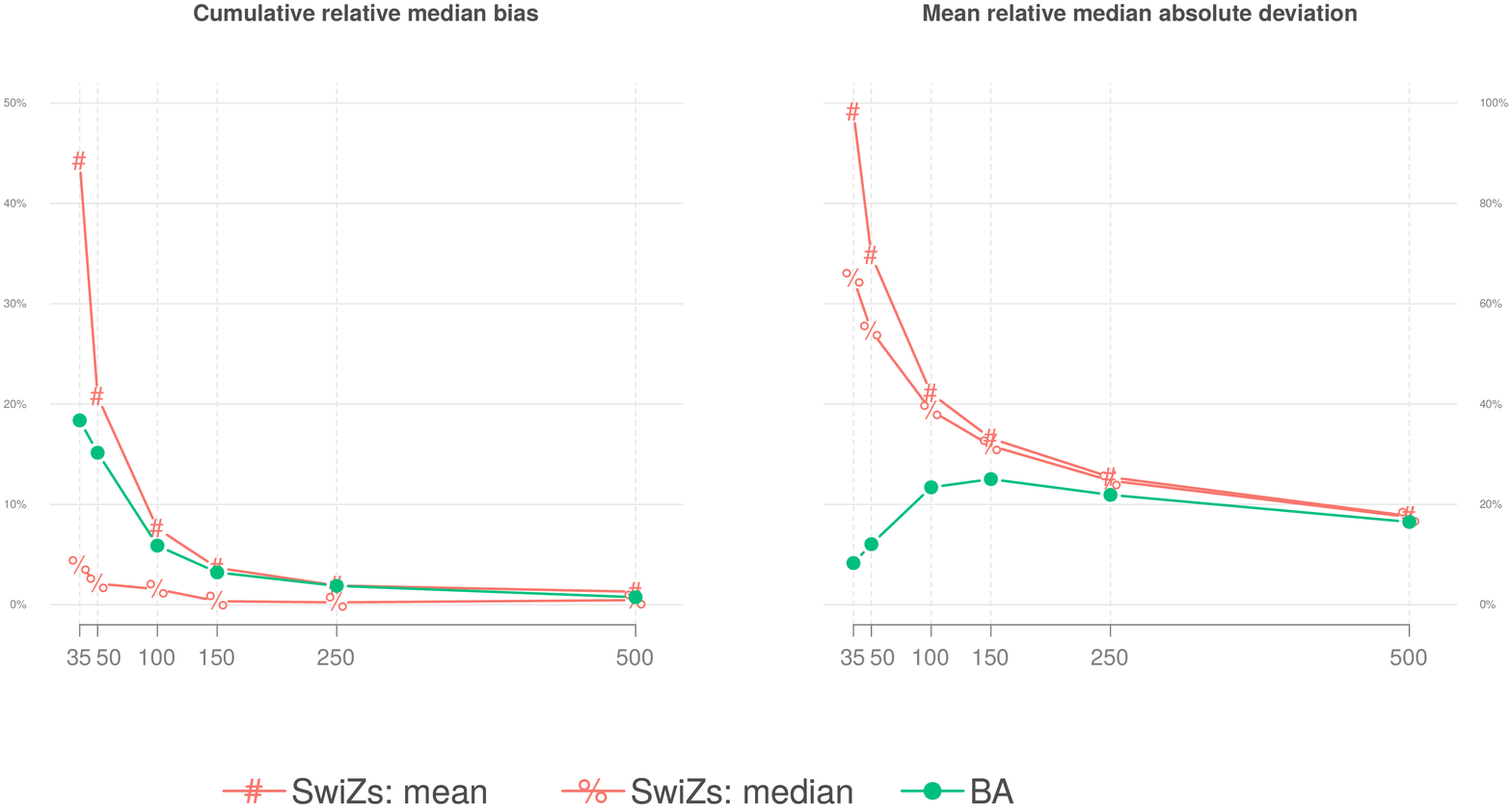}
		\end{center}
		\caption{\textit{On the left panel}: the sum of absolute value of the median bias for the two estimators
			divided by their respective true values
			for the mean of SwiZs distribution, the median of the SwiZs distribution and the bias-adjustment (BA)
			proposal of~\cite{giles2013bias} evaluated on the different sample sizes.
			\textit{On the right panel}: likewise the \textit{left panel}, but for a different measure: 
			the average of the median absolute deviation for the two estimators divided by their respective true values.
			The lower is the better.
			}
		\label{ch2:fig:lomax3}
	\end{figure}
	All the detailed results of simulation are in Appendix~\ref{ch2:app:lomax}.
	In Figure~\ref{ch2:fig:lomax1}, we discover that the SwiZs has very accurate 
	coverage probabilities at all levels and all sample sizes which seems in accordance
	with Theorem~\ref{ch2:thm:freq} and the subsequent verification analysis for this example.
	For sample sizes greater or equal to 250, the parametric bootstrap and the bias-adjustment
	proposal of~\cite{giles2013bias} meet the performance of the SwiZs at almost every levels.
	However, below a sample of 150, the performance of the bias-adjustment are catastrophic.
	This may only be explained by the following phenomenon: the maximum likelihood is adjusted
	too severely for small values of $n$, and for a large proportion of the time the resulting
	bias-adjusted estimator is out of the parameter space $\bT$. We report
	in Table~\ref{ch2:tab:lomax0} our empirical findings. This phenomenon affects not only the
	coverage probabilities but also the variation of this estimator (Figure~\ref{ch2:fig:lomax3})
	and the length of the confidence intervals (Figure~\ref{ch2:fig:lomax2}). Here 
	we opted for discarding the inadmissible values (negative), thereby reducing 
	artificially the variance and the length of the confidence intervals of the bias-adjustment.
	All the other methods considered do not suffer from the positivity constrain on $\bt$
	and thus we do not attempt to tackle this limitation of the bias-adjustment method.
	\begin{figure}[htbp]
		\begin{center}
			\includegraphics[width=0.9\textwidth, keepaspectratio]{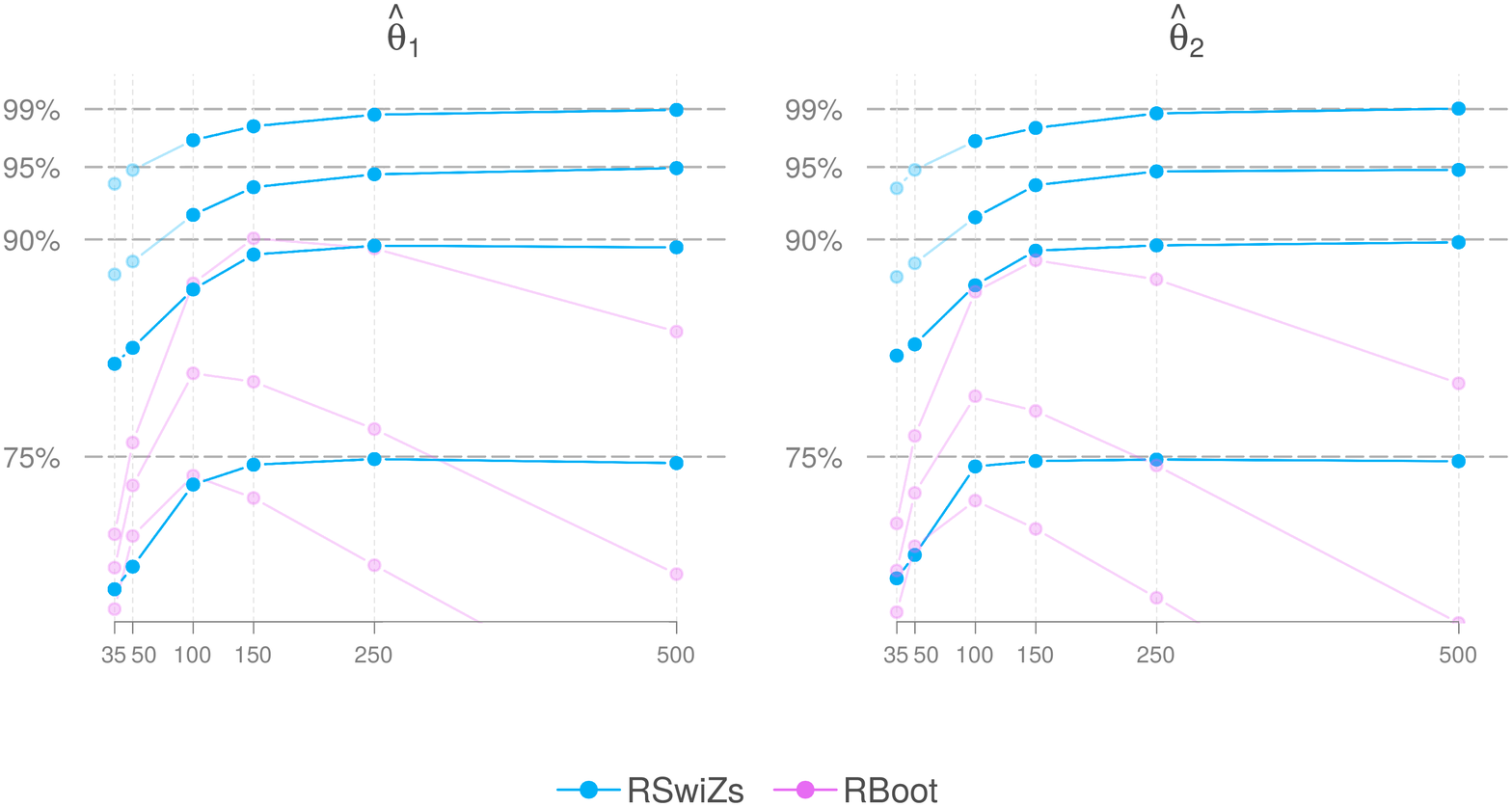}
		\end{center}
		\caption{Coverage probabilities for different sample sizes 
			of the SwiZs (RSwiZs) and the parametric bootstrap (RBoot)
			when taking the weighted maximum likelihood as auxiliary estimator.
			\textit{On the left panel} is the coverage for the first estimator, and the second is on the \textit{right}.
			The gray horizontal dotted-lines indicate the perfect coverage probabilities. The closer
			to these lines is the better.}
		\label{ch2:fig:lomax4}
	\end{figure}
	\begin{figure}[htbp]
		\begin{center}
			\includegraphics[width=0.9\textwidth, keepaspectratio]{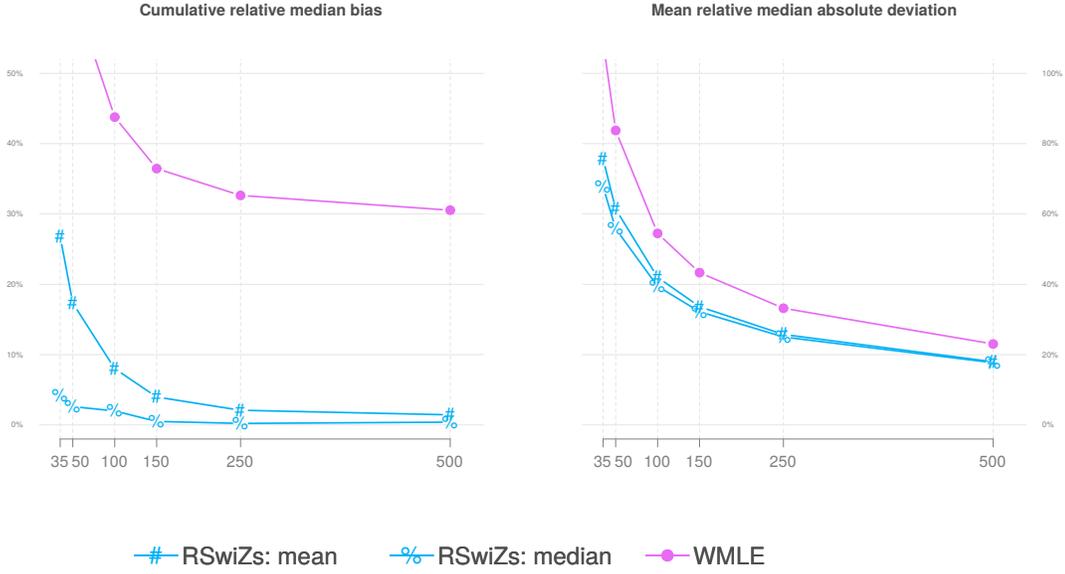}
		\end{center}
		\caption{\textit{On the left panel}: 
			the sum of absolute value of the median bias for the two estimators
			divided by their respective true values for different sample sizes 
			for the mean of SwiZs distribution (RSwiZs: mean), the median of the SwiZs distribution (RSwiZs: median)
			when considering the weighted maximum likelihood (WMLE) as the auxiliary estimator.
			\textit{On the right panel}: likewise the \textit{left panel}, but for a different measure: 
			the average of the median absolute deviation for the two estimators divided by their respective true values.
			The lower is the better.
			}
		\label{ch2:fig:lomax5}
	\end{figure}
	\begin{table}[hbpt]
		\centering
		\begin{tabular}{@{}cccc@{}}
			\toprule
			$n=35$&$n=50$&$n=100$&$n=150$\\
			\midrule
			38.78\%&21.94\%&3.02\%&0.40\%\\
			\bottomrule
		\end{tabular}
		\caption{Empirical proportion of times the bias-adjusted maximum likelihood estimator is jointly out of the 
		parameter space $\bT$.}
		\label{ch2:tab:lomax0}
	\end{table}
	The SwiZs has shorter uncertainty intervals than the parametric bootstrap, however it is more demanding
	in computational efforts (Figure~\ref{ch2:fig:lomax2}). The computational comparison is not 
	entirely fair in disfavor of the SwiZs as here we take advantage that the maximum likelihood estimator 
	can be optimized directly on the log-likelihood, which is numerically easier to evaluate than the likelihood
	scores that constitues the estimating function. An unexpected good surprise emerges from Figure~\ref{ch2:fig:lomax3}
	where it seems that taking the median of the SwiZs leads to almost median unbiased point estimators.
	The same may be said when using the weighted maximum likelihood as the auxiliary estimator (Figure~\ref{ch2:fig:lomax5}).
	However, using a robust estimator as the auxiliary parameter do not offer interesting coverage probabilities
	in small samples (Figure~\ref{ch2:fig:lomax4}), which seems to indicate that Assumption~\ref{ch2:ass:bvp}
	may not be easily relaxed. The parametric bootstrap unsurprisingly fails completely when considering an inconsistent
	estimator. Eventually, the empirical distributions in Figure~\ref{ch2:fig:lomax6} reminds us of the difficulty
	of estimating confidence regions.
	\begin{figure}[htbp]
		\begin{center}
			\includegraphics[width=0.9\textwidth, keepaspectratio]{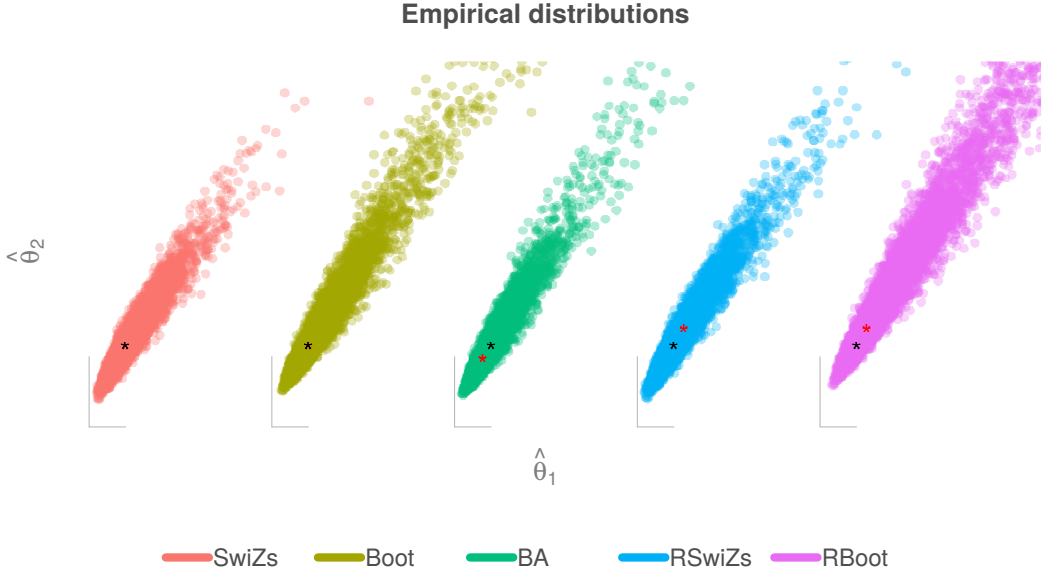}
		\end{center}
		\caption{Empirical conditional distribution for a given $\hbpi_n$ and a sample size of $n=100$
			of the SwiZs, the parametric bootstrap (Boot),
			the bias-adjustment proposal of~\cite{giles2013bias} (BA) when considering the maximum likelihood
			as the auxiliary estimator and the SwiZs (RSwiZs) and the parametric bootstrap (RBoot)
			when considering the weighted maximum likelihood as the auxiliary estimator.
			The \textit{black star} represents $\bto={[2\;2.3]}^T$ whereas the \textit{red stars}
			indicate the values of $\hbpi_n$: the maximum likelihood estimator for SwiZs and Boot,
			the bias-adjustment for BA and the weighted maximum likelihood estimator for RSwiZs and RBoot.
			The ``\textit{try square}'' at \textit{bottom-left-corner} of each distribution has both sides of 
			length 2 and has its corner exactly at the $(0,0)$-coordinate.}
		\label{ch2:fig:lomax6}
	\end{figure}
\end{example}
\newpage\newpage
Third, we investigate a linear mixed-model. These models are very common in statistics
as they incorporate both parameters associated with an entire population and parameters
associated with individual experimental units facilitating thereby the study of, for examples,
longitudinal data, multilevel data and repeated measure data. Although being widespread,
the inference on the parameters remain a formidable task. We study a rather simple model,
namely the random intercept and random slope model when data is balanced. 
\begin{example}[random intercept and random slope linear mixed model]\label{ch2:ex:mlm}
	Consider the following balanced Gaussian mixed linear model expressed for the $i$th individual
	as
	\begin{equation*}
		\y_i = (\beta_0 + \alpha_i) \1_m + (\beta_1 + \gamma_i) \x_i + \be_i, \quad i=1,\cdots,n,
	\end{equation*}
	where $\be_i, \alpha_i$ and $\gamma_i$ are \iid\;according to centered Gaussian distributions
	with respective variances $\sigma^2_{\epsilon}\bI_m, \sigma^2_{\alpha}$ and $\sigma^2_{\gamma}$,
	$m$ being the number of replicates, the same for each individual, and $\1_m$ is
	a vector of $m$ ones. The vector of parameters of interest is $\bt = {\left(\beta_0,\beta_1,\sigma^2_{\epsilon},
	\sigma^2_{\alpha},\sigma^2_{\gamma}\right)}^{T}$.
	Let $\bpi={\left(\pi_0,\ldots,\pi_4\right)}^{T}$ be the corresponding 
	vector of auxiliary parameters.
	We take the MLE as the auxiliary estimator and thus consider the likelihood score function as the 
	estimating function. With this setup, the parameter spaces $\bT$ and $\bPi$ are equivalent,
	and the parametric bootstrap may be employed. 
	Denote by $N=nm$ the total sample size.
	The negative log-likelihood may be expressed as
	\begin{equation*}
		\bm{\ell}\left(\y,\bt\right)=k+\frac{1}{2N}\sum_{i=1}^n 
		\log\left(\det\left(\bO_i(\bt)\right)\right) + 
		{\left(\y_i-\beta_0\1_m-\beta_1\x_i\right)}^{T}{\bO_i}^{-1}(\bt)
		{\left(\y_i-\beta_0\1_m-\beta_1\x_i\right)},
	\end{equation*}
	for some constant $k$ and
	where $\bO_i(\bt)=\sigma^2_{\epsilon}\bI_m + \sigma^2_{\alpha}\1_m\1^T_m+\sigma^2_{\gamma}\x_i\x^T_i$
	is clearly a symmetric positive definite matrix.
	Taking the derivatives with respect to $\bt$, then substituing $\bt$ by $\bpi$
	and $\y_i$ by $\g(\bt,\u_i)$ leads to
	\begin{equation*}
		\bP_{N}\left(\bt,\u,\bpi\right) =
		\begin{pmatrix*}[l]
			\frac{-1}{N}\sum_{i=1}^n\z^T(\bt,\u_i,\bpi)\bO^{-1}_i(\bpi)\1_m \\[1em]	
			\frac{-1}{N}\sum_{i=1}^n\z^T(\bt,\u_i,\bpi)\bO^{-1}_i(\bpi)\x_i \\[1em]
			\frac{1}{2N}\sum_{i=1}^n\trace\left(\bO^{-1}_i(\bpi)\frac{\partial}{\partial\pi_j}\bO_i(\bpi)\right) \\
			\quad - \z^T(\bt,\u_i,\bpi)\bO^{-1}_i(\bpi)\frac{\partial}{\partial\pi_j}\bO_i(\bpi) \\
			\quad \times\bO^{-1}_i(\bpi)\z(\bt,\u_i,\bpi),\quad j=2,3,4
		\end{pmatrix*},
	\end{equation*}
	where $\z(\bt,\u_i,\bpi)=\g(\bt,\u_i)-\pi_0\1_m-\pi_1\x_i$ (see also~\cite{jiang2007linear} for more details
	on these derivations). The derivatives of $\bO_i(\bpi)$ are easily obtained: $(\partial/\partial\pi_2) \bO_i(\bpi) = \bI_m$,
	$(\partial/\partial\pi_3)\bO_i(\bpi)=\1_m\1^T_m$ and $(\partial/\partial\pi_4)\bO_i(\bpi)=\x_i\x^T_i$.
	Since they do not depend on parameters, let denotes $(\partial/\partial\pi_j)\bO_i(\bpi)\equiv\D_{ij}$.
	
	We now motivate the possibility to employ Theorem~\ref{ch2:thm:freq} by verifying Assumption~\ref{ch2:ass:bvp}.
	First, we suppose that a random variable $\w$ of the same dimension as $\bt$ exists. Then,
	we assume that the estimating function may be re-expressed as follows:
	\begin{equation*}
		\bvP_{\hbpi_N}\left(\bt,\w\right) = 
		\begin{pmatrix*}[l]
			\frac{-1}{N}\sum_{i=1}^n\z^T_i(\bt,w_0,\hbpi_N)\bO^{-1}_i(\hbpi_N)\1_m \\[1em]	
			\frac{-1}{N}\sum_{i=1}^n\z^T_i(\bt,w_1,\hbpi_N)\bO^{-1}_i(\hbpi_N)\x_i \\[1em]
			\frac{1}{2N}\sum_{i=1}^n\trace\left(\bO^{-1}_i(\hbpi_N)\D_{ij}\right) \\
			\quad - \z^T_i(\bt,w_j,\hbpi_N)\bO^{-1}_i(\hbpi_N)\D_{ij} \\
			\quad \times\bO^{-1}_i(\hbpi_N)\z_i(\bt,w_j,\hbpi_N),\quad j=2,3,4
		\end{pmatrix*},
	\end{equation*}
	where $\z_i(\bt,w_j,\hbpi_N)=\g(\bt,w_j)-\hat{\pi}_0\1_m-\hat{\pi}_1\x_i$, $j=0,1,2,3,4$, and $\hbpi_N$ is fixed.
	The Jacobian matrix with respect to $\bt$ is given by
	\begin{equation*}
		D_{\bt}\bvP_{\hbpi_N}\left(\bt,\w\right) = 
		\begin{pmatrix*}[l]
			\frac{-1}{N}\sum_{i=1}^nD_{\bt}\g^T(\bt,w_0)\bO^{-1}_i(\hbpi_N)\1_m \\[1em]	
			\frac{-1}{N}\sum_{i=1}^nD_{\bt}\g^T(\bt,w_1)\bO^{-1}_i(\hbpi_N)\x_i \\[1em]
			\frac{-1}{N}\sum_{i=1}^nD_{\bt}\g^T(\bt,w_j)\bO^{-1}_i(\hbpi_N)\D_{ij} \\
			\quad \times\bO^{-1}_i(\hbpi_N)\g(\bt,w_j),\quad j=2,3,4
		\end{pmatrix*}.
	\end{equation*}
	Substituing $D_{\bt}\g^T$ by $D_{\w}\g^T$ in the above delivers immediately the Jacobian matrix
	with respect to $\w$. Note that this second Jacobian is a diagonal matrix. Clearly, the
	differentiability and continuity of $\bvP_{\hbpi_N}$ depends exclusively upon the 
	differentiability and continuity of $\g$. Ergo, if $D_{\bt}\g$ and $D_{\w}\g$ exist and are continuous,
	then Assumption~\ref{ch2:ass:bvp} (\textit{i}) holds.

	These Jacobian matrices may have a null determinant under two circumstances:
	whether the generating function $\g$ is flat on $\bt$ and/or $\w$, and/or 
	whether they are linearly dependent. Since the Normal distribution is absolutely continuous,
	$\g$ may be flat only on extreme cases. The Jacobian $D_{\w}\bvP_{\hbpi_N}$
	is a diagonal matrix, so its determinant is null if and only if one of its diagonal element is
	null. Since both the design and $\hbpi_N$ are fixed,
	situations where $D_{\bt}\bvP_{\hbpi_N}$ is linearly dependent may occur if 
	the vectors $(\partial/\partial\theta_j)\g(\bt,\w) = k(\partial/\partial\theta_{j'})\g(\bt,\w), j\neq j',$
	for some constant $k\in\R$. But because $\w$ is random, this situation is unlikely to occur,
	and, depending on $\g$, Assumption~\ref{ch2:ass:bvp} (\textit{ii}) is plausible.
	
	Eventually, it clearly holds that
	\begin{equation*}
		\lim_{\lVert(\bt,\w)\rVert\to\infty}\left\lVert\bvP_{\hbpi_N}(\bt,\w)\right\rVert=\infty
	\end{equation*}
	if $\lVert\g(\bt,\w)\rVert\to\infty$ as $\lVert(\bt,\w)\rVert\to\infty$, so Assumption~\ref{ch2:ass:bvp} (\textit{iii})
	is satisfied given that $\g$ fulfills the requirement.

	Once again, the plausibility of Assumption~\ref{ch2:ass:bvp} is up to the choice of the generating function. 
	A popular choice is the following:
	\begin{equation*}
		\g(\bt,\u_i) = \beta_0\1_m + \beta_1\x_i + \bC_i(\bt)\u_i,\quad \u_i\sim\Nor\left(\0,\bI_m\right),
	\end{equation*}
	where $\bC_i(\bt)$ is the lower triangular Cholesky factor such that $\bC_i(\bt)\bC_i^T(\bt)=\bO_i(\bt)$. 
	It is straightforward to remark that $\g$ is once continuously differentiable in $\beta_0,\beta_1$ and $\u_i$. For the variances
	components, the partial derivatives of the Cholesky factor is given by
	Theorem A.1 in~\cite{sarkka2013bayesian}:
	\begin{equation*}
		\frac{\partial}{\partial\theta_j}\bC_i(\bt) = \bC_i(\bt)L\left(\bC^{-1}_i(\bt)\frac{\partial}
		{\partial\theta_j}\bO_i(\bt)\bC_i^{-T}(\bt)\right), \quad j=2,3,4,
	\end{equation*}
	where the function $L$ returns the lower triangular and half of the diagonal elements
	of the inputed matrix, that is:
	\begin{equation*}
		L_{ij}(\bA) = \left\{
			\begin{array}{lr}
				\bA_{ij}, & i>j,\\
				\frac{1}{2}\bA_{ij}, & i=j, \\
				0, & i<j.
			\end{array}\right.
	\end{equation*}
	The partial derivatives of the covariance matrix are given by: $(\partial/\partial\sigma^2_{\epsilon})\bO_i(\bt) = \bI_m$,
	$(\partial/\partial\sigma^2_{\alpha})\bO_i(\bt) = \1_m\1_m^T$ and $(\partial/\partial\sigma^2_{\gamma})\bO_i(\bt) = \x_i\x_i^T$.
	Hence, $\bC_i(\bt)$ is once differentiable. For the continuity of the partial derivative of $\bC_i(\bt)$, note
	that $\bC_i(\bt)$ and $\bC^{-1}_i(\bt)$ are once differentiable and thus continuous. Indeed,
	$(\partial/\partial\theta_j)\bC^{-1}_i(\bt) = -\bC^{-1}_i(\bt)[(\partial/\partial\theta_j)\bC_i(\bt)]\bC_i^{-1}(\bt)$.
	Eventually, $(\partial/\partial\theta_j)\bO_i(\bt)$ is constant in $\bt$, and therefore continuous.
	Since matrix product preserves the continuity, the Cholesky factor is once continuously differentiable.
	The partial derivatives of $\g$ may be zero if the design is null or if the pivotal quantity is zero,
	two extreme situations unlikely encountered. It is straightforward to remark that the estimating function
	diverges as $\bt$ and $\u_i$ tends to infinity. 
	All these findings make usage of Theorem~\ref{ch2:thm:freq} highly plausible.
	
	Let us turn our attention to simulations. We set $\bto = (1, 0.5, 0.5^2, 0.5^2, 0.2^2)^T$
	and considered $n=m=\{5,10,20,40\}$ such that $N=nm=\{25,\;100,\;400,\;1,600\}$.
	The detailed results of simulations may be found in the tables of Appendix~\ref{ch2:app:mlm}.
	In Figure~\ref{ch2:fig:mlm1}, we can observe the outstanding performances of the SwiZs in terms
	of coverage probabilities, which supports our analysis and the possibility of using
	Theorem~\ref{ch2:thm:freq}. The parametric bootstrap meets the performance of the SwiZs
	as the sample size increases, however, when the sample size is small, it is off the ideal level
	for the variance components. 
	The length of the marginal intervals of uncertainty are comparable
	between the two methods, except for the smallest sample size considered where it is 
	anyway harder to interpret the size of the interval of the parametric bootstrap
	since it is off the confidence level. 
	We also bear the comparison with profile likelihood confidence intervals
	which are based on likelihood ratio test. The coverage probabilities
	are almost undistinguishable from the SwiZs whereas interval lengths for
	variance components are the shortest. We interpret such good performances
	as follows: first, as shown in Example~\ref{ch2:ex:reg} on linear regression, asymptotic
	and finite sample distributions coincides in theory, coincidance that may be 
	still hold in the present case with balanced linear mixed model; second,
	larger intervals accounts for the fact that no simulations are needed.
	A good surprise appears in Figure~\ref{ch2:fig:mlm2}
	where the median of the SwiZs shows good performances in terms of relative
	median bias.
	\begin{figure}[htbp]
		\begin{center}
			\includegraphics[width=0.9\textwidth, keepaspectratio]{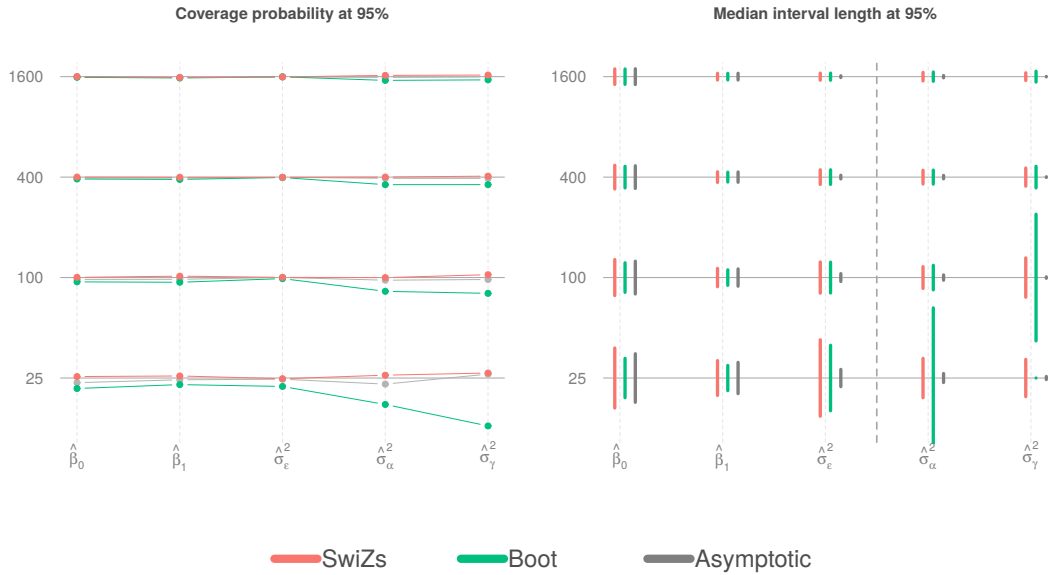}
		\end{center}
		\caption{\textit{On the left panel}:
			Representation of the coverage probabilities for different sample sizes of the 
			SwiZs, the parametric bootstrap (Boot) and the confindence intervals 
			based on the likelihood ratio test (Asymptotic) for the five estimators. The gray line 
			represents the ideal level of 95\% coverage probabilitiy. \textit{On the right panel}: 
			median length of the marginal intervals of uncertainty at a level of 95\%. For graphical reason,
			the lengths corresponding to $\hat{\sigma}^2_{\alpha}$ and $\hat{\sigma}^2_{\gamma}$ \textit{on the right}
			is downsized by a factor of 5 compared to the lengths corresponding to the other estimators.
			}
		\label{ch2:fig:mlm1}
	\end{figure}
	\begin{figure}[htpb]
		\begin{center}
			\includegraphics[width=0.9\textwidth, keepaspectratio]{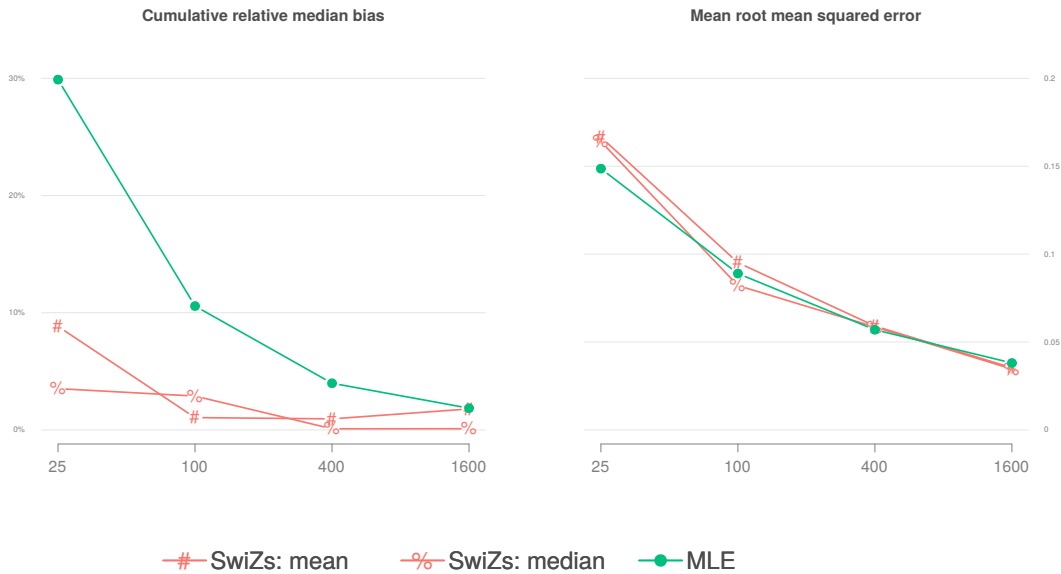}
		\end{center}
		\caption{\textit{On the left panel}: 
			the sum of absolute value of the median bias for the five estimators
			divided by their respective true values for different sample sizes 
			for the mean of SwiZs distribution (SwiZs: mean), the median of the SwiZs distribution (SwiZs: median)
			and the maximum likelihood estimator (MLE).
			\textit{On the right panel}: likewise the \textit{left panel}, but for a different measure: 
			the average of root mean squared error for the five estimators.
			For both panels, the lower is the better.
			}
		\label{ch2:fig:mlm2}
	\end{figure}
\end{example}
Fourth, we study inference in queueing theory models (see~\cite{shortle2018fundamentals} for a monograph).
In particular, we re-investigate the M/G/1 model studied by~\cite{heggland2004estimating,blum2010non,fearnhead2012constructing}.
Although the underlying process is relatively simple, there is no known closed-form for the likelihood function
and inference is not easy to conduct. 
\begin{example}[M/G/1-queueing model]\label{ch2:ex:gg1}
	Consider the following stochastic process
	\begin{equation*}
		x_i = \left\{\begin{array}{lr}
			v_i, & \text{if } \sigma^{\varepsilon}_i \leq \sigma^x_{i-1}, \\
			v_i + \sigma^{\varepsilon}_i - \sigma^x_{i-1}, & \text{if } \sigma^{\varepsilon}_i>\sigma^x_{i-1},
		\end{array}\right.
	\end{equation*}
	for $i=1,\cdots,n,$ where $\sigma^{\varepsilon}_i = \sum_{j=1}^i\varepsilon_j$, $\sigma^x_i = \sum_{j=1}^ix_j$, $v_i$ is
	\iid~according to a uniform distribution $\mathcal{U}(\theta_1,\theta_2)$, $0\leq\theta_1<\theta_2<\infty$
	and $\varepsilon_i$ is \iid~according to an exponential distribution $\mathcal{E}(\theta_3)$, $\theta_3>0$.
	In queueing theory, random variables have special meaning, for the $i$th customer:
	$x_i$ represents interdeparture time, 
	$v_i$ is service time and $\varepsilon_i$ corresponds to interarrival time.
	Only the interdeparture times $x_i$ are observed, $v_i$ and $\varepsilon_i$ are latent.
	All past information influence the current observation and therefore this process is not Markovian.
	Finding an ``appropriate'' auxiliary estimator is challenging as we now discuss.

	In this context, semi-automatic ABC approaches by~\cite{blum2010non} and~\cite{fearnhead2012constructing}
	use several quantiles as summary statistics for the auxiliary estimator. This method cannot be employed here
	for the SwiZs because, first, the restriction that $\dim(\bt)=\dim(\bpi)$ would be violated, and second,
	the quantiles are non-differentiables with respect to $\g$ and consequently, as already discussed, 
	Assumptions~\ref{ch2:ass:bvp} and~\ref{ch2:ass:bvp2} would not hold.
	However, \cite{heggland2004estimating} present different choices and motivate a particular auxiliary model with the following
	closed-form:
	\begin{equation*}
		f(x_i,\bpi) = \left\{\begin{array}{lr}
			0, & \text{if } x_i\leq\pi_1,\\
			{\left(\pi_2-\pi_1\right)}^{-1}
			\left[1 - \alpha\exp\left(-\pi_3^{-1}(x_i-\pi_1)\right)\right],
			& \text{if } \pi_1<x_i\leq\pi_2, \\
			\frac{\alpha}{\pi_2-\pi_1}\left[\exp\left(-\pi_3^{-1}(x_i-\pi_2)\right) - 
				\exp\left(-\pi_3^{-1}(x_i-\pi_1)\right)\right],
				& \text{if } x_i>\pi_2,
		\end{array}\right.
	\end{equation*}
	where $-1\leq\alpha\leq1$ is some constant. Motivations for this auxiliary model 
	are based on a graphical analysis of the sensitivity of $\hbpi_n(\bt)$ with respect
	to $\bt$ and the root mean squared errors performances of $\hbt_n$ on simulations.
	Unfortunately, Assumption~\ref{ch2:ass:bvp} is not satisfied with this choice. 
	Indeed, by taking the likelihood scores of the auxiliary model as the estimating 
	equation, one can realize that the score relative to $\pi_2$ is
	\begin{equation*}
		\Phi_{n,2}(\bt,\u,\bpi) = \left\{\begin{array}{lr}
			0, &\text{if } g(\bt,\u) < \pi_1,\\
			\frac{1}{\pi_2-\pi_1},&\text{if }\pi_1\leq g(\bt,\u)<\pi_2, \\
			\frac{1}{\pi_2-\pi_1} - \frac{\pi_3^{-1}e^{\pi_2/\pi_3}}{e^{\pi_2/\pi_3}-e^{\pi_1/\pi_3}},&
			\text{if } g(\bt,\u)\geq\pi_2,
		\end{array}\right.
	\end{equation*}
	hence, it does not depend on $\bt$!
	This result implies directly that all the partial derivatives with respect to $\bt$ and $\w$
	are null and $\det(\bvP_{\hbpi_n})=0$ for all $(\bt,\w)\in(\bT_n\times W_n)$.
	Assumption~\ref{ch2:ass:bvp2} is also violated and Theorem~\ref{ch2:thm:freq} cannot be invoked.
	Worse, the behaviour of this score does not depend on $n$ and 
	the identifiability condition in Assumption~\ref{ch2:ass:sw} (\textit{ii}) does not hold 
	since $\Phi_{2}(\bt_1,\bpi)=\Phi_2(\bt_2,\bpi)$ for all $(\bt_1,\bt_2)\in\bT$, so 
	using this auxiliary model does not lead to a consistent estimator. 
	It is however not clear whether Assumption~\ref{ch2:ass:iie}, the alternative to Assumption~\ref{ch2:ass:sw},
	holds or not because the quantities to verify are unknown. 
	Note however that in view of the equivalence theorem between the SwiZs and the indirect
	inference estimator (Theorem~\ref{ch2:thm:equiv}), it would appear as a contradiction
	for Assumption~\ref{ch2:ass:sw} not to hold but Assumption~\ref{ch2:ass:iie} to be satisfied.

	\cite{heggland2004estimating} idea is to select an auxiliary model where $\hbpi_n(\bt)$ is 
	both sensitive to $\bt$ and efficient for a given $\bt$. Since they justify their choice on a
	graphical analysis with simulated samples, one may wonder whether the authors were unlucky or
	misleaded by the graphics on this particular example. 
	In fact, although $\hbpi_n(\bt)$ is unknown in an explicit form,
	its Jacobian may be derived explicitly by mean of an implicit function theorem, so
	for a given $\bt_1\in\bT$ we have:
	\begin{equation*}
		D_{\bt}\hbpi_n(\bt_1) = -{\left[D_{\bpi}\bP_n\left(\bt_1,\u,\bpi\right)\Big\vert_{\bpi=\hbpi_n(\bt_1)}\right]}^{-1}
		D_{\bt}\bP_n\left(\bt_1,\u,\hbpi_n(\bt_1)\right).
	\end{equation*}
	The Jacobian $D_{\bpi}\bP_n$ is non zero. Yet, as already discussed, the second partial derivative
	of $\bP_n$ with respect to $\bt$ is null. Because only the second row of $D_{\bt}\bP_n$ has zero
	entries, there is no reason to believe that $D_{\bt}\hbpi_n(\bt)$ has zero entries.
	Consequently, the authors were not misleaded by the gaphics or unlucky, 
	it is the criterion itself that is misleading.

	We now face ourselves to the delicate task of choosing an auxiliary model which non-only
	respects the constraint $\dim(\bt)=\dim(\bt)$, but also makes Assumption~\ref{ch2:ass:bvp}
	plausible. 
	In view of this particular M/G/1 stochastic process, 
	using the convolution between a gamma with shape parameter $n$ and
	unknown rate parameter and a uniform
	distributions may be a ``natural'' choice, 
	yet, terms computationally complicated to evaluate readily appear.
	We propose instead of using Fr\'echet's three parameters extreme value distribution, whose
	density is given, for $i=1,\ldots,n$, by:
	\begin{equation*}
		f(x_i,\bpi) = \frac{\pi_1}{\pi_2}{\left(\frac{x_i-\pi_3}{\pi_2}\right)}^{-1-\pi_1}
		\exp\left\{-{\left(\frac{x_i-\pi_3}{\pi_2}\right)}^{-\pi_1}\right\}, \quad\text{if } x_i>\pi_3,
	\end{equation*}
	where $\pi_1>0$ is a shape parameter, $\pi_2>0$ is a scale parameter and $\pi_3\in\R$ is a parameter
	representing the location of the minimum. The relationship between $\pi_3$ and $\theta_1$ 
	as the minimum of the distribution seems natural and we thus further constrain 
	here $\pi_3$ to be non-negative, so $\bpi>0$. However, the existence of a potential link between 
	${(\theta_2,\theta_3)}^T$ and ${(\pi_1,\pi_2)}^T$ is not self-evident, but 
	certainly that the shape ($\pi_1$) and scale ($\pi_2$) parameters offer enough flexibility
	to ``encompass'' the distribution of the M/G/1 stochastic process as illustrated 
	in Figure~\ref{ch2:fig:gg11}. Note that the ``closeness'' between M/G/1 and Fr\'echet
	models is also dependent on the parametrization.
 	\begin{figure}[htbp]
 		\begin{center}
 			\includegraphics[width= 0.8\textwidth, keepaspectratio]{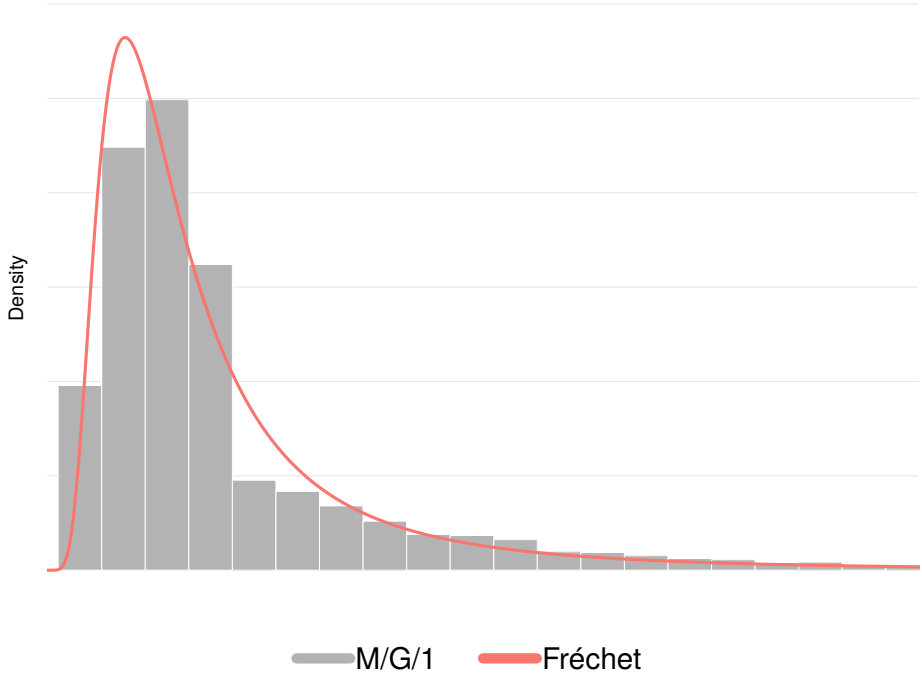}
 		\end{center}
 		\caption{Histogram of a simulated M/G/1 stochastic process of size $n=10^4$ on which
 		the density (solid line) of Fr\'echet distribution has been added. The true parameter is 
		$\bto={[0.3\;0.9\;1]}^T$, the auxiliary estimator we obtain here is approximately
		$\hbpi_{n}={[0.02\;0.60\;2.05]}^T$.}
 		\label{ch2:fig:gg11}
 	\end{figure}
	It remains to advocate this choice in the light of Assumption~\ref{ch2:ass:bvp}.
	We take the maximum likelihood estimator of Fr\'echet's distribution  as the
	auxiliary estimator and thus the likelihood score as the estimating function, which is given by:
	\begin{equation*}
		\bP_n\left(\bt,\u,\bpi\right) = 
		\begin{pmatrix}
			\frac{-1}{\pi_1} + \frac{1}{n}\sum_{i=1}^n\log\left(\frac{\g(\bt,\u_i)-\pi_3}{\pi_2}\right)
			\left[1-{\left(\frac{\g(\bt,\u_i)-\pi_3}{\pi_2}\right)}^{-\pi_1}\right] \\
			-\frac{\pi_1}{\pi_2}\frac{1}{n}\sum_{i=1}^n\left[1-{\left(\frac{\g(\bt,\u_i)-\pi_3}{\pi_2}\right)}^{-\pi_1}\right] \\
			\frac{-1}{n}\sum_{i=1}^n\frac{1+\pi_1}{\g(\bt,\u_i)-\pi_3} + \frac{\pi_1}{\pi_2}\frac{1}{n}
			\sum_{i=1}^n{\left(\frac{\g(\bt,\u_i)-\pi_3}{\pi_2}\right)}^{-\pi_1-1}
		\end{pmatrix}.
	\end{equation*}
	Let us assume that a random variable $\w$ with the same dimension as $\bt$ exists such that
	the estimating function may be expressed as follows:
	\begin{equation*}
		\bvP_{\hbpi_n}\left(\bt,\w\right) = 
		\begin{pmatrix}
			\frac{-1}{\hat{\pi}_1} + \log\left(z_1\right)
			\left[1-{z_1}^{-\hat{\pi}_1}\right] \\
			-\frac{\hat{\pi}_1}{\hat{\pi}_2}\left[1-{z_2}^{-\hat{\pi}_1}\right] \\
			-\frac{\left(1+\hat{\pi}_1\right)z_3^{-1}}{\hat{\pi}_2} + \frac{\hat{\pi}_1}{\hat{\pi}_2}
			{z_3}^{-\hat{\pi}_1-1}
		\end{pmatrix},
	\end{equation*}
	where $\hbpi_n$ is fixed and $z_i\equiv\frac{\g(\bt,w_i)-\hat{\pi}_3}{\hat{\pi}_2}$, $i=1,2,3$.
	The Jacobian matrix with respect to $\bt$ is give by:
	\begin{equation*}
		D_{\bt}\bvP_{\hbpi_n}\left(\bt,\w\right) =
		\begin{pmatrix}
			D_{\bt^T}g(\bt,w_1)\left[\frac{z_1^{-1}}{\hat{\pi}_2}\left(1-z_1^{-\hat{\pi}_1}\right)
				+\frac{\hat{\pi}_1}{\hat{\pi}_2}\log\left(z_1\right)z_1^{-\hat{\pi}_1-1}\right]\\
			D_{\bt^T}g(\bt,w_2)\left[-\frac{\hat{\pi}_1^2}{\hat{\pi}_2^2}z_2^{-\hat{\pi}_1-1}\right]\\
			D_{\bt^T}g(\bt,w_3)\left[\frac{\left(\hat{\pi}_1-1\right)}{\hat{\pi}_2^2}z_3^{-2}
				-\frac{\hat{\pi}_1(\hat{\pi}_1+1)}{\hat{\pi}_2^2}z_3^{-\hat{\pi}_1-2}\right]
		\end{pmatrix}.
	\end{equation*}
	Substituing $D_{\bt}\g^T$ by $D_{\w}\g^T$ in the above equation gives the Jacobian matrix
	with respect to $\w$, a matrix which is diagonal. It is straightforward to remark
	that the differentiability and continuity depends exclusively on the smoothness of $\g$.
	Thus, if $\g$ is once continuously differentiable in both $\bt$ and $\w$, then
	Assumption~\ref{ch2:ass:bvp} (\textit{i}) holds.

	Concerning the determinant of these Jacobian matrices, they may be null only
	on unlikely situations: first, if $\g$ equals $\hat{\pi}_3$ then $z_i$ is zero for $i=1,2,3$,
	second, if $D_{\bt}\g$ or $D_{\w}\g$ are zeros. The choice of $\g$ may be guided
	by this restriction so typically the determinants may be null, but only on a countable set,
	and Assumption~\ref{ch2:ass:bvp} (\textit{ii}) is verified. For Assumption~\ref{ch2:ass:bvp}
	(\textit{iii}), it is straightforward to remark that
	\begin{equation*}
		\lim_{\lVert(\bt,\w)\rVert\to\infty}\left\lVert\bvP_{\hbpi_n}(\bt,\w)\right\rVert,
	\end{equation*}
	as long as $\lim_{\lVert(\bt,\w)\rVert\to\infty}\lVert\g(\bt,\w)\rVert=\infty$, since
	$\log(z_1)$ would diverge. Depending on $g$, Assumption~\ref{ch2:ass:bvp} (\textit{iii}) is satisfied.
	
	Therefore, the plausibility of Assumption~\ref{ch2:ass:bvp} is up to the choice of the generating
	equation $g$. Here, the choice is quasi immediate as it is driven by the form of the process:
	\begin{equation*}
		g(\bt,\u_i) = \left\{\begin{array}{lr}
			v_i(\bt), & \text{if } \sigma^{\varepsilon}_i(\bt) \leq \sigma^g_{i-1}(\bt), \\
			v_i(\bt) + \sigma^{\varepsilon}_i(\bt) - \sigma^g_{i-1}(\bt), & \text{if } \sigma^{\varepsilon}_i(\bt)>\sigma^g_{i-1}(\bt),
		\end{array}\right.
	\end{equation*}
	where $\u_i={(u_{1i},u_{2i})}^T$, $u_{ji}\sim\U(0,1)$, $j=1,2$, $u_{1i}$ and $u_{2i}$ are
	independent, $v_i(\bt) \eqd \theta_1 + (\theta_2-\theta_1)u_{1i}$, $\sigma^{\varepsilon}_i(\bt) = \sum_{j=1}^i{\varepsilon}_j(\bt)$,
	$\varepsilon_j(\bt)=-\theta_3^{-1}\log(u_{2j})$ and $\sigma^g_i=\sum_{j=1}^ig(\bt,\u_j)$.
	Let $E_i$ corresponds to the event $\{\sigma^{\varepsilon}_i(\bt) \leq \sigma^g_{i-1}(\bt)\}$
	and $\bar{E}_i$ be the contrary. The partial derivatives may be found recursively
	as follows:
	\begin{equation*}
		\frac{\partial}{\partial\theta_1}g(\bt,\u_i) = \left\{\begin{array}{lr}
			1-u_{1i}, & \text{if } i=1, \\
			1-u_{1i}, & \text{if } i>1 \text{ and } E_i, \\
			1-u_{1i}-\sum_{j=1}^{i-1}\frac{\partial}{\partial\theta_1}g(\bt,\u_j),
			& \text{if } i>1 \text{ and } \bar{E}_i.
		\end{array}\right.
	\end{equation*}
	\begin{equation*}
		\frac{\partial}{\partial\theta_2}g(\bt,\u_i) = \left\{\begin{array}{lr}
			u_{1i}, & \text{if } i=1, \\
			u_{1i}, & \text{if } i>1 \text{ and } E_i, \\
			u_{1i}-\sum_{j=1}^{i-1}\frac{\partial}{\partial\theta_2}g(\bt,\u_j),
			& \text{if } i>1 \text{ and } \bar{E}_i.
		\end{array}\right.
	\end{equation*}
	\begin{equation*}
		\frac{\partial}{\partial\theta_3}g(\bt,\u_i) = \left\{\begin{array}{lr}
			0, & \text{if } i=1, \\
			0, & \text{if } i>1 \text{ and } E_i, \\
			-\frac{1}{\theta_3^2}\sum_{j=1}^i\log(u_{2j})-\sum_{j=1}^{i-1}\frac{\partial}{\partial\theta_3}g(\bt,\u_j),
			& \text{if } i>1 \text{ and } \bar{E}_i.
		\end{array}\right.
	\end{equation*}
	\begin{equation*}
		\frac{\partial}{\partial u_{1}}g(\bt,\u_i) = \left\{\begin{array}{lr}
			\theta_2 - \theta_1, & \text{if } i=1, \\
			\theta_2 - \theta_1, & \text{if } i>1 \text{ and } E_i, \\
			\theta_2 - \theta_1 - \sum_{j=1}^{i-1}\frac{\partial}{\partial u_1}g(\bt,\u_j),
			& \text{if } i>1 \text{ and } \bar{E}_i.
		\end{array}\right.
	\end{equation*}
	\begin{equation*}
		\frac{\partial}{\partial u_2}g(\bt,\u_i) = \left\{\begin{array}{lr}
			0, & \text{if } i=1, \\
			0, & \text{if } i>1 \text{ and } E_i, \\
			-\theta_3^{-1}\sum_{j=1}^i\frac{1}{u_{2j}}-\sum_{j=1}^{i-1}\frac{\partial}{\partial u_2}g(\bt,\u_j),
			& \text{if } i>1 \text{ and } \bar{E}_i.
		\end{array}\right.
	\end{equation*}
	Clearly $g$ is once continuously differentiable in both its arguments with non-zero derivatives.
	Eventually, we have that $v_i(\bt)$ goes to $\infty$ when $\theta_1\to\infty$,
	$\theta_2\to\infty$ and $u_{1i}\to1$, whereas $\varepsilon_i(\bt)$ tends to zero
	whenever $\theta_3\to\infty$ and $u_{2i}\to1$. It is not clear 
	whether $v_i(\bt)+\sigma^{\varepsilon}_i(\bt)-\sigma^{g}_i(\bt)$ diverges 
	or converges to 0 when $\lVert(\bt,\u_i)\rVert\to\infty$, but in any case
	$\lVert g(\bt,\u_i)\rVert$ tends to $\infty$ since $v_i(\bt)$ diverges.
	As a consequence, Assumption~\ref{ch2:ass:bvp} is highly plausible and thus Theorem~\ref{ch2:thm:freq}
	seems invokable. 
	
	For the simulation, we set $\bto={[0.3\;0.9\;1]}^T$ and $n=100$ as in~\cite{heggland2004estimating}.
	We compare the SwiZs with indirect inference in Definition~\ref{ch2:def:iie}
	and the parametric bootstrap using the indirect inference with $B=1$ as
	the initial consistent estimator (see Definition~\ref{ch2:def:pb}).
	By Theorem~\ref{ch2:thm:equiv}, the SwiZs and the indirect inference are equivalent,
	but as argued, the price for obtaining the inidirect inference is higher so here 
	we seek empirical evidence, and Table~\ref{ch2:tab:gg10} speaks for itself,
	the difference is indeed monstrous. The parametric bootstrap is even worse
	in terms of computational time. 
	It is maybe good to remind the reader that the comparison is fair:
	all three methods benefits from the same level of implementation and uses the very
	same technology.
	\begin{table}
		\centering
		\begin{tabular}{@{}cccc@{}}
			\toprule
			&SwiZs&indirect inference&parametric bootstrap\\
			\midrule
			Average time $[seconds]$&0.97&134.18&197.15\\
			Total time $[hours]$&2.7&372.5&547.4\\
			\bottomrule
		\end{tabular}
		\caption{Average time in seconds to estimate a conditional distribution on $S=10,000$ points
		and total time in hours for the $M=10,000$ independent trials.}
		\label{ch2:tab:gg10}
	\end{table}
	\begin{figure}[htbp]
		\begin{center}
			\includegraphics[width=0.9\textwidth, keepaspectratio]{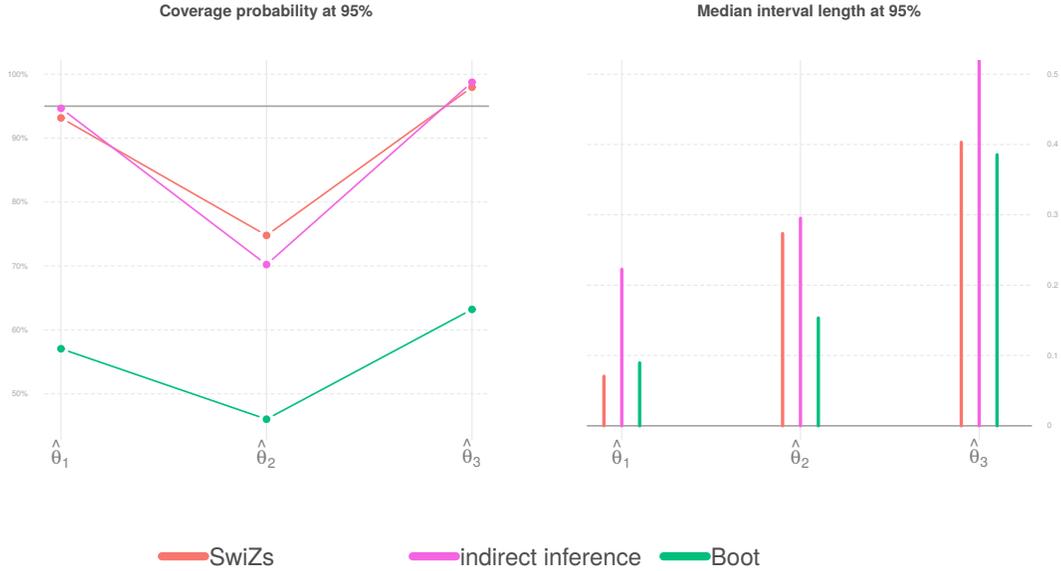}
		\end{center}
		\caption{\textit{On the left panel}:
			Representation of the 95\% coverage probability (ideal is
			gray line) of the SwiZs, the indirect inference and the parametric bootstrap with indirect
			inference as initial estimator. The closer to the gray line is the better. \textit{On the right panel}:
			Illustration of the median interval lengths at a target level of 95\%. The shorter is the
			better.
			}
		\label{ch2:fig:gg12}
	\end{figure}
	\begin{figure}[htbp]
		\begin{center}
			\includegraphics[width=0.9\textwidth, keepaspectratio]{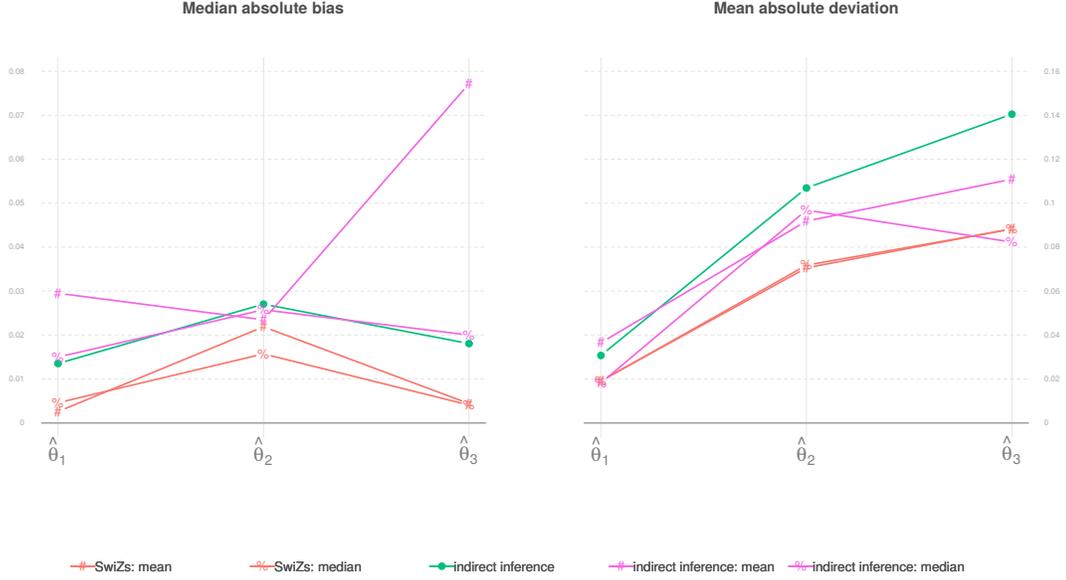}
		\end{center}
		\caption{\textit{On the left panel}:
			 Median absolute bias of point estimators: mean and
			 median on the SwiZs and indirect inference distributions plus the indirect inference with
			 $B = 1$. \textit{On the right panel}: same as \textit{left panel} with a different measure: mean absolute
			 deviation. For both panel, the lower is the better.
			}
		\label{ch2:fig:gg13}
	\end{figure}
	The complete results may be found in Appendix~\ref{ch2:app:gg1}. 
	In Figure~\ref{ch2:fig:gg12} we can realize that the SwiZs do not offer
	an exact coverage in this case, it is even far from ideal for $\hat{\theta}_2$.
	It is nonetheless better than the parametric bootstrap. Especially the coverage
	of $\hat{\theta}_1$ and $\hat{\theta}_3$ are close to the ideal level.
	Considering the context of this simulation: moderate sample size, no closed-form
	for the likelihood, the results are very encouraging. A good surprise appears from
	Figure~\ref{ch2:fig:gg13} where the SwiZs demonstrates better performances of
	its point estimates (mean and median) compared to indirect inference approaches in
	termes of absolute median bias and mean absolute deviation.

	It is however not clear which one, if not both, we should blame
	for failure of missing exact coverage probability between our analysis
	on the applicability of Theorem~\ref{ch2:thm:freq} to this case or 
	the numerical optimization procedure. The previous examples seem
	to indicate for the latter. To this end, we re-run the same experiment
	only for the SwiZs (for pure operational reason) by changing the starting
	values to be the true parameter $\bto$ to measure the implication. Indeed,
	starting values are a sensitive matter for quasi-Newton routine and since
	$\hbpi_n$ is not a consistent estimator of $\bto$, using it as a starting
	value might have a persistent influence on the sequence $\{\hbt_n^{(s)}:s\in\N^+_S\}$.
	Results are reported in Table in Appendix~\ref{ch2:app:gg1}. The coverage
	probabilities of $\hat{\theta}_1$ and $\hat{\theta}_3$ becomes nearly
	perfect, which shows that indeed good starting values may reduce the numerical
	error in the coverage probabilities. However, coverage probability for $\hat{\theta}_2$
	persistently shows result off the desired levels, which seems rather to indicate
	a problem related to the applicability of Theorem~\ref{ch2:thm:freq}.
	Increasing the sample size to $n=1,000$ (see Table~\ref{ch2:tab:gg13})
	makes the coverage of all three parameters nearly perfect.
\end{example}
Fifth and last, we consider logistic regression. This is certainly one of the most widely
used statistical model in practice. This case is challenging at least on two aspects.
First, the random variable is discrete and the finite sample theory in Section~\ref{ch2:sec:fs}
does not hold. Second, the generating function is non-differentiable with respect to $\bt$,
therefore gradient-based optimization routines cannot be employed. 
In what follows, we circumvent this inconvenient by smoothing the generating function.
To this end, we start by introducing the continuous latent representation of the logistic regression.
\begin{example}
	Suppose we have the model
	\begin{equation*}
		\yy=\X\bt + \be,
	\end{equation*}
	where $\be=\left(\epsilon_1,\cdots,\epsilon_n\right)^T$ and
	$\epsilon_i$, $i=1,\cdots,n$, are \iid\;according to 
	a logistic distribution with mean 0 and unity variance.
	This distribution belongs to symmetric location-scale families.
	It is similar to the Gaussian distribution with heavier tails.
	The unknwon parameters $\bt$ of this model could be easily estimated
	by the ordinary least squares:
	\begin{equation*}
		\hbpi_n = \left(\X^T\X\right)^{-1}\X^T\yy.	
	\end{equation*}
	The corresponding estimating function is:
	\begin{equation*}
		\bP_n\left(\bt,\u,\bpi\right) = \X^T\X\bpi - \X^T\g\left(\bt,\u\right).
	\end{equation*}
	A straightforward generating function is $\g(\bt,\u)=\X\bt+\u$ where 
	$u_i\sim\emph{Logistic}(0,1)$. 
	Evaluating this function at $\bpi=\hbpi_n$ leads to 
	\begin{equation*}
		\bP_n\left(\bt,\u,\hbpi_n\right) = \X^T\yy -\X^T\X\bt-\X^T\u.
	\end{equation*}
	Solving the root of this function in $\bt$ gives the following explicit solution:
	\begin{equation}\label{ch2:eq:ideal}
		\hbt_n = \left(\X^T\X\right)^{-1}\X^T\left(\yy-\u\right).
	\end{equation}
	Following Example~\ref{ch2:ex:reg} on linear regression, it is easy
	to show that inference based on the distribution of this estimator
	leads to exact frequentist coverage probabilities.

	Let us turn our attention to logistic regression.
	In this case, $\yy$ is not observed. 
	Instead, we observe a binary random variable $\y$,
	whose elements are:
	\begin{equation*}
		y_i = \left\{\begin{array}{lr}
			1,&\X_i\bt+\epsilon_i\geq0,\\
			0,&\X_i\bt+\epsilon_i<0,
		\end{array}\right.
	\end{equation*}
	where $\X_i$ is the $i$th row of $\X$.
	Saying it differently, this consideration implies that the generating function is modified to
	the following indicator function:
	\begin{equation*}
		\g\left(\bt,u_i\right) = \1\left\{\X_i\bt+u_i\geq0\right\}.
	\end{equation*}
	Clearly, this change implies that $\bP_n$ has a flat Jacobian matrix and 
	Assumptions~\ref{ch2:ass:bvp} and~\ref{ch2:ass:bvp2} do not hold.
	Moreover, this problem becomes numerically more invloved, especially
	if we want to pursue with a gradient-based optimization routine.
	As mentionned, in practice we seek the solution of the following problem:
	\begin{equation}\label{ch2:eq:log}
		\argmin_{\bt\in\bT}\left\lVert\X^T\y - \X^T\g(\bt,\u)\right\rVert^2_2\equiv
		\argmin_{\bt\in\bT}f(\bt).
	\end{equation}
	Note that $\X^T\y$ is the sufficient statistic for a logistic regression
	(see Chapter 2 in~\cite{mccullagh1989generalized}).
	The gradient of $f(\bt)$ is 
	\begin{equation*}
		-D_{\bt}\g(\bt,\u)\X\left[\X^T\y-\X^T\g(\bt,\u)\right].
	\end{equation*}
	However, the Jacobian $D_{\bt}\g(\bt,\u)$ is 0 almost everywhere and alternatives
	are necessary for using gradient-based methods.
	A possibility is to smooth $\g(\bt,\u)$ by using for example a sigmoid function:
	\begin{equation*}
		\g(\bt,u_i) = \lim_{t\to0}\frac{1}{1+\exp\left(-\nicefrac{\left(\X_i\bt+u_i\right)}{t}\right)}.
	\end{equation*}
	The value of $t$ tunes the approximation and the value of the gradient.
	However, from our experience, large values of $t$, say $t>0.1$, leads to poor results
	and small values, say $t<0.1$, leads to numerical instability.
	We thus prefer to use a different strategy by taking $-f(\bt)$ as the gardient.
	This strategy corresponds to the iterative bootstrap procedure (\cite{guerrier2018simulation}).
	In Figure~\ref{ch2:fig:logistic}, we illustrate the difference between these two approximations
	and the ``ideal'' distribution we would have obtained by observing the
	continuous underlying latent process. 
	\begin{figure}[htpb]
		\begin{center}
			\includegraphics[width=0.9\textwidth, keepaspectratio]{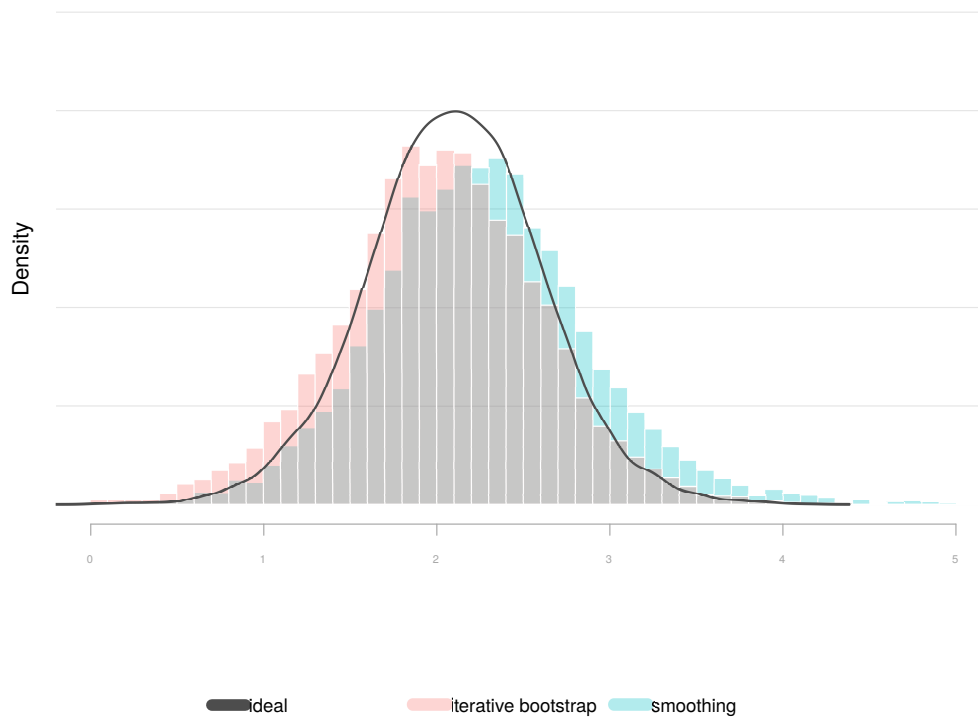}
		\end{center}
		\caption{Simulated SwisZ distribution of a single logistic regression
			with coefficient $\bt=2$ and sample size of 10. ``Ideal'' is~\eqref{ch2:eq:ideal}.
			``Smoothing'' approximates the gradient with a sigmoid function and $t=0.01$.
			``Iterative bootstrap'' uses $-f(\bt)$ as the gradient.}
		\label{ch2:fig:logistic}
	\end{figure}
	Clearly, the loss of information
	induced from the possibility of only observing a binary outcome results
	in an increase of variability. Nonetheless, the difference is not
	enormous. 
	Both approximations leads to similar distributions in terms of shapes.
	We can notice a little difference in their modes.
	Since the iterative bootstrap approximation is numerically advantageous,
	we use it in the next study.
	
	For simulation, we setup $\bto = {(0,5,5,-7,-7,\underbrace{0,\ldots,0}_{15})}^T$
	and sample size $n=200$. We compare coverage probabilities of 95\%
	confidence intervals obtained by the SwiZs and by asymptotic theory.
	We report results in Table~\ref{ch2:tab:logistic}.
	We can clearly see that the SwiZs have the most precise confidence intervals
	for all coefficients with coverage close to the target level of 95\%.
	\begin{table}
		\centering
		\begin{tabular}{@{}ccc@{}}
			\toprule
			&SwiZs&asymptotic\\
			\midrule
			$\theta_1$&0.9442&0.9187 \\
			$\theta_2$&0.9398&0.8115 \\
			$\theta_3$&0.9382&0.8121 \\
			$\theta_4$&0.9432&0.7688 \\
			$\theta_5$&0.9450&0.7737 \\
			$\theta_6$&0.9397&0.9233 \\
			$\theta_7$&0.9357&0.9170 \\
			$\theta_8$&0.9398&0.9237 \\
			$\theta_9$&0.9391&0.9218 \\
			$\theta_{10}$&0.9400&0.9208 \\
			$\theta_{11}$&0.9424&0.9208 \\
			$\theta_{12}$&0.9375&0.9214 \\
			$\theta_{13}$&0.9368&0.9204 \\
			$\theta_{14}$&0.9389&0.9210 \\
			$\theta_{15}$&0.9400&0.9207 \\
			$\theta_{16}$&0.9400&0.9183 \\
			$\theta_{17}$&0.9361&0.9183 \\
			$\theta_{18}$&0.9449&0.9241 \\
			$\theta_{19}$&0.9412&0.9218 \\
			$\theta_{20}$&0.9427&0.9240 \\
			\bottomrule
		\end{tabular}
		\caption{95\% coverage probabilities of confidence intervals 
		from the SwiZs and asymptotic theory.}
		\label{ch2:tab:logistic}
	\end{table}
\end{example}
%


\begin{appendix}
\section{Technical results}
\begin{lemma}\label{ch2:thm:diffeo_invers}
	Let $X$ and $Y$ be open subsets of $\R^n$.
	If $\f:X\to Y$ is a $\mathcal{C}^1$-diffeomorphism, then the 
	Jacobian matrices of the maps $x\mapsto\f$ and
	$y\mapsto\f^{-1}$ are invertible, and 
	the derivatives at the points
	$a\in X$ and $b\in Y$, are given by:
	\begin{equation*}
		D_x\f(a) = {\left[D_y\f^{-1}\vert_{y=\f(a)}\right]}^{-1},
		\quad D_y\f(b) = {\left[D_x\f\vert_{x=\f^{-1}(b)}\right]}^{-1}.
	\end{equation*}
\end{lemma}
\begin{proof}
	By assumption, $\f$ is invertible, once continuously differentiable
	and $\f^{-1}$ is once continuously differentiable.
	
	We have $\f^{-1}\circ\f=\id_X$, where $\id_X$ is the identity function
	on the set $X$.
	Fix $a\in X$. By the chain rule, the derivative at $a$ is the following:
	\begin{equation*}
		D_{y}\f^{-1}\left(\f(a)\right) D_x\f(a) = \mathbf{I}_n,
	\end{equation*}
	where $\mathbf{I}_n$ is the identity matrix. Since $D_y\f^{-1}$ and 
	$D_x\f$ are square matrices, we have:
	\begin{equation*}
		\det\left(D_y\f^{-1}(\f(a))\right)\det\left(D_x\f(a)\right) = 1.
	\end{equation*}
	The determinants cannot be 0, there are either 1 or -1 for both matrices,
	ergo, the Jacobian are invertible and we can write
	\begin{equation*}
		D_x\f(a) = {\left[D_y\f^{-1}(\f(a))\right]}^{-1}.
	\end{equation*}
	The proof for $\f\circ\f^{-1}=\id_Y$ follows by symmetry.
\end{proof}
\begin{lemma}\label{ch2:thm:cov}
	Let $\bT$ and $W$ be open subsets of $\R^p$.
	If there exists a $\mathcal{C}^1$-diffeomorphic mapping $\a:W\to\bT$,
	that is, $\w\mapsto\a$ is continuously once differentialbe in $\bT\times W$
	and the inverse map $\bt\mapsto\a^{-1}$ is continuously once differentiable
	in $\bT\times W$,
	then the cumulative distribution function of $\{\hbt_n^{(s)}:s\in\N\}$ is given
	by:
	\begin{equation*}
		\int_{\bT_n} f_{\hbt_n}\left(\bt\vert\hbpi_n\right)\d\bt 
		= \int_{W} f_{\w}\left(\a(\w)\vert\hbpi_n\right)
		\frac{1}{\left\rvert \det\left(D_{\w}\a(\w)\right)\right\rvert}\d\w,
	\end{equation*}
	provided that $f$ is a nonnegative Borel function and $\Pr\left(\hbpi_n\neq\emptyset\right)=1$.
\end{lemma}
\begin{proof}[\textbf{Proof of Lemma~\ref{ch2:thm:cov}}]
	By assumption, $\w\mapsto\a$ is a $\mathcal{C}^1$-diffeomorphism
	so by Lemma~\ref{ch2:thm:diffeo_invers} the Jacobian
	of $\a$ and $\a^{-1}$ are invertible. 
	All the conditions of the change-of-variable formula for multidimensional 
	Lebesgue integral in~\cite[Theorem 17.2, p.239]{billingsley2012probability}
	are satisfied, so we obtain
	\begin{equation*}
		\int_{\bT_n}f_{\hbt_n}\left(\bt\vert\hbpi_n\right)\;\d\bt = 
		\int_{\a^{-1}(\bT_n)}f_{\w}\left(\a^{-1}(\bt)\vert\hbpi_n\right)\det\left(D_{\bt}\a^{-1}(\bt)\right) \;\d\bt
	\end{equation*}
	By Lemma~\ref{ch2:thm:diffeo_invers}, we have that 
	$D_{\bt}\a^{-1} = {\left[D_{\w}\a\right]}^{-1}$.
	Taking the determinant ends the proof.
\end{proof}

\section{Finite sample}
\begin{proof}[\textbf{Proof of Theorem~\ref{ch2:thm:equiv}}]
	We proceed by showing first that $\bT^{(s)}_{\text{II},n}\subset\bT^{(s)}_n$, and 
	second that $\bT^{(s)}_{\text{II},n}\supset\bT^{(s)}_n$.

	It follows from Assumption~\ref{ch2:ass:uniq} that $\hbpi_n$ is the unique solution of $\argzero_{\bpi\in\bPi}\bP_n(\bto,\uo,\bpi)$, 
	ergo $\bPi_n$ in the Definition~\ref{ch2:def:swizs2} is a singleton. 
	
	\textit{(1)}. Fix $\bt_1\in\bT^{(s)}_{\text{II},n}$. By Definition~\ref{ch2:def:iie}, it holds that
	\begin{equation*}
		\hbpi_n = \hbpi^{(s)}_{\text{II},n}\left(\bt_1\right),\quad \bP_n\left(\bt_1,\u_s,\hbpi^{(s)}_{\text{II},n}(\bt_1)\right)=\0,
	\end{equation*}
	where $\hbpi_{\text{II},n}^{(s)}$ is the unique solution of $\argzero_{\bt\in\bPi}\bP_n(\bt_1,\u_s,\bpi)$. 
	Ergo, it holds as well that 
	\begin{equation*}
		\bP_n\left(\bt_1,\u_s,\hbpi_n\right)=\0,
	\end{equation*}
	implying that $\bt_1\in\bT^{(s)}_n$ by Definition~\ref{ch2:def:swizs2}. Thus $\bT^{(s)}_{\text{II},n}\subset\bT_n^{(s)}$.

	\textit{(2)}. Fix $\bt_2\in\bT_n$. By Definition~\ref{ch2:def:swizs2} we have
	\begin{equation*}
		\bP_n\left(\bt_2,\u_s,\hbpi_n\right)=\0.
	\end{equation*}
	By Definition~\ref{ch2:def:iie}, we also have
	\begin{equation*}
		\bP_n\left(\bt_2,\u_s,\hbpi^{(s)}_{\text{II},n}(\bt_2)\right)=\0,
	\end{equation*}
	where $\hbpi^{(s)}_{\text{II},n}(\bt_2)$ is the unique solution of $\argzero_{\bpi\in\bPi}\bP_n(\bt_2,\u_s,\bpi)$.
	It follows that $\hbpi_n = \hbpi^{(s)}_{\text{II},n}\left(\bt_2\right)$ uniquely,
	implying that $\bt_2\in\bT^{(s)}_{\text{II},n}$ by Definition~\ref{ch2:def:iie}. 
	Thus $\bT^{(s)}_{\text{II},n}\supset\bT^{(s)}_n$, which concludes the proof.
\end{proof}
\begin{proof}[\textbf{Proof of Theorem~\ref{ch2:thm:pb}}]
	We proceed by showing first that (A) $\bT^{(s)}_n=\bT^{(s)}_{\text{Boot},n}$ implies
	(B) $\bP_n(\bt,\u_s,\bpi)=\bP_n(\bpi,\u_s,\bt)=\0$, then that (B) implies (A).

	1. Suppose (A) holds. Fix $\bt_1\in\bT^{(s)}_n$ and $\hbpi_n\in\bPi_n$. 
	We have by the Definition~\ref{ch2:def:swizs2}
	\begin{equation*}
		\bP_n\left(\bt_1,\u_s,\hbpi_n\right) = \0.
	\end{equation*}
	By (A), we also have that $\bt_1\in\bT^{(s)}_{\text{Boot},n}$ so by the Definition~\ref{ch2:def:pb}
	\begin{equation*}
		\bP_n\left(\hbpi_n,\u_s,\bt_1\right)=\0.
	\end{equation*}
	Since both estimating equations equal zero, we have
	\begin{equation*}
		\bP_n\left(\hbpi_n,\u_s,\bt_1\right)=\bP_n\left(\bt_1,\u_s,\hbpi_n\right) = \0.
	\end{equation*}
	Hence (A) implies (B).

	2. Suppose now that (B) holds. Fix $\bt_1\in\bT^{(s)}_n$ and $\hbpi_n\in\bPi_n$ so $\bP_n(\bt_1,\u_s,\hbpi_n)=\0$.
	By (B), we have
	\begin{equation*}
		\bP_n\left(\bt_1,\u_s,\hbpi_n\right)=\bP_n\left(\hbpi_n,\u_s,\bt_1\right)=\0,
	\end{equation*}
	so $\bt_1\in\bT^{(s)}_{\text{Boot},n}$ and thus $\bT^{(s)}_n\subset\bT^{(s)}_{\text{Boot},n}$.
	The same argument shows that $\bT^{(s)}_n\supset\bT^{(s)}_{\text{Boot},n}$ which ends the proof.
\end{proof}
\begin{proof}[\textbf{Proof of Proposition~\ref{ch2:thm:loc}}]
	Since $\hat{\pi}_n=\bar{\x}=\frac{1}{n}\sum_{i=1}^n x_i$, the sample average, we can write the following estimating equation
	\begin{equation*}
		\hat{\pi}_n = \argzero_{\pi\in\Pi}\left(\bar{\x}-\pi\right)=\argzero_{\pi\in\Pi}\Phi_n\left(\theta_0,\uo,\pi\right),
	\end{equation*}
	where $x\eqd g(\theta_0,u_0)$. Since $x$ follows a location family, we have that $x\eqd \theta_0 + g(0,u_0)\eqd\theta_0+y$.
	
	The SwiZs is defined as 
	\begin{equation*}
		\hat{\theta}^{(s)}_n = \argzero_{\theta\in\Theta}\Phi_n\left(\theta,\u_s,\hat{\pi}_n\right).
	\end{equation*}
	On the other hand, the parametric bootstrap estimator is
	\begin{equation*}
		\hat{\theta}^{(s)}_{\text{Boot},n} = \argzero_{\theta\in\Theta}\Phi_n\left(\hat{\pi}_n,\u_s,\theta\right).
	\end{equation*}
	Eventually, we obtain that
	\begin{align*}
		\Phi_n\left(\hat{\theta}^{(s)}_n,\u_s,\hat{\pi}_n\right) &= \hat{\theta}_n^{(s)} + \bar{\y} - \hat{\pi}_n = 0,\\
		\Phi_n\left(\hat{\pi}_n,\u_s,\hat{\theta}_n^{(s)}\right) &= \hat{\pi}_n + \bar{\y} - \hat{\theta}^{(s)}_{\text{Boot},n} \\
		&= -\hat{\pi}_n + \bar{\y} + \hat{\theta}^{(s)}_{\text{Boot},n} \\
		&= \Phi_n\left(\hat{\theta}^{(s)}_{\text{Boot},n},\u_s,\hat{\pi}_n\right) = 0,
	\end{align*}
	where we use the fact that $\bar{\y}\eqd-\bar{\y}$. Therefore, $\hat{\theta}^{(s)}_n=\hat{\theta}^{(s)}_{\text{Boot},n}$,
	or equivalently $\Phi_n\left(\theta,\u_s,\pi\right)=\Phi_n\left(\pi,\u_s,\theta\right)=0$,
	which ends the proof.
\end{proof}
\begin{proof}[\textbf{Proof of Theorem~\ref{ch2:thm:equiv2}}]
	Fix $\varepsilon=0$. The Theorem~\ref{ch2:thm:equiv} is satisfied so $\bT_n^{(s)}=\bT_{\text{II},n}^{(s)}$
	for any $s$.
	It is sufficient then to prove $\bT^{(s)}_{\text{ABC,n}}(0)=\bT^{(s)}_{\text{II},n}$ for any
	$s\in\N^+_S$.  We proceed by verifying that first 
	$\bT^{(s)}_{\text{ABC},n}(0)\subset\bT^{(s)}_{\text{II},n}$, and second that
	$\bT^{(s)}_{\text{ABC},n}(0)\supset\bT^{(s)}_{\text{II},n}$.

	\textit{(1)}. Fix $\bt_1\in\bT^{(s)}_{\text{ABC},n}(0)$. By the Assumption~\ref{ch2:ass:prior},
	$\bt_1$ is also a realization from the prior distribution $\mathcal{P}$. By Definition~\ref{ch2:def:abc},
	we have
	\begin{equation*}
		d\left(\hbpi_n,\hbpi^{(s)}_{\text{II},n}(\bt_1)\right) = 0.
	\end{equation*}
	By Definition~\ref{ch2:def:iie}, $\bt_1\in\bT^{(s)}_{\text{II},n}$, thus $\bT^{(s)}_{\text{ABC},n}(0)\subset\bT^{(s)}_{\text{II},n}$.

	\textit{(2)}. Fix $\bt_2\in\bT^{(s)}_{\text{II},n}$. By Definition~\ref{ch2:def:iie}, we have
	\begin{equation*}
		d\left(\hbpi_n,\hbpi^{(s)}_{\text{II},n}(\bt_2)\right) = 0.
	\end{equation*}
	By Assumption~\ref{ch2:ass:prior} and Definition~\ref{ch2:def:abc}, $\bt_2\in\bT^{(s)}_{\text{ABC},n}(0)$,
	ergo $\bT^{(s)}_{\text{ABC},n}(0)\supset\bT^{(s)}_{\text{II},n}$, which ends the proof.
\end{proof}
\begin{proof}[\textbf{Proof of Proposition~\ref{ch2:thm:exa}}]
	Fix $\alpha_1,\alpha_2>0$ such that $\alpha_1+\alpha_2=\alpha\in(0,1)$.
	Since we consider an exact $\alpha$-credible set $C_{\hbpi_n}$, we have
	\begin{align*}
		1-\alpha &= \Pr\left(\bt\in C_{\hbpi_n}\vert\hbpi_n\right) \\
		&= \Pr\left(\bt\in\bT_n\setminus\{\uQ_{\alpha_1}\cup\oQ_{\alpha_2}\}\right) \\
		&= \Pr\left(F_{\hbt_n\vert\hbpi_n}(\bt)\in(\alpha_1,1-\alpha_2)\right).
	\end{align*}
	Consider the event $E=\{u\in(\alpha_1,1-\alpha_2)\}$ taking value one
	with probability $p$ if $u$ is inside the interval and 0 otherwise. 
	Let $u = F_{\hbt_n\vert\hbpi_n}(\bto)$
	so at each trial there is one such event. Now consider indefinitely many trials,
	so we have $\{E_i:i\in\N^+\}$ where $\E(E_i)=\Pr(E_i=1)=p_i$. 
	Denote by $N$ is the number of trials. The frequentist coverage probability is
	given by
	\begin{equation*}
		\lim_{N\to\infty} \frac{\sum_{i=1}^{N}E_i}{N}.
	\end{equation*}
	By assumption, $u$ is an independent standard uniform variable, so 
	the events are independent and $p_i = 1-\alpha$ for all $i\geq1$ and for every $\alpha\in(0,1)$.
	It follows that $\{E_i:i\in\N^+\}$ are \iid\;Bernoulli random variables.
	The proof follows by Borel's strong law of large numbers (see~\cite{wen1991analytic}). 
\end{proof}
\begin{proof}[\textbf{Proof of Lemma~\ref{ch2:thm:uniq}}]
	Fix $\uo$. Fix $\bt_1\in\bT$. By definition we have
	\begin{equation*}
		\hbpi_n = \argzero_{\bpi\in\bPi}\bP_n\left(\bt_1,\uo,\bpi\right).
	\end{equation*}
	By assumption, the following equation
	\begin{equation*}
		\bP_n\left(\bt_1,\uo,\hbpi_n\right) = \0
	\end{equation*}
	is uniquely defined. Now fix $\bpi_1\in\bPi$.
	By definition we have
	\begin{equation*}
		\hbt_n=\argzero_{\bt\in\bT}\bP_n\left(\bt,\uo,\bpi_1\right),
	\end{equation*}
	and by assumption
	\begin{equation*}
		  \bP_n\left(\hbt_n,\uo,\bpi_1\right)=\0
	\end{equation*}
	is uniquely defined. It follows that
	$\bt_1=\hbt_n$ if and only if $\bpi_1=\hbpi_n$.
\end{proof}
\begin{proof}[\textbf{Proof of Theorem~\ref{ch2:thm:freq}}]
	We gives the demonstration under the Assumptions~\ref{ch2:ass:bvp} and~\ref{ch2:ass:bvp2}
	separately.
	
	1. We proceed by showing that we have a $\mathcal{C}^1$-diffeomorphism
	which is unique so Lemma~\ref{ch2:thm:cov} and Lemma~\ref{ch2:thm:uniq}
	apply. We then demonstrate that the obtained cumulative
	distribution function evaluated at $\bto\in\bT$ is a realization
	from a standard uniform random variable. The conclusion is eventually reached
	by the Proposition~\ref{ch2:thm:exa}.

	Let $\pi_1:\bT_n\times W_n\to\bT_n$ and $\pi_2:\bT_n\times W_n\to W_n$ be the projections
	defined by $\pi_1(\bt,\w)=\bt$ and $\pi_2(\bt,\w)=\w$ if $(\bt,\w)\in\bT_n\times W_n$.
	By Assumption~\ref{ch2:ass:bvp} the conditions of the global implicit function theorem 
	of~\cite[Theorem 1]{cristea2017global} are satisfied, so it holds that there exists a unique
	(global) continuous implicit function $\a:W_n\to\bT_n$ such that $\a(\wo)=\bto$
	and $\bvP_{\hbpi_n}(\w,\a(\w))=\0$ for every $\w\in W$. In addition, the mapping is 
	continuously differentiable on $W_n\setminus\pi_2(K_n)$ with derivative given by
	\begin{equation*}
		D_{\w}\a = -{\left[D_{\bt}\bvP_p\vert_{\bt=\a(\w)}\right]}^{-1}D_{\w}\bvP_p
	\end{equation*}
	for every $\w\in W_n\setminus\pi_2(K_n)$. Clearly the map $\a$ is invertible with a continuous inverse.
	Since the derivative $D_{\w}\bvP_p$ is continuous and invertible for $(\bt,\w)\in\bT_n\times W_n\setminus K_n$, 
	we immediately have that $\a$ is a $\mathcal{C}^1$-diffeomorphism with deriative of the inverse
	given by
	\begin{equation*}
		D_{\bt}\a^{-1} = -{\left[D_{\w}\bvP_p\vert_{\w=\a^{-1}(\bt)}\right]}^{-1}D_{\bt}\bvP_p
	\end{equation*}
	for $\bt\in\bT_n\setminus\pi_1(K_n)$.
	The conditions of Lemma~\ref{ch2:thm:cov} are satisfied and we obtain 
	the cumulative distribution function 
	\begin{equation*}
		F_{\hbt_n\vert\hbpi_n} = \int_{W_n} f_{\w}\left(\a^{-1}(\bt)\vert\hbpi_n\right)
		\frac{\det\left(D_{\bt}\bvP_p\right)}{\det\left(D_{\w}\bvP_p\right)}\;\d\w\;
		= F_{\w\vert\hbpi_n},
	\end{equation*}
	proving point \textit{(i)}. 
	Since $\hbpi_n$ is the unique zero of $\bP_n(\bto,\uo,\bpi)$, and hence of $\bvP_p(\bto,\wo,\bpi)$,
	and $\bt=\a(\w)$ is the unique zero of $\bvP_p(\bt,\w,\hbpi_n)$, we have by Lemma~\ref{ch2:thm:uniq}
	that $\bto=\a(\wo)$, and therefore that $\wo=\a^{-1}(\bto)$. In consequence, evaluating
	the above distribution at $\bto$ leads to
	\begin{equation*}
		F_{\hbt_n\vert\hbpi_n}(\bto) = F_{\w\vert\hbpi_n}(\wo) = u \sim\mathcal{U}(0,1),
	\end{equation*}
	that is, the distribution evaluated at $\bto$ is a realization from a standard uniform random
	variable. The conclusion follows by the Proposition~\ref{ch2:thm:exa}.

	2. Fix $\bto\in\bT_n$ and $\wo\in W_n$. Fix $\hbpi_n\in\bPi_n$, the point such that 
	$\bvP_p(\bto,\wo,\hbpi_n)=\0$.
	Let $\pi_1:W_n\times\bPi_n\to W_n$ and $\pi_2:W_n\times\bPi_n\to\bPi_n$ be the projections
	such that $\pi_1(\w,\bpi)=\w$ and $\pi_2(\w,\bpi)=\bpi$ if $(\w,\bpi)\in W_n\times\bPi_n$.
	By Assumption~\ref{ch2:ass:bvp2} (\textit{(i), (iii), (v)}), the Theorem 1 in~\cite{cristea2017global}
	is satisfied, as a consequence it holds that $\bvP_{\bto}$ admits a unique
	global implicit function $\bpi_{\bto}:W_n\to\bPi_n$ such that $\bvP_{\bto}(\w,\bpi_{\bto}(\w))=\0$
	for every $\w\in W_n$, $\bpi_{\bto}(\wo)=\hbpi_n$, and $\bpi_{\bto}$ is once continuously
	differentiable on $W_n\setminus\pi_1(K_{1n})$ with derivative given by 
	\begin{equation*}
		D_{\w}\bpi_{\bto} = -{\left[D_{\bpi}\bvP_{\bto}\right]}^{-1}D_{\w}\bvP_{\bto}.
	\end{equation*}
	Clearly $\w\mapsto\bpi_{\bto}$ is a homeomorphism. Since $D_{\w}\bvP_{\bto}$ is continuous
	and invertible on $W_n\times\bPi\setminus K_{1n}$, we have that $\bpi_{\bto}$ is a 
	$\mathcal{C}^1$-diffeomorphism with differentiable inverse function on $\bPi\setminus\pi_2(K_{1n})$
	given by Lemma~\ref{ch2:thm:diffeo_invers}:
	\begin{equation*}
		D_{\bpi}\bpi_{\bto}^{-1} = {\left[D_{\w}\bpi_{\bto}\right]}^{-1} 
		= -{\left[D_{\w}\bvP_{\bto}\right]}^{-1}D_{\bpi}\bvP_{\bto}.
	\end{equation*}
	Let $\pi_3:\bT_n\times\bPi_n\to\bT_n$ and $\pi_4:\bT_n\times\bPi_n\to\bPi_n$ denotes the projections 
	such that $\pi_3(\bt,\bpi)=\bt$ and $\pi_4(\bt,\bpi)=\bpi$.
	By using the same argument presented above, the Assumption~\ref{ch2:ass:bvp2} (\textit{(ii), (iv), (vi)})
	permits us to have an implicit $\mathcal{C}^{1}$-diffeomorphism $\bpi_{\wo}:\bT_n\to\bPi_n$
	with the following continuous derivatives:
	\begin{align*}
		D_{\bt}\bpi_{\wo} &= -{\left[D_{\bpi}\bvP_{\wo}\right]}^{-1}D_{\bt}\bvP_{\wo},\quad \bt\in\bT\setminus\pi_3(K_2), \\
		D_{\bpi}\bpi_{\wo}^{-1} &= -{\left[D_{\bt}\bvP_{\wo}\right]}^{-1}D_{\bpi}\bvP_{\wo},\quad \bpi\in\bPi\setminus\pi_4(K_2).
	\end{align*}
	Now define the function $\bx(\bt) = \bpi_{\bto}^{-1}\circ\bpi_{\wo}(\bt)$. It is trivial to show 
	that this mapping $\bt\mapsto\bx$ is a $\mathcal{C}^{1}$-diffeomorphism.
	We have from the preceding results and the chain rule that
	\begin{equation*}
		D_{\bt}\bx = {\left[D_{\wo}\bvP_{\bto}\right]}^{-1}D_{\bpi}\bvP_{\bto}{\left[D_{\bpi}\bvP_{\wo}\right]}^{-1}
		D_{\bt}\bvP_{\wo}.
	\end{equation*}
	We make the following remarks.
	First, note that  all these derivatives are square matrices of dimension $p\times p$.
	Second, we have that $D_{\bpi}\bvP_{\bto}(\wo,\hbpi_n)=D_{\bpi}\bvP_p(\bto,\wo,\hbpi_n)=D_{\bpi}\bvP_{\wo}(\bto,\hbpi_n)$
	so $D_{\bpi}\bvP_{\bto}{\left[D_{\bpi}\bvP_{\wo}\right]}^{-1}=\mathbf{I}_p$.
	Third, it holds that $D_{\w}\bvP_{\bto}(\wo,\hbpi_n)=D_{\w}\bvP_{\hbpi_n}(\bto,\wo)$
	and $D_{\bt}\bvP_{\wo}(\bto,\hbpi_n)=D_{\bt}\bvP_{\hbpi_n}(\bto,\wo)$.
	As a consequence, we obtain that
	\begin{equation*}
		\det\left(D_{\bt}\bx\right) = \frac{\det\left(D_{\bt}\bvP_{\hbpi_n}(\wo,\bto)\right)}
		{\det\left(D_{\w}\bvP_{\hbpi_n}(\wo,\bto)\right)}.
	\end{equation*}
	Using Lemma~\ref{ch2:thm:cov} ends the proof of point \textit{(i)} in Theorem~\ref{ch2:thm:freq}.
	From the above display, we have that the relation $\bpi_{\bto}(\wo)=\hbpi_n=\bpi_{\wo}(\bto)$ 
	is uniquely defined, so $\bx(\bto) = \bpi_{\bto}^{-1}(\hbpi_n) = \wo$. Since
	$\bx$ is a diffeomorphism, then $\bx^{-1}(\wo)=\bto$, which finishes the proof.
\end{proof}
\begin{proof}[\textbf{Proof of Proposition~\ref{ch2:thm:freq3}}]
	This is a special case of the Theorem~\ref{ch2:thm:freq}.
	Let define $\bvP_{\hbpi_n}(\w,\bt)=\h(\xo)-\gc(\bt,\w)$, where $\h(\xo)=\hbpi_n$ is fixed.	
	Following the proof of Theorem~\ref{ch2:thm:freq}, we have
	by assumption that $\a:W_n\to\bT_n$ is a $\mathcal{C}^1$-diffeomorphism
	with derivatives
	\begin{align*}
		D_{\w}\a &= -{\left[D_{\bt}\gc\vert_{\bt=\a(\w)}\right]}^{-1}D_{\w}\gc,\quad \w\in W_n\setminus\pi_2(K_n), \\
		D_{\bt}\a^{-1} &= -{\left[D_{\w}\gc\vert_{\w=\a^{-1}(\bt)}\right]}^{-1}D_{\bt}\gc,\quad \bt\in\bT_n\setminus\pi_1(K_n).
	\end{align*}
	The rest of the proof is identical to the proof of Theorem~\ref{ch2:thm:freq}.
\end{proof}
\section{Asymptotics}
\begin{proof}[\textbf{Proof of Theorem~\ref{ch2:thm:con}}]
	We start by showing the claim 1: the pointwise convergence of $\hbpi_n$.
	Then we demonstrate the claim 2 with two different approaches corresponding respectively to
	the Assumptions~\ref{ch2:ass:sw} and~\ref{ch2:ass:iie}.

	1. Fix $\bpio\in\bPi$. Since $\{\bP_n(\bt,\u,\bpi)\}$ is stochastically Lipschitz in $\bpi$, it is stochastically
	equicontinuous by the Lemma~\ref{thm:equicontinuity}.
	In addition, $\bPi$ is compact and $\{\bP_n\}$
	is pointwise convergent by assumption, so by the Lemma~\ref{thm:uniform} $\{\bP_n\}$ converges
	uniformly and the limit $\bP$ is uniformly continuous.
	By $\bPi$ compact and the continuity of the norm,
	the infimum of the norm of $\bP$ exists. The infimum of $\bP$ is well-separated by the 
	bijectivity of the function. Therefore, all the conditions of Lemma~\ref{thm:lemma1}
	are satisfied and $\{\hbpi_n\}$ converges pointwise to $\bpio$.

	2 (\textit{i}). For this proof, we consider $\bt$ and $\bpi$ jointly. 
	Let $\Ks=\bT\cap\bPi$ be the set for both $\bt$ and $\bpi$.
	Fix $(\bto,\bpio)\in\Ks$.
	Since $\bPi\subset\R^p$ and $\bT\subset\R^p$ are compact subsets of a metric space,
	they are closed (see the Theorem 2.34 in~\cite{rudin1976principles}), and
	$\Ks$ is compact (see the Corollary to the Theorem 2.35 in~\cite{rudin1976principles})
	and nonempty (Theorem 2.36 in~\cite{rudin1976principles}). Having $\Ks$ compact,
	it is now sufficient to show that $\{\bP_n\}$ is jointly stochastically Lipschitz
	as the rest of the proof follows exactly the same steps as the claim 1.
	
	For every $(\bt_1,\bpi_1),(\bt_2,\bpi_2)\in\Ks$,
	$n$ and $\u\sim F_{\u}$, we have by the triangle inequality that
	\begin{align*}
		\left\lVert\bP_n(\bt_1,\u,\bpi_1)-\bP_n(\bt_2,\u,\bpi_2)\right\rVert
		&= \big\lVert\bP_n(\bt_1,\u,\bpi_1)-\bP_n(\bt_1,\u,\bpi_2) \\
		&\qquad+\bP_n(\bt_1,\u,\bpi_2)-\bP_n(\bt_2,\u,\bpi_2)\big\rVert \\
		&\leq 
		\left\lVert\bP_n(\bt_1,\u,\bpi_1)-\bP_n(\bt_1,\u,\bpi_2)\right\rVert\\
		&\qquad+\left\lVert\bP_n(\bt_1,\u,\bpi_2)-\bP_n(\bt_2,\u,\bpi_2)\right\rVert \\
		&\leq
		D_n\left(\left\lVert\bpi_1-\bpi_2\right\rVert+
		\left\lVert\bt_1-\bt_2\right\rVert\right),
	\end{align*}
	where for the last inequality we make use of the marginal stochastic Lipschitz assumptions 
	and $D_n=\max(A_n,B_n)$. Let $a=\lVert\bt_1-\bt_2\rVert$ and $b=\lVert\bpi_1-\bpi_2\rVert$.
	Now remark that for the $\ell_2$-norm we have
	\begin{equation*}
		\left\lVert\begin{pmatrix}\bt_1\\ \bpi_1\end{pmatrix}-\begin{pmatrix}\bt_2\\ \bpi_2\end{pmatrix}\right\rVert
			= \sqrt{a^2+b^2}.
	\end{equation*}
	Since $a,b$ are positive real numbers, a direct application of the inequality of arithmetic and geometric means
	gives
	\begin{equation*}
		\sqrt{2}\sqrt{a^2+b^2}\geq a+b.
	\end{equation*}
	Therefore, we have that
	\begin{equation*}
		D_n\left(\left\lVert\bpi_1-\bpi_2\right\rVert+
		\left\lVert\bt_1-\bt_2\right\rVert\right) 
		\leq D_n^\star 
		\left\lVert\begin{pmatrix}\bt_1\\ \bpi_1\end{pmatrix}
			-\begin{pmatrix}\bt_2\\ \bpi_2\end{pmatrix}\right\rVert,
	\end{equation*}
	where $D_n^\star = \sqrt{2}D_n$. Consequently, $\{\bP_n\}$ is jointly
	stochastically Lipschitz, and following the proof of claim 1
	we have that $\hbt_n\wcp\bto$. More precisely, we even have
	that $(\hbt_n,\hbpi_n)\wcp(\bto,\bpio)$.

	2 (\textit{ii}). This proof is different from 2 (\textit{i}) since 
	$\hbpi_{\text{II},n}$ is considered as a function of $\bt$.
	Fix $\bpio\in\bPi$.
	Since $\{\hbpi_{\text{II},n}\}$ is stochastically Lipschitz in $\bt$, it is stochastically
	equicontinuous by the Lemma~\ref{thm:equicontinuity}. In addition, $\bT$ is compact and $\{\hbpi_{\text{II},n}\}$
	is pointwise convergent by the claim 1, so by the Lemma~\ref{thm:uniform} $\{\hbpi_{\text{II},n}\}$ converges
	uniformly and the limit $\bpi$ is uniformly continuous in $\bt$.
	Let the stochastic and deterministic objective functions be $Q_n(\bt)=\lVert\hbpi_n-\hbpi_{\text{II},n}(\bt)\rVert$
	and $Q(\bt)=\lVert\bpio-\bpi(\bt)\rVert$, for any norms.
	Now, we have by using successively the reverse and the regular triangle inequalities
	\begin{align*}
		\left\lvert Q_n(\bt)-Q(\bt)\right\rvert 
		&= \left\lvert\left\lVert\hbpi_n-\hbpi_{\text{II},n}(\bt)\right\rVert-\left\lVert\bpio-\bpi(\bt)\right\rVert\right\rvert\\
		&\leq
		\left\lVert\hbpi_n-\hbpi_{\text{II},n}(\bt)-\bpio+\bpi(\bt)\right\rVert \\
		&\leq
		\left\lVert\hbpi_n-\bpio\right\rVert + \left\lVert\bpi(\bt)-\hbpi_{\text{II},n}(\bt)\right\rVert.
	\end{align*}
	By the convergence of $\{\hbpi_n\}$ and the uniform convergence of $\{\hbpi_{\text{II},n}\}$, we have
	\begin{equation*}
		\lim_{n\to\infty}\Pr\left(\sup_{\bt\in\bT}\left\lvert Q_n(\bt)-Q(\bt)\right\rvert\right) = \op(1).
	\end{equation*}
	By $\bPi$ compact and the continuity of the norm,
	the infimum of the norm of $\bP$ exists. The infimum of $\bP$ is well-separated by the 
	bijectivity of the function. Therefore, all the conditions of Lemma~\ref{thm:lemma1}
	are satisfied and $\{\hbpi_n\}$ converges pointwise to $\bpio$.
\end{proof}
\begin{proof}[\textbf{Proof of Theorem~\ref{ch2:thm:asynorm}}]
	We first demonstrate the asymptotic distribution of the auxiliary estimator,
	then separately shows the result for $\hbt_n$ using independentely the
	Assumption~\ref{ch2:ass:sw2} and~\ref{ch2:ass:iie2}.

	1. The result on $\hbpi_n$ is a special case of $\hbpi_{\text{II},n}(\bt)$.
	Fix $\bto\in\Xs$ and denote $\bpi(\bto)\equiv\bpio$. 
	By assumptions, the conditions for the delta method in Lemma~\ref{thm:fot} are satisfied so
	we have
	\begin{equation}\label{eq1}
		\bP_n\left(\bto,\u_s,\hbpi_{\text{II},n}(\bto)\right) - \bP_n\left(\bto,\u_s,\bpio\right)
		= D_{\bpi}\bP_n\left(\bto,\u_s,\bpio\right)\cdot\left(\hbpi_{\text{II},n}(\bto)-\bpio\right) 
		+ \op\left(\left\lVert\hbpi_{\text{II},n}(\bto)-\bpio\right\rVert\right).
	\end{equation}
	By the Definition~\ref{ch2:def:iie}, we have $\bP_n\left(\bto,\u_s,\hbpi_{\text{II},n}(\bto)\right)=\0$.
	By the Theorem~\ref{ch2:thm:con}, $\op\left(\left\lVert\hbpi_{\text{II},n}(\bto)-\bpio\right\rVert\right)=\op(1)$.
	By assumptions, $D_{\bpi}\bP_n\left(\bto,\u_s,\bpio\right)\wcp\K$, $\K$ nonsingular. 
	Multiplying by square-root $n$, the proof results from the central limit theorem assumption on $\bP_n$ and 
	the Slutsky's lemma.

	2 \textit{(i)}. From the delta method in Lemma~\ref{thm:fot}, we obtain
	\begin{equation*}
		\bP_n\left(\hbt_n,\u_s,\hbpi_n\right) - \bP_n\left(\bto,\u_s,\hbpi_n\right)
		= D_{\bt}\bP_n\left(\bto,\u_s,\hbpi_n\right)\cdot\left(\hbt_n-\bto\right)
		+ \op\left(\left\lVert\hbt_n-\bto\right\rVert\right).
	\end{equation*}
	By definition we have $\bP_n\left(\hbt_n,\u_s,\hbpi_n\right)=\0$.
	Using again the delta method
	on the non-zero left-hand side element, we obtain from~\eqref{eq1}
	\begin{align*}
		& \0 - \left[\bP_n\left(\bto,\u_s,\bpio\right) + 
			D_{\bpi}\bP_n\left(\bto,\u_s,\bpio\right)\cdot\left(\hbpi_n-\bpio\right)
			+ \op\left( \left\lVert\hbpi_n-\bpio\right\rVert\right)\right] \\
		&\quad = D_{\bt}\bP_n\left(\bto,\u_s,\hbpi_n\right)\cdot\left(\hbt_n-\bto\right)
		+ \op\left( \left\lVert\hbt_n-\bto\right\rVert \right).
	\end{align*}
	Since $\{D_{\bt}\bP_n(\bto,\u_s,\bpi)\}$ is stochastically Lipschitz in $\bpi$, it is stochastically
	equicontinuous by the Lemma~\ref{thm:equicontinuity}. In addition, $\bPi$ is compact and $\{D_{\bt}\bP_n\}$
	is pointwise convergent by assumption, so by the Lemma~\ref{thm:uniform} $\{D_{\bt}\bP_n\}$ converges
	uniformly and the limit $\J$ is uniformly continuous in $\bpi$.

	Next, we obtain the following
	\begin{align*}
		\left\lVert D_{\bt}\bP_n(\hbpi_n)-\J(\bpio)\right\rVert
		& \leq\left\lVert D_{\bt}\bP_n(\hbpi_n)-\J(\hbpi_n)\right\rVert + \left\lVert \J(\hbpi_n)-\J(\bpio)\right\rVert \\
		& \leq\sup_{\bpi\in\bPi}\left\lVert D_{\bt}\bP_n(\bpi)-\J(\bpi)\right\rVert + \left\lVert \J(\hbpi_n)-\J(\bpio)\right\rVert.
	\end{align*}
	By uniform convergence $\sup_{\bpi\in\bPi}\left\lVert D_{\bt}\bP_n(\bpi)-\J(\bpi)\right\rVert = \op(1)$ and
	by the continuous mapping theorem $\left\lVert\J(\hbpi_n)-\J(\bpio)\right\rVert=\op(1)$.
	
	The central limit theorem is satisfied for the estimating equation thus $n^{1/2}\bP_n\rightsquigarrow\Nor\left(\0,\Q\right)$.
	Let $\y$ be a random variable identically and independently distributed according to $\Nor(\0,\Q)$.
	Therefore, by multiplying by square-root $n$ we obtain
	\begin{equation*}
		-\y - \K n^{1/2}\left(\hbpi_n-\bpio\right) - \op\left(\left\lVert\hbpi_n-\bpio \right\rVert \right)
		= \J n^{1/2}\left(\hbt_n-\bto\right) + \op\left(\left\lVert\hbt_n-\bto\right\rVert \right).
	\end{equation*}
	By the Theorem~\ref{ch2:thm:con}, we have $\op\left(\left\lVert\hbpi_n-\bpio \right\rVert \right)=\op(1)$
	and $\op\left(\left\lVert\hbt_n-\bto\right\rVert \right)=\op(1)$. 
	By the result of the claim 1 and the nonsingularity of $\J$, we have
	\begin{equation*}
		n^{1/2}\left(\hbt_n-\bto\right) = -\J^{-1}\left(\y + \K\cdot\K^{-1}\y + \op(1)\right) + \op(1).
	\end{equation*}
	Slutsky's lemma ends the proof.

	2 (\textit{ii}). Let $\g_n(\bt)=\hbpi_n-\hbpi_{\text{II},n}(\bt)$.
	The conditions for the delta method in Lemma~\ref{thm:fot} are satisfied by
	assumption so we have
	\begin{equation}\label{eq2}
		\g_n(\hbt_n)-\g_n(\bto) = D_{\bt}\g_n(\bto)\cdot\left(\hbt_n-\bto\right) + 
		\op\left(\left\lVert\hbt_n-\bto\right\rVert\right).
	\end{equation}
	Since $\hbt_n=\argzero_{\bt}d(\hbt_n,\hbt_{\text{II},n}(\bt))$, we have
	$\hbt_n-\hbt_{\text{II},n}(\hbt_n)=\0$ and thus $\g_n(\hbt_n)=\0$.
	By the Theorem~\ref{ch2:thm:con}, we have $\op\left(\left\lVert\hbt_n-\bto\right\rVert\right)=\op(1)$.
	We have $D_{\bt}\g_n(\bto)=-D_{\bt}\hbpi_{\text{II},n}(\bto)$ which, by assumption
	converges pointwise to $D_{\bt}\bpi(\bto)$.
	By the claim 1, we have $n^{1/2}(\hbpi_n-\bpio)\eqd n^{1/2}(\hbpi_{\text{II},n}(\bt)-\bpio)\eqd\K^{-1}\y$
	as $n\to\infty$. Hence, multiplying the Equation~\ref{eq2} by square-root $n$, gives the following
	\begin{equation*}
		\K^{-1}(\y-\y)=D_{\bt}\bpi(\bto)\cdot n^{1/2}\left(\hbt_n-\bto\right) + \op(1),
	\end{equation*}
	for sufficiently large $n$. Remark that the mapping $\bt\mapsto\bpi$ is implicitely defined by
	\begin{equation*}
		\bP\left(\bt,\bpi(\bt)\right) = \0.
	\end{equation*}
	Since $\bP$ is once continuously differentiable in $(\bt,\bpi)$ and the partial derivatives
	are invertibles, the conditions for invoking an implicit function theorem are satisfied 
	(see for example the Theorem 9.28 in~\cite{rudin1976principles}) and one of the conclusion is that
	\begin{equation*}
		D_{\bt}\bpi(\bto) = -\K^{-1}\J.
	\end{equation*}
	Since $\J$ is invertible, the conclusion follows by Slutsky's lemma.
\end{proof}
\begin{proof}[\textbf{Proof of Proposition~\ref{ch2:thm:asynorm2}}]
	The proof follows essentially the same steps as the proof of Theorem~\ref{ch2:thm:asynorm}.
	From the proof of Theorem~\ref{ch2:thm:asynorm}, the following holds:
	$n^{1/2}\left(\hbpi_n-\bpio\right)\overset{\d}{=}\K^{-1}\y_0$ and 
	$n^{1/2}\bP_n\left(\bto,\u_s,\bpio\right)\overset{\d}{=}\y_s$ as $n\to\infty$ where
	$\y_j\sim\Nor\left(\0,\Q\right)$, $j\in\N^+$, $D_{\bpi}\bP_n\left(\bto,\uo,\bpio\right)$ converges in probability
	to $\K$ and $D_{\bt}\bP_n\left(\bto,\u_s,\bpi\right)$ converges	uniformly in probability to $\J$. 
	The $\{\u_j:j\in\N_S\}$ are assumed independent and so are $\{\y_j:j\in\N_S\}$.
	
	From the delta method in Lemma~\ref{thm:fot}, we obtain
	\begin{equation*}
		\frac{1}{S}\sum_{s\in\N^+_S}\bP_n\left(\hbt_n^{(s)},\u_s,\hbpi_n\right) 
		- \frac{1}{S}\sum_{s\in\N^+_S}\bP_n\left(\bto,\u_s,\hbpi_n\right)
		= \frac{1}{S}\sum_{s\in\N^+_S} D_{\bt}\bP_n\left(\bto,\u_s,\hbpi_n\right)
		\cdot\left(\hbt_n^{(s)}-\bto\right) + \op(1).
	\end{equation*}
	By definition $\frac{1}{S}\sum_{s\in\N^+_S}\bP_n\left(\hbt_n^{(s)},\u_s,\hbpi_n\right)=\0$.
	Using the delta method on $\frac{1}{S}\sum_{s\in\N^+_S}
	\bP_n\left(\bto,\u_s,\hbpi_n\right)$,
	multiplying by square-root $n$, we obtain from the results of
	Theorem~\ref{ch2:thm:asynorm}:
	\begin{equation*}
		-\frac{1}{S}\sum_{s\in\N^+_S}\y_s -\K\K^{-1}\y_0-\op(1)
		= \J n^{1/2}\left(\bar{\bt}_n-\bto\right) + \op(1).
	\end{equation*}
	Clearly $\frac{1}{S}\sum_{s\in\N^+_S}\y_s\sim\Nor\left(\0,\frac{1}{S}\Q\right)$.
	The conclusion follows from Slutsky's lemma.
\end{proof}
\newpage
\section{Additional simulation results}
\subsection{Lomax distribution}\label{ch2:app:lomax}
\begin{table}[hbt]
	\centering
	\begin{tabular}{@{}cccccc@{}}
		\toprule
		&SwiZs&Boot&AB&RSwiZs&RBoot\\
		\midrule
		$n=35$&0.1430&0.0222&0.0197&0.5613&0.0998\\
		$n=50$&0.2002&0.0293&0.0268&0.7889&0.1320\\
		$n=100$&0.3826&0.0526&0.0504&1.3520&0.2314\\
		$n=150$&0.5580&0.0753&0.0736&1.7792&0.3291\\
		$n=250$&0.8998&0.1228&0.1211&2.3141&0.5174\\
		$n=500$&1.7763&0.2364&0.2398&3.2132&0.9848\\
		\bottomrule
	\end{tabular}
	\caption{Average computationnal time in seconds to approximate a distribution on $S=10,000$ points.}
	\label{ch2:tab:lomax7}
\end{table}
\begin{table}
	\centering
	\resizebox{\columnwidth}{!}{%
	\begin{tabular}{@{}ccccccccccccc@{}}
		\toprule
		&\multicolumn{2}{c}{SwiZs} & \multicolumn{2}{c}{Boot} & \multicolumn{2}{c}{BA} 
		&\multicolumn{2}{c}{RSwiZs} & \multicolumn{2}{c}{RBoot}\\
		$\alpha$ & $\theta_1$ & $\theta_2$ & $\theta_1$ & $\theta_2$ & $\theta_1$ & $\theta_2$ & $\theta_1$ & $\theta_2$ & $\theta_1$ & $\theta_2$\\
		\midrule
		\multicolumn{11}{c}{$n=35$} \\[.2em]
		50\%&49.48&50.07&43.10&44.26&0.00&0.00&42.73&44.07&36.72&36.84\\
		75\%&74.49&75.14&65.82&65.39&0.00&0.00&65.84&66.59&55.00&55.06\\
		90\%&89.31&89.39&80.64&78.74&0.00&0.00&81.41&81.97&64.47&64.26\\
		95\%&94.27&94.34&86.71&84.28&0.03&0.00&87.58&87.41&67.33&67.13\\
		99\%&98.26&98.43&91.23&91.07&0.75&0.00&93.84&93.53&69.64&70.39\\[.4em]
		\multicolumn{11}{c}{$n=50$} \\[.2em]
		50\%&49.59&49.88&44.48&45.30&0.01&0.00&45.70&46.93&37.37&37.64\\
		75\%&74.73&76.67&68.43&67.84&0.08&0.00&67.40&68.21&57.44&56.73\\
		90\%&89.89&90.62&83.15&81.57&0.76&0.00&82.51&82.75&69.52&68.81\\
		95\%&94.67&94.94&89.26&87.11&1.92&0.00&88.47&88.35&73.01&72.49\\
		99\%&98.40&98.46&95.19&93.69&10.86&0.00&94.79&94.80&75.97&76.43\\[.4em]
		\multicolumn{11}{c}{$n=100$}\\[.2em]
		50\%&49.86&49.95&47.52&48.04&20.52&27.75&49.44&49.80&36.19&35.48\\
		75\%&75.37&75.88&72.00&71.59&44.13&57.82&73.07&74.32&57.01&55.61\\
		90\%&90.20&90.42&86.69&85.86&69.68&81.85&86.54&86.83&73.68&71.96\\
		95\%&95.41&95.67&92.06&90.96&81.89&91.13&91.69&91.52&80.75&79.17\\
		99\%&98.85&98.91&97.32&96.42&94.93&98.74&96.85&96.79&86.96&86.38\\[.4em]
		\multicolumn{11}{c}{$n=150$}\\[.2em]
		50\%&50.12&49.80&48.36&48.58&47.05&49.78&49.80&49.82&33.94&33.00\\
		75\%&74.85&75.32&72.41&72.63&70.68&72.58&74.44&74.69&55.12&53.45\\
		90\%&90.31&90.32&87.58&86.85&86.94&89.18&88.95&89.22&72.14&70.01\\
		95\%&95.08&95.35&93.03&92.11&93.26&94.89&93.60&93.74&80.17&78.15\\
		99\%&99.08&99.10&97.92&97.43&98.72&99.28&97.81&97.69&90.07&88.56\\[.4em]
		\multicolumn{11}{c}{$n=250$}\\[.2em]
		50\%&49.46&49.84&48.60&49.01&47.61&47.09&49.55&49.90&29.16&28.45\\
		75\%&75.02&74.49&73.59&72.75&72.09&72.63&74.83&74.80&49.94&47.56\\
		90\%&89.55&89.81&88.05&88.11&89.54&90.13&89.56&89.58&67.50&65.25\\
		95\%&94.77&94.79&93.56&93.34&94.79&95.68&94.50&94.70&76.90&74.39\\
		99\%&99.02&99.03&98.46&97.92&99.18&99.50&98.61&98.70&89.37&87.24\\[.4em]
		\multicolumn{11}{c}{$n=500$}\\[.2em]
		50\%&50.08&49.89&49.29&49.81&48.76&48.67&50.26&49.64&20.51&18.95\\
		75\%&74.73&74.36&73.90&73.64&73.68&73.85&74.55&74.68&37.76&34.96\\
		90\%&89.53&89.75&88.86&88.69&89.03&89.22&89.45&89.80&56.15&52.68\\
		95\%&94.92&94.86&94.11&94.22&94.33&94.77&94.92&94.80&66.89&63.51\\
		99\%&98.97&98.99&98.62&98.40&99.01&99.07&98.94&99.03&83.63&80.06\\
		\bottomrule
	\end{tabular}%
	}
	\caption{Estimated coverage probabilities.}
	\label{ch2:tab:lomax1}
\end{table}
\begin{table}
	\centering
	\begin{tabular}{@{}cccccc@{}}
		\toprule
		& SwiZs & Boot & BA & RSwiZs & RBoot \\
		$\alpha$ &\multicolumn{5}{c}{Gini index}\\ 
		\midrule
		\multicolumn{6}{c}{$n=35$}\\[.2em]
		50\%&50.22&44.26&0.02&44.27&36.84\\
		75\%&76.03&65.44&0.72&67.12&55.06\\
		90\%&91.07&78.96&68.11&83.07&64.36\\
		95\%&96.76&84.35&100.00&89.43&67.19\\
		99\%&98.84&91.10&100.00&93.88&70.41\\[.4em]
		\multicolumn{6}{c}{$n=50$}\\[.2em]
		50\%&49.89&45.30&0.00&46.94&37.64\\
		75\%&76.86&67.84&0.00&68.26&56.73\\
		90\%&90.83&81.58&41.20&82.68&68.82\\
		95\%&95.17&87.16&71.42&88.40&72.49\\
		99\%&98.92&93.76&99.82&95.14&76.45\\[.4em]
		\multicolumn{6}{c}{$n=100$}\\[.2em]
		50\%&49.95&48.04&32.96&49.80&35.48\\
		75\%&75.88&71.59&59.90&74.32&55.61\\
		90\%&90.42&85.86&82.63&86.83&71.96\\
		95\%&95.74&90.98&91.44&91.64&79.19\\
		99\%&98.85&96.46&98.73&96.83&86.43\\[.4em]
		\multicolumn{6}{c}{$n=150$}\\[.2em]
		50\%&49.80&48.58&46.30&49.82&33.00\\
		75\%&75.32&72.63&72.68&74.69&53.45\\
		90\%&90.32&86.85&89.18&89.22&70.01\\
		95\%&95.35&92.12&94.87&93.73&78.15\\
		99\%&99.06&97.47&99.27&97.71&88.60\\[.4em]
		\multicolumn{6}{c}{$n=250$}\\[.2em]
		50\%&49.84&49.01&46.99&49.90&28.45\\
		75\%&74.49&72.75&72.41&74.80&47.56\\
		90\%&89.81&88.11&88.95&89.58&65.25\\
		95\%&94.81&93.34&94.99&94.69&74.43\\
		99\%&99.04&97.93&99.48&98.68&87.34\\[.4em]
		\multicolumn{6}{c}{$n=500$}\\[.2em]
		50\%&49.89&49.81&48.67&49.64&18.95\\
		75\%&74.36&73.64&73.85&74.68&34.96\\
		90\%&89.75&88.69&89.22&89.80&52.68\\
		95\%&94.86&94.22&94.77&94.79&63.57\\
		99\%&98.98&98.41&99.03&99.02&80.28\\
		\bottomrule
	\end{tabular}
	\caption{Estimated coverage probabilities of Gini index.}
	\label{ch2:tab:lomax2}
\end{table}
\begin{table}
	\centering
	\begin{tabular}{@{}cccccc@{}}
		\toprule
		& SwiZs & Boot & BA & RSwiZs & RBoot \\
		$\alpha$ & \multicolumn{5}{c}{95\% value-at-risk}\\ 
		\midrule
		\multicolumn{6}{c}{$n=35$}\\[.2em]
		50\%&47.30&46.08&20.92&45.34&41.13\\
		75\%&73.76&67.53&55.77&70.38&61.00\\
		90\%&90.05&80.35&93.73&88.08&73.92\\
		95\%&95.67&85.36&98.92&94.80&79.41\\
		99\%&99.17&91.63&99.97&99.25&87.26\\[.4em]
		\multicolumn{6}{c}{$n=50$}\\[.2em]
		50\%&48.14&47.23&31.76&46.40&41.27\\
		75\%&73.39&69.40&63.30&70.22&61.47\\
		90\%&89.63&82.24&91.60&87.07&74.72\\
		95\%&94.89&87.41&97.72&93.60&80.20\\
		99\%&99.23&93.17&99.90&99.27&87.87\\[.4em]
		\multicolumn{6}{c}{$n=100$}\\[.2em]
		50\%&49.75&48.90&48.33&49.18&39.94\\
		75\%&74.68&72.61&75.68&72.93&61.39\\
		90\%&89.48&86.38&91.97&87.16&75.97\\
		95\%&95.07&91.17&96.79&94.17&82.45\\
		99\%&99.23&96.31&99.75&99.11&90.45\\[.4em]
		\multicolumn{6}{c}{$n=150$}\\[.2em]
		50\%&50.10&49.19&49.47&49.91&37.43\\
		75\%&74.13&73.17&75.42&73.57&59.31\\
		90\%&89.77&87.25&91.21&88.49&75.26\\
		95\%&94.76&92.57&96.18&93.31&81.76\\
		99\%&98.89&97.34&99.61&98.46&91.00\\[.4em]
		\multicolumn{6}{c}{$n=250$}\\[.2em]
		50\%&50.28&49.52&50.02&50.24&34.09\\
		75\%&75.29&74.25&74.87&74.75&55.55\\
		90\%&89.43&88.10&90.27&89.13&72.28\\
		95\%&94.66&93.26&95.15&94.14&80.35\\
		99\%&98.89&97.85&99.10&98.67&90.11\\[.4em]
		\multicolumn{6}{c}{$n=500$}\\[.2em]
		50\%&49.15&48.63&49.00&49.22&27.45\\
		75\%&74.88&74.01&74.63&74.53&45.61\\
		90\%&90.02&89.46&90.37&89.93&62.84\\
		95\%&94.97&94.45&95.18&94.85&72.65\\
		99\%&98.92&98.32&98.87&98.96&86.63\\[.4em]
		\bottomrule
	\end{tabular}
	\caption{Estimated coverage probabilities of value-at-risk at 95\%.}
	\label{ch2:tab:lomax3}
\end{table}
\begin{table}
	\centering
	\begin{tabular}{@{}cccccc@{}}
		\toprule
		& SwiZs & Boot & BA & RSwiZs & RBoot \\
		$\alpha$ &\multicolumn{5}{c}{95\% expected shortfall}\\ 
		\midrule
		\multicolumn{6}{c}{$n=35$}\\[.2em]
		50\%&50.33&48.55&0.02&50.08&47.38\\
		75\%&74.97&72.60&0.72&74.70&71.28\\
		90\%&89.61&87.63&68.11&89.24&86.35\\
		95\%&94.65&92.87&100.00&94.37&92.23\\
		99\%&98.80&97.97&100.00&98.72&97.48\\[.4em]
		\multicolumn{6}{c}{$n=50$}\\[.2em]
		50\%&49.48&48.24&0.00&49.28&47.06\\
		75\%&74.81&72.74&0.00&74.45&71.28\\
		90\%&89.76&88.07&41.20&89.25&86.85\\
		95\%&94.74&93.32&71.42&94.48&92.16\\
		99\%&98.89&97.92&99.82&98.62&97.48\\[.4em]
		\multicolumn{6}{c}{$n=100$}\\[.2em]
		50\%&49.94&49.16&32.96&49.64&47.22\\
		75\%&74.47&74.12&59.90&74.37&72.21\\
		90\%&90.13&89.15&82.63&89.99&87.57\\
		95\%&95.10&94.23&91.44&95.00&93.13\\
		99\%&98.98&98.55&98.73&98.91&98.10\\[.4em]
		\multicolumn{6}{c}{$n=150$}\\[.2em]
		50\%&49.91&49.49&46.30&49.81&48.13\\
		75\%&75.03&74.25&72.68&74.95&72.45\\
		90\%&89.82&89.31&89.18&89.74&87.76\\
		95\%&95.05&94.37&94.87&94.98&93.15\\
		99\%&98.91&98.62&99.27&98.86&98.14\\[.4em]
		\multicolumn{6}{c}{$n=250$}\\[.2em]
		50\%&50.53&50.64&46.99&50.44&47.94\\
		75\%&75.01&74.97&72.41&74.91&72.31\\
		90\%&89.96&89.72&88.95&89.98&87.75\\
		95\%&95.11&94.58&94.99&95.13&93.16\\
		99\%&99.04&98.70&99.48&99.06&98.14\\[.4em]
		\multicolumn{6}{c}{$n=500$}\\[.2em]
		50\%&49.25&49.34&48.67&49.48&46.61\\
		75\%&74.50&74.29&73.85&74.28&70.91\\
		90\%&90.02&89.56&89.22&89.99&86.47\\
		95\%&95.05&94.77&94.77&95.13&92.52\\
		99\%&99.01&99.01&99.03&99.04&98.23\\
		\bottomrule
	\end{tabular}
	\caption{Estimated coverage probabilities of expected shortfall at 95\%.}
	\label{ch2:tab:lomax4}
\end{table}
\begin{table}
	\centering
	\footnotesize	
	\begin{tabular}{@{}ccccccccccccc@{}}
		\toprule
		&\multicolumn{2}{c}{SwiZs} & \multicolumn{2}{c}{Boot} & \multicolumn{2}{c}{BA} 
		&\multicolumn{2}{c}{RSwiZs} & \multicolumn{2}{c}{RBoot}\\
		$\alpha$ & $\theta_1$ & $\theta_2$ & $\theta_1$ & $\theta_2$ & $\theta_1$ & $\theta_2$ & $\theta_1$ & $\theta_2$ & $\theta_1$ & $\theta_2$\\
		\midrule
		\multicolumn{11}{c}{$n=35$}\\[.2em]
		50\%&2.19&1.85&7.52&6.04&0.26&0.34&1.89&1.64&7.97&6.73\\
		75\%&4.79&3.92&216.08&179.86&0.46&0.54&3.84&3.37&27.20&23.62\\
		90\%&11.18&8.56&9710.48&8673.53&1.31&1.09&6.96&5.97&86.42&75.85\\
		95\%&24.30&18.00&2.55$\times10^4$&2.18$\times10^4$&8.99&8.89&9.89&7.92&161.13&142.10\\
		99\%&2488.08&1849.66&1.19$\times10^5$&1.05$\times10^5$&3.18$\times10^9$&3.30$\times10^9$&22.62&17.10&435.99&401.28\\[.4em]
		\multicolumn{11}{c}{$n=50$}\\[.2em]
		50\%&1.78&1.51&3.61&2.98&0.39&0.42&1.56&1.34&5.04&4.20\\
		75\%&3.60&2.97&10.55&8.78&0.66&0.68&3.11&2.65&14.89&12.37\\
		90\%&6.78&5.41&642.67&551.95&1.22&0.94&5.57&4.83&44.67&38.37\\
		95\%&10.78&8.38&7.40$\times10^3$&6.31$\times10^3$&6.13&5.27&7.80&6.70&84.42&73.24\\
		99\%&54.20&39.06&5.57$\times10^4$&4.82$\times10^4$&1.09$\times10^7$&1.04$\times10^7$&15.60&12.65&231.61&202.96\\[.4em]
		\multicolumn{11}{c}{$n=100$}\\[.2em]
		50\%&1.26&1.06&1.69&1.39&0.64&0.60&1.19&1.01&2.73&2.27\\
		75\%&2.32&1.92&3.32&2.74&1.08&1.02&2.23&1.87&6.01&5.01\\
		90\%&3.74&3.04&6.28&5.20&1.55&1.36&3.67&3.03&13.00&10.88\\
		95\%&4.92&3.94&10.30&8.58&1.93&1.54&4.89&4.00&22.10&18.69\\
		99\%&8.58&6.63&181.34&153.63&20.11&16.79&8.41&6.95&64.18&55.35\\[.4em]
		\multicolumn{11}{c}{$n=150$}\\[.2em]
		50\%&1.02&0.86&1.21&1.01&0.71&0.62&1.00&0.85&2.02&1.68\\
		75\%&1.82&1.52&2.24&1.88&1.23&1.08&1.80&1.52&4.00&3.35\\
		90\%&2.78&2.30&3.71&3.11&1.78&1.59&2.80&2.32&7.50&6.28\\
		95\%&3.52&2.89&5.05&4.26&2.12&1.90&3.58&2.95&11.12&9.34\\
		99\%&5.38&4.35&10.59&8.97&2.86&2.27&5.62&4.52&26.58&22.47\\[.4em]
		\multicolumn{11}{c}{$n=250$}\\[.2em]
		50\%&0.78&0.66&0.85&0.72&0.64&0.55&0.79&0.66&1.45&1.21\\
		75\%&1.36&1.15&1.52&1.29&1.13&0.96&1.38&1.16&2.68&2.24\\
		90\%&2.01&1.69&2.34&1.99&1.68&1.44&2.07&1.72&4.41&3.68\\
		95\%&2.48&2.08&2.97&2.52&2.07&1.78&2.56&2.12&5.94&4.97\\
		99\%&3.56&2.92&4.72&4.01&2.96&2.55&3.69&3.01&10.84&9.10\\[.4em]
		\multicolumn{11}{c}{$n=500$}\\[.2em]
		50\%&0.55&0.46&0.57&0.48&0.50&0.42&0.56&0.47&0.97&0.81\\
		75\%&0.94&0.80&0.99&0.84&0.87&0.74&0.96&0.81&1.71&1.43\\
		90\%&1.37&1.16&1.47&1.25&1.27&1.08&1.41&1.18&2.63&2.20\\
		95\%&1.66&1.40&1.80&1.53&1.54&1.32&1.71&1.43&3.31&2.78\\
		99\%&2.27&1.90&2.55&2.16&2.16&1.83&2.35&1.95&5.05&4.22\\
		\bottomrule
	\end{tabular}
	\caption{Estimated median interval length.} 
	\label{ch2:tab:lomax5}
\end{table}
\begin{landscape}
\begin{table}
	\centering
	\footnotesize
	\begin{tabular}{@{}cccccccccccccccccc@{}}
		\toprule
		&\multicolumn{2}{c}{SwiZs: mean} &\multicolumn{2}{c}{SwiZs: median} &\multicolumn{2}{c}{MLE}&\multicolumn{2}{c}{AB}
		&\multicolumn{2}{c}{RSwiZs: mean}&\multicolumn{2}{c}{RSwiZs: median} &\multicolumn{2}{c}{WMLE} \\ 
		&$\theta_1$&$\theta_2$&$\theta_1$&$\theta_2$&$\theta_1$&$\theta_2$&$\theta_1$&$\theta_2$&$\theta_1$&$\theta_2$&$\theta_1$&$\theta_2$&$\theta_1$&$\theta_2$\\
		\midrule
		\multicolumn{15}{c}{\textbf{Mean bias}}\\[.2em]
		$n=35$&2511.13&2226.09&2504.27&2230.19&2492.15&2241.82&-1.38$\times10^{12}$&-1.34$\times10^{12}$&13.33&11.50&13.38&11.53&13.78&12.10\\
		$n=50$&832.02&739.28&829.87&739.77&827.45&742.50&-1.54$\times10^{11}$&-1.55$\times10^{11}$&5.99&5.19&6.07&5.22&6.52&5.70\\
		$n=100$&45.96&37.47&45.71&37.28&45.81&37.48&-6.65$\times10^8$&-5.22$\times10^8$&1.20&1.03&1.26&1.05&1.72&1.47\\
		$n=150$&1.03&0.91&0.96&0.82&1.06&0.92&-1.60$\times10^4$&-1.48$\times10^4$&0.48&0.42&0.52&0.43&0.96&0.82\\
		$n=250$&0.17&0.15&0.15&0.12&0.21&0.18&-0.02&-0.02&0.20&0.18&0.21&0.17&0.62&0.53\\
		$n=500$&0.08&0.07&0.07&0.06&0.10&0.08&0.00&0.00&0.08&0.08&0.08&0.06&0.45&0.39\\[.4em]
		\multicolumn{15}{c}{\textbf{Median bias}}\\[.2em]
		$n=35$&0.4583&0.4894&0.0538&0.0276&0.5885&0.4654&-1.5551&-1.2966&0.2523&0.3257&0.0561&0.0309&0.9571&0.7846\\
		$n=50$&0.2083&0.2374&0.0250&0.0197&0.3684&0.3008&-1.1319&-0.9168&0.1691&0.2039&0.0335&0.0213&0.7112&0.5986\\
		$n=100$&0.0801&0.0824&0.0191&0.0135&0.1770&0.1389&-0.4093&-0.3267&0.0813&0.0905&0.0228&0.0195&0.5025&0.4289\\
		$n=150$&0.0358&0.0434&0.0051&0.0021&0.1011&0.0851&-0.2259&-0.1848&0.0385&0.0470&0.0063&0.0041&0.4140&0.3623\\
		$n=250$&0.0151&0.0265&-0.0022&0.0028&0.0541&0.0521&-0.1255&-0.1011&0.0184&0.0268&-0.0017&0.0029&0.3686&0.3268\\
		$n=500$&0.0129&0.0150&0.0050&0.0046&0.0331&0.0275&-0.0560&-0.0473&0.0145&0.0163&0.0049&0.0034&0.3449&0.3056\\[.4em]
		\multicolumn{15}{c}{\textbf{Root mean squared error}}\\[.2em]
		$n=35$&17263.26&15552.08&17223.34&15587.83&17137.54&15667.69&2.97$\times10^{13}$&2.95$\times10^{13}$&59.16&50.35&59.00&50.44&58.45&50.95\\
		$n=50$&7996.07&7382.94&7982.45&7395.00&7957.28&7418.68&5.15$\times10^{12}$&5.62$\times10^{12}$&27.55&24.08&27.52&24.13&27.32&24.35\\
		$n=100$&1331.57&1055.16&1330.24&1056.18&1328.51&1057.59&4.41$\times10^{10}$&3.36$\times10^{10}$&6.15&5.21&6.22&5.27&6.26&5.37\\
		$n=150$&36.30&32.42&36.27&32.44&36.24&32.48&1.11$\times10^{6}$&1.06$\times10^6$&2.46&2.13&2.56&2.20&2.70&2.34\\
		$n=250$&0.77&0.66&0.75&0.63&0.78&0.66&0.58&0.49&0.92&0.79&1.01&0.85&1.26&1.07\\
		$n=500$&0.46&0.39&0.46&0.38&0.47&0.40&0.42&0.35&0.49&0.41&0.50&0.42&0.77&0.66\\[.4em]
		\multicolumn{15}{c}{\textbf{Mean absolute deviation}}\\[.2em]
		$n=35$&2.1893&2.0002&1.5119&1.2537&2.0914&1.7082&0.5845&0.3672&1.7446&1.4744&1.5891&1.2890&2.5445&2.0878\\
		$n=50$&1.5636&1.4044&1.2510&1.0720&1.5649&1.3200&0.4261&0.3293&1.3908&1.2241&1.2901&1.0831&1.9384&1.6231\\
		$n=100$&0.9693&0.8220&0.8979&0.7479&1.0042&0.8306&0.5443&0.4800&0.9576&0.8300&0.9091&0.7685&1.2615&1.0552\\
		$n=150$&0.7571&0.6546&0.7291&0.6191&0.7807&0.6627&0.5752&0.4942&0.7685&0.6633&0.7396&0.6308&0.9975&0.8454\\
		$n=250$&0.5871&0.4942&0.5737&0.4782&0.5991&0.4995&0.5058&0.4256&0.5959&0.4984&0.5810&0.4827&0.7737&0.6368\\
		$n=500$&0.4084&0.3440&0.4041&0.3390&0.4130&0.3456&0.3818&0.3200&0.4127&0.3516&0.4076&0.3452&0.5295&0.4502\\
		\bottomrule
	\end{tabular}
	\caption{Performances of point estimators.}
	\label{ch2:tab:lomax6}
\end{table}
\end{landscape}
\newpage\newpage
\subsection{Random intercept and random slope linear mixed model}\label{ch2:app:mlm}
\begin{table}[hbt]
	\centering
	\begin{tabular}{@{}ccc@{}}
		\toprule
		&SwiZs&Parametric bootstrap\\
		\midrule
		$N=25$&1.87&0.20\\
		$N=100$&6.49&0.73\\
		$N=400$&35.60&4.58\\
		$N=1,600$&245.59&37.80\\
		\bottomrule
	\end{tabular}
	\caption{Average computational time in seconds to approximate a distribution on $S=10,000$ points.}
	\label{ch2:tab:mlm0}
\end{table}
\begin{table}[hbt]
	\centering
	\begin{tabular}{@{}cccccccccccc@{}}
		\toprule
		&\multicolumn{5}{c}{SwiZs} & \multicolumn{5}{c}{parametric bootstrap}\\
		$\alpha$ & $\beta_0$ & $\beta_1$ & $\sigma^2_{\epsilon}$&$\sigma^2_{\alpha}$&$\sigma^2_{\gamma}$
			 & $\beta_0$ & $\beta_1$ & $\sigma^2_{\epsilon}$&$\sigma^2_{\alpha}$&$\sigma^2_{\gamma}$ \\
		\midrule
		\multicolumn{11}{c}{$n=5$ $m=5$} \\[.2em]
		50\%&51.78&53.87&48.54&54.18 &70.38&42.37&43.61&44.60&32.27&28.10\\
		75\%&76.89&78.87&73.58&81.67 &89.09&64.17&66.19&66.20&48.35&41.80\\
		90\%&91.87&92.93&88.89&94.10 &98.80&78.38&81.94&81.07&61.72&46.87\\
		95\%&96.45&97.04&94.32&97.83 &99.98&84.58&88.45&86.61&68.68&47.30\\
		99\%&99.54&99.71&98.73&99.87&100.00&91.93&95.40&93.54&79.03&47.61\\[.4em]
		\multicolumn{11}{c}{$n=10$ $m=10$}\\[.2em]
		50\%&50.10&51.20&50.70&50.65&62.48&46.25&45.37&50.05&40.01&39.84\\
		75\%&75.16&77.08&74.92&75.64&85.74&69.81&68.68&74.48&60.54&59.68\\
		90\%&90.38&92.03&90.20&90.61&95.49&84.81&84.32&88.65&75.01&73.29\\
		95\%&95.23&96.40&95.23&94.96&97.86&90.71&90.32&93.95&81.30&79.29\\
		99\%&99.16&99.54&99.25&99.09&99.64&96.45&96.76&98.41&89.37&84.71\\[.4em]
		\multicolumn{11}{c}{$n=20$ $m=20$}\\[.2em]
		50\%&50.78&49.10&49.97&49.74&49.85&49.03&47.58&49.63&45.40&45.75\\
		75\%&75.28&74.45&75.24&74.89&75.88&73.08&71.87&75.06&67.66&66.98\\
		90\%&90.06&89.79&89.95&90.28&90.75&87.59&87.02&89.73&81.76&81.83\\
		95\%&95.05&94.83&94.79&95.06&95.97&93.10&92.69&94.59&87.48&87.52\\
		99\%&98.96&98.97&98.93&98.90&99.50&97.77&97.82&98.75&94.20&94.15\\[.4em]
		\multicolumn{11}{c}{$n=40$ $m=40$}\\[.2em]
		50\%&49.52&48.48&49.80&52.42&53.19&49.41&48.92&49.94&47.47&47.95\\
		75\%&74.70&72.86&75.27&77.89&78.39&74.22&73.34&75.63&70.93&71.46\\
		90\%&90.07&88.10&89.69&91.81&92.46&89.30&87.99&89.70&85.62&86.34\\
		95\%&95.15&94.09&94.71&96.27&96.59&94.37&93.65&94.82&91.29&91.82\\
		99\%&99.01&98.62&98.99&99.37&99.43&98.56&98.39&98.90&96.80&96.67\\
		\bottomrule
	\end{tabular}
	\caption{Estimated coverage probabilities.}
	\label{ch2:tab:mlm1}
\end{table}
\begin{table}
	\centering
	\begin{tabular}{@{}cccccccccccc@{}}
		\toprule
		&\multicolumn{5}{c}{SwiZs} & \multicolumn{5}{c}{parametric bootstrap}\\
		$\alpha$ & $\beta_0$ & $\beta_1$ & $\sigma^2_{\epsilon}$&$\sigma^2_{\alpha}$&$\sigma^2_{\gamma}$
			 & $\beta_0$ & $\beta_1$ & $\sigma^2_{\epsilon}$&$\sigma^2_{\alpha}$&$\sigma^2_{\gamma}$ \\
		\midrule
		\multicolumn{11}{c}{$n=5\;m=5$}\\[.2em]
		50\%&0.3303&0.2243&0.4976&1.2050&0.1755&0.2712&0.1728&0.4453&1.5575&0.0005\\
		75\%&0.5940&0.3882&0.8552&2.0974&0.4491&0.4606&0.2947&0.7607&3.5624&0.0012\\
		90\%&0.9314&0.5682&1.2436&3.1286&1.1761&0.6577&0.4217&1.0909&12.9753&0.0024\\
		95\%&1.1956&0.6934&1.5222&3.9149&3.7094&0.7845&0.5031&1.3051&13.9626&0.0036\\
		99\%&1.8698&1.0031&2.3468&9.8944&8.6739&1.0290&0.6623&1.7335&15.3409&0.0070\\[.4em]
		\multicolumn{11}{c}{$n=10\;m=10$}\\[.2em]
		50\%&0.2230&0.1198&0.2136&0.7311&1.0080&0.2038&0.1069&0.2099&0.7676&1.6745\\
		75\%&0.3902&0.2068&0.3638&1.2540&1.8614&0.3471&0.1818&0.3594&1.3370&8.6134\\
		90\%&0.5817&0.3008&0.5210&1.8131&2.9290&0.4953&0.2601&0.5144&1.9844&11.7988\\
		95\%&0.7162&0.3658&0.6218&2.1764&3.9196&0.5887&0.3097&0.6140&2.4462&12.6107\\
		99\%&1.0284&0.5130&0.8177&2.8992&7.9667&0.7745&0.4075&0.8055&3.6688&13.8600\\[.4em]
		\multicolumn{11}{c}{$n=20\;m=20$}\\[.2em]
		50\%&0.1547&0.0699&0.1006&0.4750&0.5665&0.1482&0.0674&0.0998&0.4733&0.6557\\
		75\%&0.2672&0.1205&0.1718&0.8065&0.9934&0.2530&0.1149&0.1708&0.8102&1.1462\\
		90\%&0.3900&0.1752&0.2455&1.1499&1.4857&0.3622&0.1643&0.2447&1.1655&1.7189\\
		95\%&0.4718&0.2117&0.2926&1.3701&1.8096&0.4311&0.1957&0.2918&1.3964&2.1535\\
		99\%&0.6436&0.2894&0.3833&1.8121&2.4686&0.5645&0.2569&0.3825&1.8686&3.4277\\[.4em]
		\multicolumn{11}{c}{$n=40\;m=40$}\\[.2em]
		50\%&0.1056&0.0452&0.0490&0.2816&0.1124&0.1056&0.0451&0.0493&0.3194&0.3628\\
		75\%&0.1810&0.0772&0.0834&0.4466&0.3469&0.1804&0.0770&0.0839&0.5429&0.6249\\
		90\%&0.2596&0.1107&0.1191&0.6923&0.6031&0.2576&0.1102&0.1197&0.7759&0.9014\\
		95\%&0.3100&0.1323&0.1420&0.8523&0.7672&0.3070&0.1313&0.1423&0.9257&1.0804\\
		99\%&0.4094&0.1747&0.1870&1.1467&1.1309&0.4020&0.1724&0.1864&1.2163&1.4420\\
		\bottomrule
	\end{tabular}
	\caption{Estimated median interval length.}
	\label{ch2:tab:mlm2}
\end{table}
\begin{landscape}
\begin{table}
	\centering
	\footnotesize
	\begin{tabular}{@{}cccccccccccccccc@{}}
		\toprule
		&\multicolumn{5}{c}{SwiZs: mean}&\multicolumn{5}{c}{SwiZs: median}&
		\multicolumn{5}{c}{Maximum likelihood}\\
		& $\beta_0$ & $\beta_1$ & $\sigma^2_{\epsilon}$&$\sigma^2_{\alpha}$&$\sigma^2_{\gamma}$
		& $\beta_0$ & $\beta_1$ & $\sigma^2_{\epsilon}$&$\sigma^2_{\alpha}$&$\sigma^2_{\gamma}$
		& $\beta_0$ & $\beta_1$ & $\sigma^2_{\epsilon}$&$\sigma^2_{\alpha}$&$\sigma^2_{\gamma}$\\
		\midrule
		\multicolumn{16}{c}{\textbf{Mean bias}$\times 100$}\\[.2em]
		$N=25$&-0.0647&-0.3827&-3.1193&1.8554&1.5502&-0.0761&-0.3732&-2.4630&6.4149&3.3175&-0.0708&-0.4203&-1.2985&-5.8224&-0.3807\\
		$N=100$&0.2843&-0.0320&-0.2911&2.4583&0.6119&1.6374&-0.1452&0.7182&-1.8475&1.8127&0.0685&0.0314&-0.0166&-2.8806&-0.6425\\
		$N=400$&0.0163&0.0374&0.0739&1.2927&0.0944&0.0149&0.0386&0.0514&0.9056&0.1565&0.0245&0.0417&0.0133&-1.3425&-0.2785\\
		$N=1,600$&0.0010&0.0385&0.0183&-0.9811&-0.2965&-0.0011&0.0394&0.0120&-1.1600&-0.2121&0.0130&0.0343&-0.0021&-0.6265&-0.1253\\[.4em]
		\multicolumn{16}{c}{\textbf{Median bias}$\times 100$}\\[.2em]
		$N=25$&-0.0341&-0.2171&-3.8669&-6.5130&-0.0876&-0.0018&-0.2114&-3.3736&-0.8483&0.0121&0.0327&-0.2932&-2.1012&-10.1138&-3.9990\\
		$N=100$&0.4345&0.0289&-0.4759&0.1208&-0.0951&5.3959&-1.4459&0.5589&-0.7598&0.0354&0.1838&0.0069&-0.1815&-4.8730&-1.2975\\
		$N=400$&0.0020&-0.0378&0.0422&0.4196&-0.1116&0.0149&-0.0286&0.0211&-0.0405&-0.0068&-0.0140&-0.0261&-0.0220&-2.1176&-0.4517\\
		$N=1,600$&-0.0332&0.0500&0.0082&-1.0639&-0.1813&-0.0060&0.0543&0.0041&-0.0818&-0.0021&-0.0098&0.0480&-0.0098&-1.1378&-0.1833\\[.4em]
		\multicolumn{16}{c}{\textbf{Root mean squared error}$\times 100$}\\[.2em]
		$N=25$&24.6914&16.0625&9.2357&27.0499&6.2389&24.7198&16.0766&8.6916&24.2014&8.3432&24.7291&16.0853&8.1605&18.5249&6.8108\\
		$N=100$&16.4663&8.8542&3.9374&14.7976&3.5251&14.3449&7.7017&4.1388&11.5080&3.3680&16.5630&8.7967&3.8703&12.0714&3.1774\\
		$N=400$&11.4174&5.2549&1.8779&9.1330&1.8623&11.4174&5.2550&1.8752&8.9859&1.7515&11.4182&5.2554&1.8689&8.2404&1.7092\\
		$N=1,600$&7.8721&3.4528&0.9119&4.7681&0.6698&7.9083&3.4524&0.9117&4.4706&0.5759&7.8981&3.4532&0.9110&5.7583&1.0216\\[.4em]
		\multicolumn{16}{c}{\textbf{Mean absolute deviation}$\times 100$}\\[.2em]
		$N=25$&24.4139&15.8780&8.2892&23.3025&0.6468&24.4872&15.9113&8.1000&17.1094&0.2293&24.4752&15.9014&7.8528&15.1530&0.0015\\
		$N=100$&16.7958&8.9936&3.8427&13.3232&2.8386&13.0610&6.2264&2.9059&8.0151&1.4453&16.9915&8.9079&3.8351&10.8654&3.0194\\
		$N=400$&11.2283&5.3202&1.8651&8.8004&1.8018&11.2417&5.3225&1.8695&8.8299&1.4204&11.2634&5.3160&1.8653&7.8895&1.6541\\
		$N=1,600$&7.9220&3.4259&0.9115&4.3804&0.5033&7.9954&3.4277&0.9108&0.2978&0.0214&7.9745&3.4325&0.9082&5.7040&0.9952\\
		\bottomrule
	\end{tabular}
	\caption{Performances of point estimators}
	\label{ch2:tab:mlm3}
\end{table}
\end{landscape}
\begin{table}[hbpt]
	\centering
	\begin{tabular}{@{}ccccccccccc@{}}
		\toprule
		&\multicolumn{5}{c}{Coverage probability} & \multicolumn{5}{c}{Median interval length}\\
		$\alpha$ & $\beta_0$ & $\beta_1$ & $\sigma^2_{\epsilon}$&$\sigma^2_{\alpha}$&$\sigma^2_{\gamma}$
			 & $\beta_0$ & $\beta_1$ & $\sigma^2_{\epsilon}$&$\sigma^2_{\alpha}$&$\sigma^2_{\gamma}$ \\
		\midrule
		\multicolumn{11}{c}{$n=5$ $m=5$} \\[.2em]
		50\%&43.16&44.94&48.66&40.42&36.49&0.2770&0.1791&0.1043&0.1868&0.0375\\
		75\%&67.51&69.17&73.83&64.17&70.73&0.4942&0.3180&0.1836&0.3625&0.0945\\
		90\%&83.68&86.75&88.79&81.88&96.33&0.7612&0.4897&0.2764&0.6358&0.2010\\
		95\%&90.37&93.23&93.83&88.93&98.88&0.9671&0.6226&0.3431&0.8982&0.3095\\
		99\%&97.04&98.93&98.54&96.95&99.75&1.4991&0.9746&0.4982&1.8138&0.7069\\
		\multicolumn{11}{c}{$n=10$ $m=10$} \\[.2em]
		50\%&46.38&45.98&50.75&45.86&44.91&0.2060&0.1082&0.0525&0.1422&0.0383\\
		75\%&70.85&71.03&75.36&70.65&68.84&0.3591&0.1888&0.0901&0.2583&0.0690\\
		90\%&87.23&87.08&90.04&86.58&85.82&0.5321&0.2806&0.1304&0.4088&0.1078\\
		95\%&93.20&93.27&95.12&92.37&93.09&0.6534&0.3449&0.1569&0.5299&0.1392\\
		99\%&98.41&98.53&98.95&98.02&99.59&0.9264&0.4903&0.2111&0.8593&0.2265\\
		\multicolumn{11}{c}{$n=20$ $m=20$} \\[.2em]
		50\%&49.20&47.62&49.92&48.00&47.31&0.1491&0.0677&0.0251&0.1048&0.0216\\
		75\%&73.66&72.54&75.09&72.49&72.86&0.2571&0.1168&0.0429&0.1845&0.0381\\
		90\%&88.70&88.34&89.97&88.33&88.10&0.3742&0.1700&0.0616&0.2774&0.0573\\
		95\%&94.09&94.02&94.81&93.80&93.72&0.4524&0.2055&0.0735&0.3445&0.0712\\
		99\%&98.56&98.61&98.94&98.40&98.59&0.6167&0.2801&0.0972&0.5019&0.1038\\
		\multicolumn{11}{c}{$n=40$ $m=40$} \\[.2em]
		50\%&49.46&49.32&49.79&48.67&49.01&0.1060&0.0452&0.0122&0.0748&0.0136\\
		75\%&74.46&73.78&75.28&73.52&74.77&0.1819&0.0776&0.0209&0.1295&0.0236\\
		90\%&89.88&88.76&89.70&88.89&89.83&0.2623&0.1119&0.0299&0.1899&0.0346\\
		95\%&94.95&94.28&94.85&94.22&94.71&0.3148&0.1343&0.0356&0.2310&0.0420\\
		99\%&98.98&98.86&98.99&98.77&98.82&0.4212&0.1797&0.0468&0.3194&0.0582\\
	\end{tabular}
	\caption{Asymptotic results}
	\label{ch2:tab:mlm4}
\end{table}
\newpage
\subsection{M/G/1 queueing model}\label{ch2:app:gg1}
\begin{table}[hbpt]
	\centering
	\begin{tabular}{@{}cccccccccc@{}}
		\toprule
		&\multicolumn{3}{c}{SwiZs}&\multicolumn{3}{c}{Indirect inference}&\multicolumn{3}{c}{Parametric bootstrap}\\
		&$\theta_1$&$\theta_2$&$\theta_3$&$\theta_1$&$\theta_2$&$\theta_3$&$\theta_1$&$\theta_2$&$\theta_3$\\
		\midrule
		50\%&46.92&38.68&56.73&40.59&9.95&54.59&18.31&10.23&20.96\\
		75\%&71.56&55.41&81.80&68.01&34.11&84.50&32.70&20.96&37.38\\
		90\%&87.55&67.77&94.47&87.62&57.13&96.04&48.62&35.24&53.71\\
		95\%&93.16&74.78&97.97&94.66&70.22&98.75&57.05&46.03&63.21\\
		99\%&98.17&90.06&99.90&98.84&94.89&99.94&71.99&65.43&77.64\\
		\bottomrule
	\end{tabular}
	\caption{Estimated coverage probabilities.}
	\label{ch2:tab:gg11}
\end{table}
\begin{table}[hbpt]
	\centering
	\begin{tabular}{@{}cccccccccc@{}}
		\toprule
		&\multicolumn{3}{c}{SwiZs}&\multicolumn{3}{c}{Indirect inference}&\multicolumn{3}{c}{Parametric bootstrap}\\
		&$\theta_1$&$\theta_2$&$\theta_3$&$\theta_1$&$\theta_2$&$\theta_3$&$\theta_1$&$\theta_2$&$\theta_3$\\
		\midrule
		50\%&0.0235&0.0805&0.1379&0.0382&0.0468&0.1368&0.0263&0.0420&0.1134\\
		75\%&0.0404&0.1467&0.2357&0.0911&0.0978&0.2389&0.0460&0.0757&0.2051\\
		90\%&0.0585&0.2207&0.3378&0.1563&0.1914&0.3835&0.0708&0.1185&0.3131\\
		95\%&0.0705&0.2733&0.4032&0.2225&0.2952&0.5432&0.0895&0.1533&0.3855\\
		99\%&0.0952&0.3934&0.5407&0.5331&0.7152&1.6084&0.1327&0.2514&0.5562\\
		\bottomrule
	\end{tabular}
	\caption{Estimated median interval length.}
	\label{ch2:tab:gg12}
\end{table}
\begin{table}[hbpt]
	\centering
	\begin{tabular}{@{}ccccccc@{}}
		\toprule
		&\multicolumn{3}{c}{SwiZs: starting value is $\bto$}&\multicolumn{3}{c}{SwiZs: sample size is $n=1,000$.}\\
		&$\theta_1$&$\theta_2$&$\theta_3$&$\theta_1$&$\theta_2$&$\theta_3$\\
		\midrule
		50\%&50.22&58.64&49.98&50.07&46.06&49.37\\
		75\%&75.24&91.25&74.24&75.24&71.82&74.77\\
		90\%&90.52&99.82&89.55&89.73&89.84&89.49\\
		95\%&95.37&100.00&94.87&94.81&95.41&94.69\\
		99\%&99.09&100.00&99.02&98.95&99.28&99.10\\
		\bottomrule
	\end{tabular}
	\caption{Estimated coverage probabilities under different conditions than Table~\ref{ch2:tab:gg11}.}
	\label{ch2:tab:gg13}
\end{table}
\begin{landscape}
\begin{table}
	\centering
	\footnotesize
	\begin{tabular}{@{}cccccccccccccccc@{}}
		\toprule
		&\multicolumn{3}{c}{SwiZs: mean}&\multicolumn{3}{c}{SwiZs: median}&
		\multicolumn{3}{c}{Indirect inference}&\multicolumn{3}{c}{Indirect inference: mean}&
		\multicolumn{3}{c}{Indirect inference: median}\\
		&$\theta_1$&$\theta_2$&$\theta_3$&$\theta_1$&$\theta_2$&$\theta_3$&
		$\theta_1$&$\theta_2$&$\theta_3$&$\theta_1$&$\theta_2$&$\theta_3$&$\theta_1$&$\theta_2$&$\theta_3$\\
		\midrule
		Mean bias&0.0037&-0.0149&0.0006&0.0057&-0.0096&0.0002&2$\times10^{90}$&3$\times10^{90}$&1.6107&0.0309&0.0254&3$\times10^{89}$&0.0157&0.0297&0.0201\\
		Median bias&0.0026&-0.0219&-0.0044&0.0046&-0.0157&-0.0041&0.0135&0.0270&0.0181&0.0295&0.0235&0.0772&0.0150&0.0257&0.0200\\
		RMSE&0.0197&0.0764&0.0890&0.0200&0.0762&0.0888&2$\times10^{92}$&3$\times10^{92}$&135.72&0.0451&0.0976&3$\times10^{91}$&0.0254&0.1041&0.0851\\
		MAD&0.0192&0.0705&0.0884&0.0190&0.0718&0.0882&0.0307&0.1069&0.1405&0.0365&0.0918&0.1109&0.0182&0.0968&0.0823\\
		\bottomrule
	\end{tabular}
	\caption{Performances of point estimator.}
	\label{ch2:tab:gg14}
\end{table}
\end{landscape}

\section{Generic results}
This chapter assembles some generic theoretical results useful for the other Chapters.

We generically denote $\{\g_n:n\geq1\}$ a sequence of a random vector-valued function
and $\bt\in\bT$ a vector of parameters.

The next Lemma is Theorem 5.9 in~\cite{van1998asymptotic}. The proof is given for the sake of completeness.
\begin{lemma}[weak consistency]
	Let $\{\g_n(\bt)\}$ be sequence of a random vector-valued function of vector parameter $\bt$
	with a deterministic limit $\g(\bt)$. If $\bT$ is compact, if the random function sequence converges 
	uniformly as $n\to\infty$
	\begin{equation}
		\sup_{\bt\in\bT}\left\lVert\g_n(\bt)-\g(\bt)\right\rVert\wcp0,
		\label{eq:uni_conv}
	\end{equation}
	and if there exist $\delta>0$ such that
	\begin{equation}		
		\inf_{\bt\notin\B(\bto,\delta)}\left\lVert\g(\bt)\right\rVert>0=\left\lVert\g(\bto)\right\rVert,
		\label{eq:unique}
	\end{equation}
	then any sequence of estimators $\{\hbt_n\}$ converges weakly in probability to $\bto$.
	\label{thm:lemma1}
\end{lemma}
\begin{proof}
	Choose $\hbt_n$ that nearly minimises $\lVert\g_n(\bt)\rVert$ so that
	\begin{equation*}
		\left\lVert\g_n(\hbt_n)\right\rVert\leq\inf_{\bt\in\bT}\left\lVert\g_n(\bt)\right\rVert + \op(1)
	\end{equation*}
	Clearly we have $\inf_{\bt}\rVert\g_n(\bt)\rVert\leq\rVert\g_n(\bto)\rVert$, and by \eqref{eq:uni_conv}
	$\lVert\g_n(\bto)\rVert\wcp\lVert\g(\bto)\rVert$ so that
	\begin{equation*}
		\left\lVert\g_n(\hbt_n)\right\rVert\leq\left\lVert\g(\bto)\right\rVert + \op(1)
	\end{equation*}
	Now, substracting both sides by $\lVert\g(\hbt_n)\rVert$, we have by the reverse triangle inequality
	\begin{equation*}
		-\left\lVert\g_n(\hbt_n)-\g(\hbt_n)\right\rVert \leq 
		\left\lVert\g(\bto)\right\rVert-\left\lVert\g(\hbt_n)\right\rVert + \op(1)
	\end{equation*}
	The left-hand side is bounded by the negative supremum, thus
	\begin{equation*}
		\left\lVert\g(\bto)\right\rVert-\left\lVert\g(\hbt_n)\right\rVert \geq
		-\sup_{\bt\in\bT}\left\lVert\g_n(\bt)-\g(\bt)\right\rVert - \op(1) 
	\end{equation*}
	It follows from~\eqref{eq:uni_conv} that the limit in probability of the right-hand side tends to 0.
	Let $\varepsilon>0$ and choose a $\delta>0$ as in~\eqref{eq:unique} so that
	\begin{equation*}
		\left\lVert\g(\bt)\right\rVert > \left\lVert\g(\bto)\right\rVert - \varepsilon
	\end{equation*}
	for every $\bt\notin\B(\bto,\delta)$. If $\hbt_n\notin\B(\bto,\delta)$, we have
	\begin{equation*}
		\left\lVert\g(\bto)\right\rVert-\left\lVert\g(\hbt_n)\right\rVert<\varepsilon
	\end{equation*}
	The probability of this event converges to 0 as $n\to\infty$.
\end{proof}
The next definition is taken from~\cite{andrews1992generic} (see also~\cite[Chapter 7.1]{pollard1984convergence})
\begin{definition}
	$\{\g_n(\bt)\}$ is stochastically uniformly equicontinuous on $\bT$
	if for every $\varepsilon>0$ there exist a real $\delta>0$ such that
	\begin{equation}
		\limsup_{n\to\infty}\Pr \left(\sup_{\bt\in\bT} 
		\sup_{\bt'\in\B(\bt,\delta)}\left\lVert \g_n(\bt') - \g_n(\bt) \right\rVert
		> \varepsilon \right) < \varepsilon
		\label{eq:equicontinuity}
	\end{equation}
	\label{def:equicontinuity}
\end{definition}
\begin{lemma}[uniform consistency]\label{thm:uniform}
	If $\bT$ is compact, if the sequence of random vector-valued function $\{\g_n(\bt)\}$
	is pointwise convergent for all $\bt\in\bT$ and is stochastically uniformly 
	equicontinuous on $\bT$, then 
	\begin{enumerate}[label=\roman*.]
		\item $\{\g_n(\bt)\}$ converges uniformly,
		\item $\g$ is uniformly continuous.
	\end{enumerate}
\end{lemma}
\begin{proof}
	\textit{(i)} (Inspired from~\cite[Theorem 7.25(b)]{rudin1976principles}). 
	Let $\varepsilon>0$, choose $\delta>0$ so to satisfy stochastic uniform 
	equicontinuity in \eqref{eq:equicontinuity}. Let $\B(\bt,\delta)=\{\bt'\in\bT:d(\bt,\bt')<\delta\}$.
	Since $\bT$ is compact, there are finitely many points $\bt_1,\dots,\bt_k$ in $\bT$
	such that 
	\begin{equation*}
		\bT\subset\B(\bt_1,\delta)\cup\dots\cup\B(\bt_k,\delta)
	\end{equation*}
	Since $\{\g_n(\bt)\}$ converges pointwise for every $\bt\in\bT$, we have
	\begin{equation*}
		\limsup_{n\to\infty}\Pr\left( \left\lVert\g_n(\bt_l)-\g(\bt_l)\right\rVert>\varepsilon \right)<\varepsilon,
	\end{equation*}
	whenever $1\leq l\leq k$. If $\bt\in\bT$, so $\bt\in\B(\bt_l,\delta)$ for some $l$, so that
	\begin{equation*}
		\limsup_{n\to\infty}\Pr\left( \left\lVert \g_n(\bt_l) - \g_n(\bt)
		\right\rVert>\varepsilon \right) \leq \limsup_{n\to\infty}\Pr\left(\sup_{\bt\in\bT}\sup_{\bt'\in\B(\bt,\delta)}\left\lVert\g_n(\bt)-\g_n(\bt') \right\rVert  \right)< \varepsilon
	\end{equation*}
	Then, by the triangle inequality we have
	\begin{align*}
		& \limsup_{n\to\infty}\Pr\left(\sup_{\bt\in\bT}\left\lVert\g_n(\bt)-\g(\bt)\right\rVert>\varepsilon\right) \\ 
		&\quad \leq \limsup_{n\to\infty}\Pr\left(\sup_{\bt\in\bT}\sup_{\bt'\in\B(\bt,\delta)}\left\lVert\g_n(\bt)-\g_n(\bt')\right\rVert>\varepsilon\right) \\
		&\qquad + \limsup_{n\to\infty}\Pr\left(\left\lVert\g_n(\bt')-\g(\bt')\right\rVert>\varepsilon\right) +
		\Pr\left(\sup_{\bt\in\bT}\sup_{\bt'\in\B(\bt,\delta)}\left\lVert\g(\bt)-\g(\bt')\right\rVert>\varepsilon\right) < 3\varepsilon
	\end{align*}
	\textit{(ii)}. The proof follows the same steps.
\end{proof}
The next Lemma is similar to~\cite[Lemma 1]{andrews1992generic}.
The result of~\cite{andrews1992generic} is on the difference between
a random and a nonrandom functions and requires the extra assumption
of absolute continuity of the nonrandom function.
The proof provided here is also different.
\begin{lemma}\label{thm:equicontinuity}
	If for all $\bt,\bt'\in\bT$, $\lVert\g_n(\bt)-\g_n(\bt')\rVert\leq B_n d(\bt,\bt')$ with $B_n=\Op(1)$, then 
	$\{\g_n(\bt)\}$ is stochastically uniformly equicontinuous.
\end{lemma}
\begin{proof}
	By $B_n=\Op(1)$, there is $M>0$ such that for all $n$, $\Pr(\lvert B_n\rvert>M)<\varepsilon$.
	Let $\varepsilon>0$ and choose a sufficiently small $\delta>0$ 
	such that for all $\bt',\bt\in\bT$, $d(\bt,\bt')<\varepsilon/M=\tau$, $\delta\leq\tau$.
	Let $\B(\bt,\delta)=\{\bt'\in\bT:d(\bt,\bt')<\delta\}$.
	Then, we have
	\begin{align*}
		& \limsup_{n\to\infty}\Pr\left(\sup_{\bt\in\bT}\sup_{\bt'\in\B(\bt,\delta)}\left\lVert\g_n(\bt)-\g_n(\bt')\right\rVert>\varepsilon\right) \\
		& \quad \leq \limsup_{n\to\infty}\Pr\left( B_n\sup_{\bt\in\bT}\sup_{\bt'\in\B(\bt,\delta)}d(\bt,\bt')>\varepsilon \right) \\
		& \qquad \leq \limsup_{n\to\infty}\Pr\left( B_n\tau>\varepsilon \right) \leq \limsup_{n\to\infty}\Pr\left(\lvert B_n\rvert>M \right) < \varepsilon
	\end{align*}
\end{proof}
The next Lemma is a special case of~\cite[Corollary 3.1]{newey1991uniform}.
\begin{lemma}
	Let $\{\x_i:i\geq1\}$ be an i.i.d.\ sequence of random variable and let 
	$\g_n(\bt)=n^{-1}\sum_{i=1}^n\g(\x_i,\bt)$. If for all $i=1,\dots,n$ and $\bt,\bt'\in\bT$,
	$\lVert\g(\x_i,\bt)-\g(\x_i,\bt')\lVert\leq b_n(\x_i)d(\bt,\bt')$ with 
	$\E[b_n(\x_i)]=\mu_n=\mathcal{O}(1)$, then $\{\g_n(\bt)\}$ is stochastically
	uniformly equicontinuous.\label{thm:uniform_lln}
\end{lemma}
\begin{proof}
	Let $B_n = n^{-1}\sum_{i=1}^n b_n(\x_i)$, so $\E[B_n]=\mathcal{O}(1)$. We have by triangle inequality
	\begin{equation*}
		\left\lVert\g_n(\bt)-\g_n(\bt') \right\rVert\leq\frac{1}{n}\sum_{i=1}^n\left\lVert\g(\x_i,\bt)-\g(\x_i,\bt')\right\rVert
		\leq B_nd(\bt,\bt')
	\end{equation*}
	The rest of the proof follows from Lemma~\ref{thm:equicontinuity}.
\end{proof}
\begin{lemma}[uniform weak law of large number]
	If, in addition to Lemma~\ref{thm:uniform_lln}, for each $\bt\in\bT$, $\g_n(\bt)$ is pointwise
	convergent, then $\{\g_n(\bt)\}$ converges uniformly.
	\label{thm:ulln}
\end{lemma}
\begin{proof}
	The proof is an immediat consequence of Lemma~\ref{thm:uniform_lln} and Lemma~\ref{thm:uniform}. 
\end{proof}
%
%
%
%
The next Lemma is essentially a combination of Theorem 4.2 and Corollary 4.3 in~\cite{lang1993real}.
The proof is given for the sake of completeness.
\begin{lemma}[mean value inequality]
	Let $U$ be a convex open set in $\bT$. Let $\bt_1\in U$ and $\bt_2\in\bT$.
	If $\g:U\to F$ is a $\mathcal{C}^{1}$-mapping, then
	\begin{enumerate}[label=\roman*.]
		\item $\g(\bt_1+\bt_2) - \g(\bt_1) = \int_0^1 D\g(\bt_1+t\bt_2)dt\cdot\bt_2$
		\item $\lVert\g(\bt_1+\bt_2)-\g(\bt_1)\rVert\leq\sup_{0\leq t\leq1}\lVert D\g(\bt_1+t\bt_2)\rVert\cdot\lVert\bt_2\rVert$
	\end{enumerate}\label{thm:mvt}
\end{lemma}
\begin{proof}
	\textit{(i)}. Fix $\bt_1\in U$, $\bt_2\in\bT$. Let $\bt_3=\bt_1+\bt_2$ and $\lambda_t = (1-t)\bt_1 + t\bt_3$.
	For $t\in[0,1]$ we have by the convexity of $U$ that $\lambda_t\in U$, and so $\bt_1+t\bt_2$
	is in $U$ as well. Put $\h(t) = \g(\bt_1+t\bt_2)$, so $D\h(t) = D\g(\bt_1+t\bt_2)\cdot\bt_2$.
	By the fundamental theorem of calcul we have that
	\begin{equation*}
		\int_0^1D\h(t)\d t = \h(1) - \h(0)
	\end{equation*}
	Since $\h(1)=\g(\bt_1+\bt_2)$, $\h(0)=\g(\bt_1)$, and $\bt_2$ is allowed to be pulled out of the integral,
	part \textit{(i)} is proven.\\
	\textit{(ii)}. We have that
	\begin{align*}
		\left\lVert\g(\bt_1+\bt_2)-\g(\bt_1)\right\rVert 
		& \leq\left\lVert\int_0^1D\g(\bt_1+t\bt_2)\d t\right\rVert\cdot\left\lVert\bt_2\right\rVert, \\
		& \leq\lvert(1-0)\rvert\sup_{0\leq t\leq1}\left\lVert D\g(\bt_1+t\bt_2)\right\rVert\cdot\left\lVert\bt_2\right\rVert,
	\end{align*}
	where we use the Cauchy-Schwarz inequality for the first inequality, and the upper bound of integral for the second.
	The supremum of the norm exists because the affine line $\bt_1+t\bt_2$ is compact and the Jacobian is continuous.
\end{proof}
\begin{lemma}[delta method]
	If conditions of Lemma~\ref{thm:mvt} holds, then 
	\begin{equation*}
		\g(\bt_1+\bt_2)-\g(\bt_1) = D\g(\bt_1)\cdot\bt_2 + o\left( \lVert\bt_2\rVert \right)
	\end{equation*}
	\label{thm:fot}
\end{lemma}
\begin{proof}
	Fix $\bt_1\in U$ and $\bt_2\in\bT$. By Lemma~\ref{thm:mvt}, we have
	\begin{equation*}
		\left\lVert\int_0^1D\g(\bt_1+t\bt_2)\d t\right\rVert\leq\sup_{0\leq t\leq1}\left\lVert D\g(\bt_1+t\bt_2)\right\rVert
	\end{equation*}
	Let $\bt_3=\bt_1+\bt_2$ so $\lambda_t=(1-t)\bt_1+t\bt_3$, $t\in[0,1]$, is in $U$ and $\bt_1+t\bt_2$ as well. 
	Let $\B^c(\bt_1,\lVert\bt_2\rVert)=\{\lVert\bt_1-\bt\rVert\leq\lVert\bt_2\rVert\}$. 
	We have
	\begin{align*}
		\left\lVert t\bt_1+(1-t)\bt_3-\bt \right\rVert 
		&\leq t\left\lVert\bt_1-\bt\right\rVert + (1-t)\left\lVert\bt_3-\bt \right\rVert \\
		&\leq t\lVert\bt_2\rVert + (1-t)\lVert\bt_2\rVert = \lVert\bt_2\rVert,
	\end{align*}
	so the line segment $\lambda_t$ is in the closed ball. Hence, we have
	\begin{equation*}
		\left\lVert\int_0^1D\g(\bt_1+t\bt_2)\d t\right\rVert\leq\sup_{\bt\in\B(\bt_1,\lVert\bt_2\rVert)}\left\lVert D\g(\bt)\right\rVert
	\end{equation*}
	Eventually, we have by continuity of the Jacobian in a neighborhood of $\bt_1$ that
	\begin{equation*}
		\sup_{\bt\in\B(\bt_1,\lVert\bt_2\rVert)}\left\lVert D\g(\bt)-D\g(\bt_1)\right\rVert\rightarrow 0
	\end{equation*}
	as $\lVert\bt_2\rVert\rightarrow 0$.
\end{proof}

\begin{lemma}[asymptotic normality]
	Let $U$ be a convex open set in $\bT$. Let $\{\hbt_n\}$ be a sequence of estimator (roots of)
	the mapping $\g_n:U\to F$. If 
	\begin{enumerate}[label=\roman*.]
		\item $\hbt_n$ converges in probability to $\bto\in U$,
		\item $\{\g_n\}$ is a $\mathcal{C}^1$-mapping,
		\item $n^{1/2}\g_n(\bto)\rightsquigarrow\mathcal{N}(\mathbf{0},\mathbf{V})$,
		\item $D\g_n(\bto)$ converges in probability to $\M$,
		\item $D\g_n(\bto)$ is nonsingular,
	\end{enumerate}
	then
	\begin{equation*}
		n^{1/2}(\hbt_n-\bto)\rightsquigarrow\mathcal{N}(\mathbf{0},\bS),
	\end{equation*}
	where $\bS = \M^{-1}\mathbf{V}\M^{-T}$.
	\label{thm:an}
\end{lemma}
\begin{proof}
	Fix $\bt_1=\bto$ and $\bt_2=\hbt_n-\bto$, from Lemma~\ref{thm:mvt} and Lemma~\ref{thm:fot} we have
	\begin{equation*}
		\g_n(\hbt_n) = \g_n(\bto) + D\g_n(\bto)\cdot(\hbt_n-\bto) + o_p(\lVert\hbt_n-\bto\rVert)
	\end{equation*}
	By definition $\g_n(\hbt_n)=\mathbf{0}$. Multiplying by square-root $n$ leads to
	\begin{equation*}
		n^{1/2}(\hbt_n-\bto) = - \left[ D\g_n(\bto) \right]^{-1}n^{1/2}\g_n(\bto)
		- n^{1/2}\left[ D\g_n(\bto) \right]^{-1}o_p\left( \lVert\hbt_n-\bto\rVert \right)
	\end{equation*}
	By the continuity of the matrix inversion $[D\g_n(\bto)]^{-1}\wcp\M^{-1}$.
	Since the central limit theorem holds for $n^{1/2}\g_n(\bto)$, the proof results from 
	Slutsky's lemma.
\end{proof}
The next Lemma is Theorem 9.4 in~\cite{loomis1968advanced} and is given without proof.
\begin{lemma}[implicit function theorem]\label{thm:ift}
	Let $\bX\times\bT$ be an open subset of $\R^m\times\R^p$.
	Let $\g:\bX\times\bT\rightarrow\R^p$ be a function of the form $\g(\bx,\bt)=k$.
	Let the solution at the points $(\bx_0,\bto)\in\bX\times\bT$ and $k_0\in\R^p$ be 
	\begin{equation*}
		\g(\bx_0,\bto)=k_0
	\end{equation*}
	If
	\begin{enumerate}[label=\roman*.]
		\item $\g$ is differentiable in $\bX\times\bT$,
		\item The partial derivative $D_{\bx}\g$ is continuous in $\bX\times\bT$,
		\item The partial derivative $D_{\bt}\g$ is invertible at the points $(\bx_0,\bto)\in\bX\times\bT$,
	\end{enumerate}
	then, there are neighborhoods $X\subset\bX$ and $O\subset\bT$ of $\bx_0$ and $\bto$
	on which the function $\hbt:O\rightarrow X$ is uniquely defined, and such that:
	\begin{enumerate}
		\item $\g(\bx,\hbt(\bx))=k_0$ for all $\bx\in X$,
		\item For each $\bx\in X$, $\hbt(\bx)$ is the unique solution lying in $O$ such that $\hbt(\bx_0)=\bto$,
		\item $\hbt$ is differentiable on $X$ and 
			\begin{equation*}
				D_{\bx}\hbt = -\left[D_{\bt}\g\right]^{-1}D_{\bx}\g
			\end{equation*}
	\end{enumerate}	
\end{lemma}

\end{appendix}
\end{document}